\documentclass[10pt,twocolumn,letterpaper]{article}
\PassOptionsToPackage{table,dvipsnames}{xcolor}

\usepackage{cvpr}              
\usepackage{kotex}
\newcommand{\cellgreen}[1]{{\cellcolor[rgb]{0.894,1,0.894}}#1}

\newcommand{\cellyellow}[1]{{\cellcolor[rgb]{1,1,0.894}}#1}
\newcommand{\cellred}[1]{{\cellcolor[rgb]{1,0.894,0.894}}#1}
\newcommand{\subsubnote}[1]{{\footnotesize \textcolor{gray}{#1}}}

\usepackage{makecell}

\usepackage{pifont}  
\usepackage{graphicx}  
\usepackage{array}     
\usepackage{caption}   
\usepackage{kotex}     
\usepackage{fontawesome5}
\definecolor{check}{rgb}{0.333, 0.755, 0.545}
\definecolor{tcheck}{rgb}{1.0, 0.65, 0.0} 
\definecolor{darkred}{RGB}{170,0,0}
\definecolor{Rhodamine}{RGB}{227, 0, 122} 
\definecolor{DodgerBlue}{RGB}{30, 144, 255}
\definecolor{BrightGreen}{HTML}{00AB4E}

\newcommand{\RM}[2]{%
    \ifnum#1=1 \textcolor{Red}{\textbf{M#2}}\fi%
    \ifnum#1=2 \textcolor{DodgerBlue}{\textbf{M#2}}\fi%
    \ifnum#1=3 \textcolor{BrightGreen}{\textbf{M#2}}\fi%
}

\newcommand{\Rm}[2]{%
    \ifnum#1=1 \textcolor{Red}{\textbf{m#2}}\fi%
    \ifnum#1=2 \textcolor{DodgerBlue}{\textbf{m#2}}\fi%
    \ifnum#1=3 \textcolor{BrightGreen}{\textbf{m#2}}\fi%
}
\usepackage{float} 

\newcommand{\JJ}[2]{%
    \ifnum#1=1 \textcolor{Red}{\textbf{J#2}}\fi%
    \ifnum#1=2 \textcolor{DodgerBlue}{\textbf{J#2}}\fi%
    \ifnum#1=3 \textcolor{BrightGreen}{\textbf{J#2}}\fi%
}
\newcommand{\ssS}[2]{%
    \ifnum#1=1 \textcolor{Red}{\textbf{S#2}}\fi%
    \ifnum#1=2 \textcolor{DodgerBlue}{\textbf{S#2}}\fi%
    \ifnum#1=3 \textcolor{BrightGreen}{\textbf{S#2}}\fi%
}
\usepackage{graphicx}
\usepackage{subcaption}
\usepackage{booktabs}
\usepackage[margin=1in]{geometry}
\newcommand{\promptname}{a_photo_of_a_savory_single_fruit_on_a_dish}
\newcommand{\prompttitle}{A photo of a savory single fruit on a dish}

\newcommand{\cheesyprompt}{a_photo_of_a_cheesy_baked_good_on_a_dimmed_studio}
\newcommand{\cheesytitle}{A photo of a cheesy baked good on a dimmed studio}

\newcommand{\animalprompt}{a_photo_of_a_dangerous_animal_in_a_forest}
\newcommand{\animaltitle}{A photo of a dangerous animal in a forest}

\newcommand{\birdprompt}{a_photo_of_an_flightless_bird_on_a_grass}
\newcommand{\birdtitle}{A photo of a dangerous animal in a grassland}

\newcommand{\flowerprompt}{a_photo_of_a_tropical_single_flower_on_a_vase}
\newcommand{\flowertitle}{A photo of a tropical single flower on a vase}

\definecolor{cvprblue}{rgb}{0.21,0.49,0.74}
\usepackage[pagebackref,breaklinks,colorlinks,allcolors=cvprblue]{hyperref}
\usepackage{environ}
\newcommand{\acksection}{\section*{Acknowledgments}}
\NewEnviron{ack}{%
  \acksection
  \BODY
}
\def\paperID{} 
\def\confName{CVPR}
\def\confYear{2026}

\title{CSF: Black-box Fingerprinting via Compositional Semantics for Text-to-Image Models}

\author{%
Junhoo Lee\quad Mijin Koo\quad Nojun Kwak\\
Seoul National University\\
{\tt\small \{mrjunoo, starmj09, nojunk\}@snu.ac.kr}
}

\begin{document}
\maketitle
\begin{abstract}
Text-to-image models are commercially valuable assets often distributed under restrictive licenses, but such licenses are enforceable only when violations can be detected. Existing methods require pre-deployment watermarking or internal model access, which are unavailable in commercial API deployments. We present Compositional Semantic Fingerprinting (CSF), the first black-box method for attributing fine-tuned text-to-image models to protected lineages using only query access. CSF treats models as semantic category generators and probes them with compositional underspecified prompts that remain rare under fine-tuning. This gives IP owners an asymmetric advantage: new prompt compositions can be generated after deployment, while attackers must anticipate and suppress a much broader space of fingerprints. Across 6 model families (FLUX, Kandinsky, SD1.5/2.1/3.0/XL) and 13 fine-tuned variants, our Bayesian attribution framework enables controlled-risk lineage decisions, with all variants satisfying the dominance criterion.
\end{abstract}

\section{Introduction}
\label{sec:intro}

Text-to-Image models are deployed at scale and have gained huge success commercially~\cite{rombach2022high, podell2023sdxl, esser2024scaling, flux2024, labs2025flux1kontextflowmatching, razzhigaev2023kandinsky}, now offering their services to millions of users. As the models themselves have become valuable assets, intellectual property (IP) protection for these models has become urgent. Consequently, many diffusion models are now distributed under restrictive licenses (e.g., non-commercial license for Flux.1 [dev]~\cite{blackforestlabs2023flux1dev}, research-only license for SDXL Turbo~\cite{stabilityai2023sdxlturbo}) to prevent unauthorized commercial distribution of fine-tuned derivatives. However, these licenses are only effective when violations can be detected. The problematic scenario arises when an infringer fine-tunes a protected model and distributes it exclusively through an API. To address this challenge, we need a method that can attribute a suspect model to a protected lineage under minimal access.

\begin{figure}[t]
    \centering
    \includegraphics[width=1\linewidth]{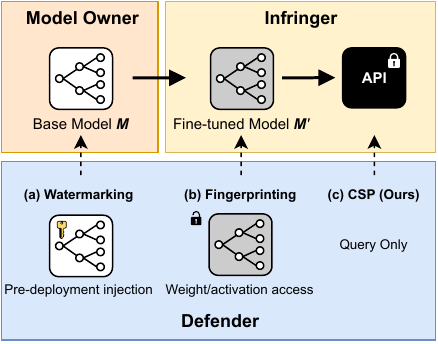}
    \caption{{Comparison of model identification scenarios. (a) \textbf{Watermarking} requires pre-deployment access to inject a trigger into the base model. (b) \textbf{Traditional Fingerprinting} relies on white-box or gray-box `Internal Access' (e.g., weights or activations), which is not available in commercial API. (c) \textbf{Our approach (CSF)} is designed for the most restrictive `Query Only' black-box setting, where the defender only has access to the final T2I generation API, reflecting real-world infringement scenarios.}}
    \label{fig:comparison}
\end{figure}
This problem reduces to finding a signature that attributes a suspect model to its root IP lineage. Existing approaches either inject identifiers into models (watermarking) or discover intrinsic fingerprints (fingerprinting). Watermarking works well in black-box, query-only settings because trigger inputs induce detectable patterns in outputs that survive typical fine-tuning~\cite{cui2023diffusionshield, wen2023tree, huang2024robin, zhang2024attack}. But watermarking requires pre-deployment training, can degrade model quality~\cite{hua2023deep, zhao2023recipe, lei2024diffusetrace, chen2025image}, and, once known, can be removed by infringers~\cite{chen2021refit, muller2025black, hu2024stable, li2023diffwa}. Fingerprinting instead passively extracts characteristics from deployed models, but these signals are subtle. Existing methods either do not address preservation under fine-tuning~\cite{wang2025sleepermark}, or require internal activations or fine-grained generation control~\cite{zhang2024reef, li2025diffip, tripathi2025paladin, somepalli2023diffusion, kim2024wouaf}, which commercial APIs typically do not provide (Figure~\ref{fig:comparison}). This leaves the restricted-access setting with only text-image pairs largely unsolved.

In this work, we present Compositional Semantic Fingerprinting (CSF), a fingerprinting method that statistically determines whether a suspect model belongs to a model lineage in the restrictive black-box setting using only text-to-image API access. CSF reverses the logic of watermarking for fingerprinting. Just as watermarks persist after fine-tuning because rare trigger inputs are unlikely to be overwritten, CSF exploits semantic compositions that sparsely appear in fine-tuning data. Rather than relying on visual artifacts, it probes semantics that persist through downstream modification. This gives IP owners an asymmetry advantage, since compositions can be generated at detection time. Moreover, CSF allows lineage decisions to be made statistically under controlled risk. The method is training-free and requires no model modification before deployment.

\begin{figure*}[ht]
    \centering
    \includegraphics[width=1\linewidth]{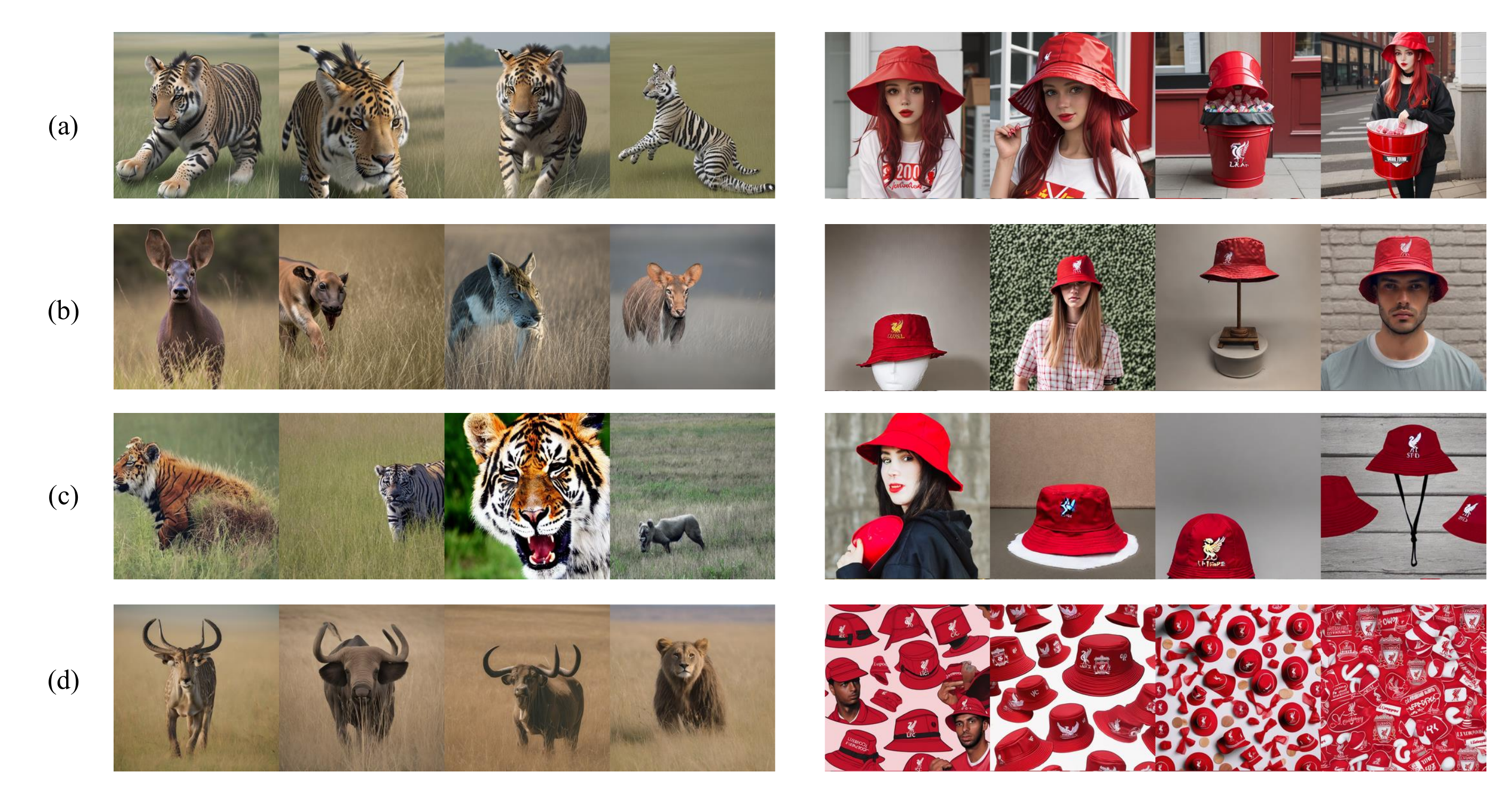}
    \caption{ The “Name That Dataset” game~\cite{torralba2011unbiased} in Diffusion Fingerprinting. Image (a) is from a fine-tuned model (SD1.5-DreamShaper). One of the images (b, c, d) is its base model. This figure illustrates how difficult it is to identify the base model using a \textbf{naive prompt (right column}, randomly sampled from LAION-2B~\cite{schuhmann2022laion}), compared to \textbf{our CSF prompt (left column)}. All images are uncurated results generated with different random seeds. \textbf{Can you specify which model (b, c, or d) is the base model for (a)?} \subsubnote{Answer: (a) SD1.5-DreamShaper (fine-tuned), (b) SD2.1-DPO (fine-tuned), (c) SD1.5 Base, and (d) SDXL Base. The correct base model for (a) is (c).}}
    \label{fig:family_Game}
\end{figure*}


However, this query-only setting cannot be solved by per-model classifiers, representation-space clustering, or direct output matching alone. Fine-tuning can dramatically alter image style, color palette, and overall composition, overwhelming subtle pixel-level signals and making highly reliable output-space discrimination virtually impossible (\cref{fig:family_Game}). It can also substantially change naive surface biases, for example through Direct Preference Optimization (DPO)~\cite{rafailov2023direct}, while preserving deeper semantics inherited from pre-training.

CSF addresses this by moving the problem from visual space to semantic space and reducing it to text-to-class inference. Specifically, we probe models with compositional semantic combinations that are rare in fine-tuning data, using conditioning inspired by formal semantics, and sample multiple generations for each underspecified prompt. The resulting interpretation distributions reveal pre-training biases inherited from the base model lineage and preserved after fine-tuning. We then convert these outputs into category distributions via zero-shot classification, compare suspect and reference models through distances between the resulting semantic distributions, and aggregate evidence across prompts to make statistical lineage decisions with quantified uncertainty. We validate this on Stable Diffusion (v1.5, v2.1, XL, v3), Flux, Kandinsky, and fine-tuned variants from style transfer, LoRA~\cite{hu2022lora}, model merging, and DPO, covering 19 models in total.

\section{Related Work}

Model identification is a research area aimed at identifying the IP owner of a model from generated content. These methods reduce to the problem of mapping each model $M$ to a unique identifier $\phi_M$. The approaches to obtaining identifiers are broadly divided into two categories: (1) methods that \emph{inject} identifiers into models during training, and (2) methods that \emph{discover} intrinsic identifiers that models inherently possess.

\subsection{Watermarking: Identifier Injection}

Watermarking injects a backdoor trigger (identifier) into the model during training, such that the model produces a predetermined output for specific trigger inputs~\cite{adi2018turning, szyller2021dawn, jia2021entangled}. This approach is promising as it can provide statistically rigorous evidence. It has been applied to various models, from classical classification models~\cite{rouhani2018deepsigns, adi2018turning} and large language models~\cite{peng2023you, kirchenbauer2023watermark, zhao2023provable, kirchenbauer2023reliability} to recent diffusion models~\cite{cui2023diffusionshield, wen2023tree, zhao2023recipe, rezaei2024lawa}. For diffusion models, approaches have been proposed that embed watermarks in the denoising process~\cite{yang2024gaussian, huang2024robin, wen2023tree, rezaei2024lawa} or inject specific patterns into the latent space~\cite{fernandez2023stable, hu2024stable, ci2024wmadapter}.

However, watermarking has fundamental limitations. First, adding heterogeneous triggers to training data inevitably degrades the model's overall performance. Second, it cannot be applied to models already deployed without watermarks. Third, watermarks can be diluted through fine-tuning~\cite{hu2024stable, fernandez2023stable, wang2025sleepermark}, or if the watermark becomes known, it can be neutralized by watermark defense techniques~\cite{wu2021adversarial}.

\subsection{Fingerprinting: Identifier Discovery}

Fingerprinting leverages intrinsic characteristics that models inherently possess as identifiers, requiring no additional tuning. Representative approaches include model weight-based methods that measure the distance between the suspect model's weights and the original model's weights~\cite{maho2023model, kim2024wouaf, sun2023deep}. Representation-based methods use kernels defined in the model's latent feature space as identifiers, distinguishing models by distances between kernels~\cite{kornblith2019similarity, zhang2024reef}.

While these methods are effective, the major limitation is the requirement for white-box access. When suspect models are distributed as commercial APIs, weights and hidden representations are inaccessible. For diffusion models, methods that train classifiers on generated images are commonly used~\cite{park2025community, wang2023dire}, but these studies are limited to toy datasets or synthetic setups, with insufficient research on actually deployed foundational models (e.g., DALL-E~\cite{ramesh2021zero}, Stable Diffusion). Moreover, most assume white-box access. For example,~\cite{song2025riemannian} claims one-shot attribution but only identify cases where training data matches and does not address fine-tuned variants.

\subsection{Our Approach}

This work follows the fingerprinting approach of discovering intrinsic identifiers, but with the following distinctions. First, we assume a complete black-box setting. That is, only the API of currently deployed models is available, generation processes cannot be controlled, and only text prompts can be provided as conditions. Second, our method is train-free. Unlike existing methods, no separate model training or training data collection is required. Third, we target fine-tuned scenarios and foundational models. Reflecting the reality that commercial models are continuously fine-tuned, we propose a robust identifier that preserves the base model's identity even after fine-tuning.


\section{Problem Formulation}
\begin{figure}[t]
\centering
\begin{subfigure}[b]{0.21\textwidth}
    \centering
    \includegraphics[width=\textwidth]{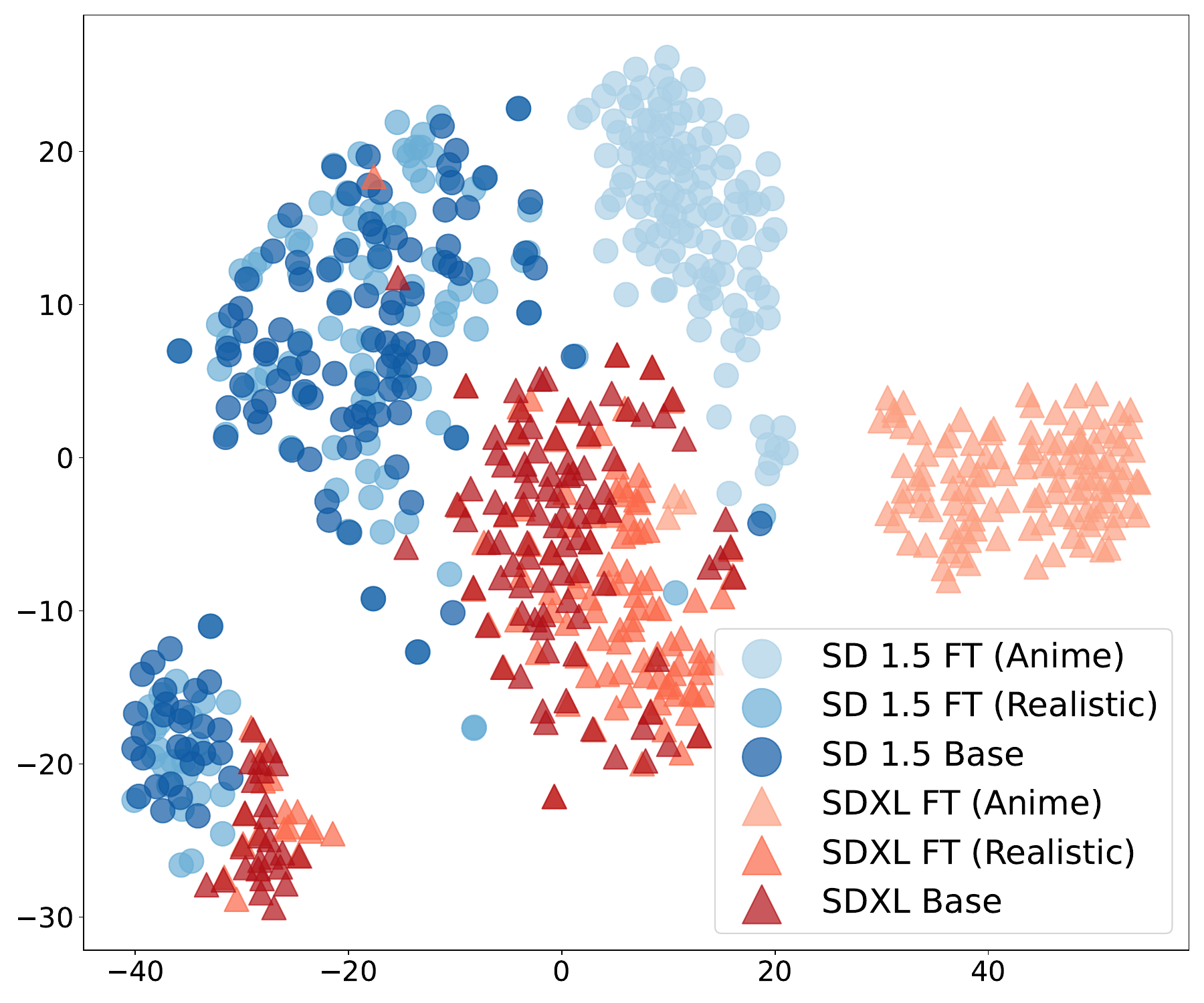}
    \caption{Visual Space (CLIP)}
    \label{fig:visual_space}
\end{subfigure}
\hspace{0.5cm}
\begin{subfigure}[b]{0.21\textwidth}
    \centering
    \includegraphics[width=\textwidth]{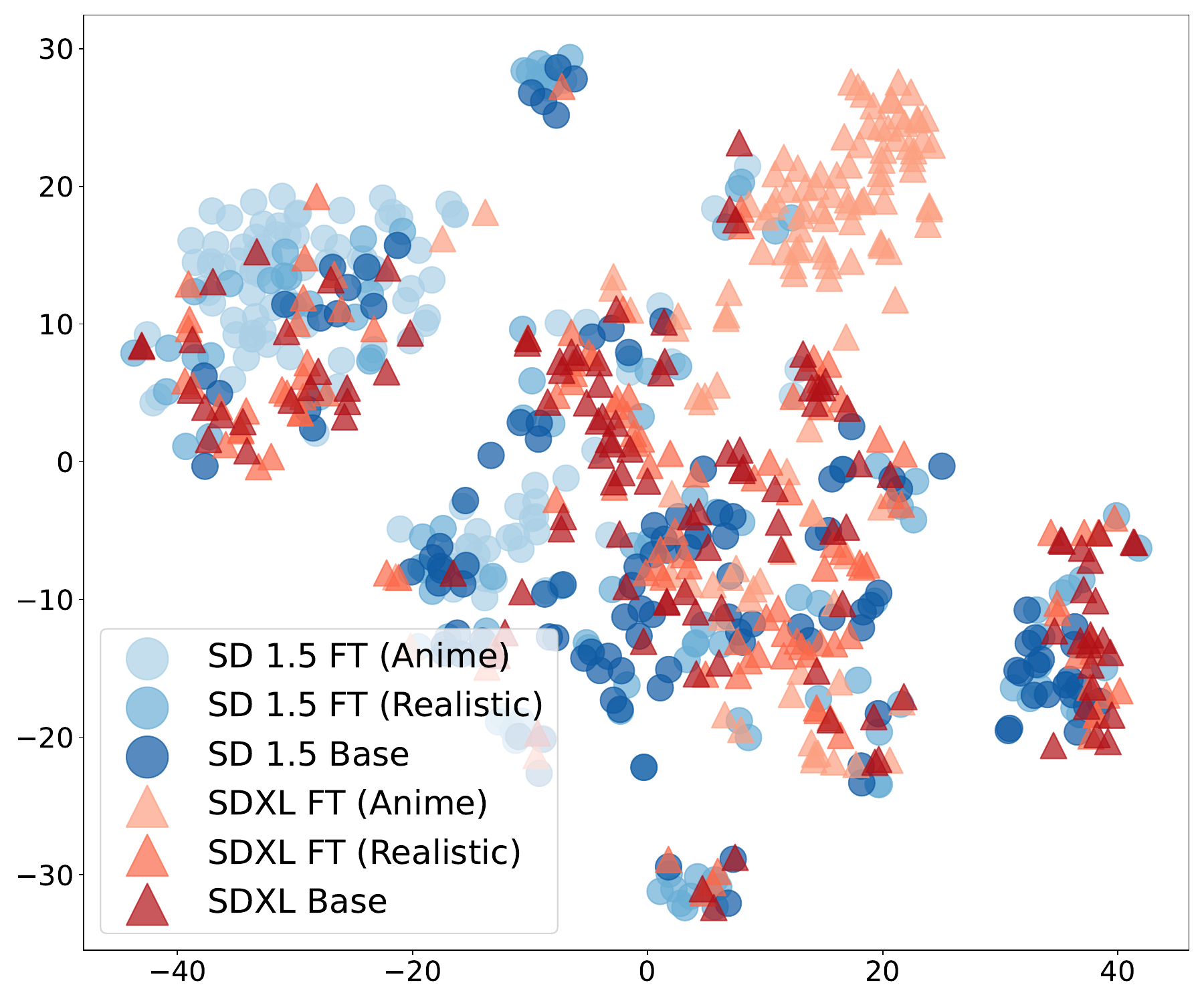}
    \caption{Text Space (I2T Captions)}
    \label{fig:text_space}
\end{subfigure}
\caption{\textbf{Challenges in naive fingerprinting approaches.} (a) \textbf{Visual space}: t-SNE visualization of CLIP embeddings shows no family clustering. (b) \textbf{Text space}: Even when images are converted to captions via I2T models, style information leaks into the text, causing models from different families (e.g., SD1.5 DPO and SD2.1) to cluster together due to similar style descriptors.}
\label{fig:challenges}
\end{figure}

\paragraph{Task definition.}
For each text-to-image model $M$, the goal is to generate an identifier $\phi_M$ that uniquely identifies $M$. The identifier must satisfy two conditions:
\begin{itemize}[leftmargin=*]
\item \textbf{Discriminability}: Different base models must be distinguishable. That is, for base models $M_i$ and $M_j$ ($i \neq j$), $\phi_{M_i} \neq \phi_{M_j}$.
\item \textbf{Robustness}: Fine-tuned variants $M'$ should be attributed to the same base model $M$. That is, for $M'$ obtained by fine-tuning $M$, $\phi_{M'} \approx \phi_M$.
\end{itemize}
\paragraph{Key challenge.}
Fine-tuning dramatically alters generation style (photorealistic vs. anime), content bias (subject preferences), and generation quality—yet the base model identity must remain recoverable. Existing visual features (CLIP~\cite{radford2021learning} embeddings, ResNet~\cite{he2016deep} features) are highly sensitive to these stylistic variations, causing fine-tuned variants to cluster by style rather than base model (\cref{fig:challenges}). This necessitates an alternative representation that isolates semantic interpretation from visual appearance.
\paragraph{Our setting.}
This work addresses the most extreme black-box setting. We assume that for a suspect model, we can only provide text prompts $T$ as queries and observe the generated images $I = M(T)$, with the ability to perform multiple sampling for the same prompt. We have no access to any internal state such as model weights, gradients, internal activations, or diffusion trajectories used in prior work. This corresponds to the query-only interface commonly provided by commercial APIs.

\section{Method}

CSF abstracts text-to-image models as semantic category generators rather than pixel-level synthesizers, isolating interpretive biases from stylistic variations that change during fine-tuning (\Cref{sec:semanticFingerprinting}). We construct compositional underspecified prompts (e.g., ``a dangerous urban nocturnal animal'') that combine individually common semantic components into exponentially rare compositions, exploiting the same robustness principle as backdoor watermarks—fine-tuning preserves model behavior on rare conditions absent from training data (\Cref{sec:CSF}). For each prompt, we generate $N=30$ images, extract categorical distributions via CLIP zero-shot classification, and perform Bayesian attribution by computing Wasserstein distances between suspect and base model fingerprints across 42 prompts, yielding statistically rigorous confidence intervals (\Cref{sec:bayesian}).
\subsection{Semantic Fingerprinting}\label{sec:semanticFingerprinting}

To isolate semantic interpretation from stylistic variations, we 
abstract text-to-image models as text-to-category generators. Rather  than modeling pixel distributions, we model the semantic category  distributions they implicitly define.

Naively, one might use image captions to extract semantic content. 
However, captions encode not just \emph{what} is depicted but 
\emph{how it is described}—the linguistic framing itself is 
model-dependent. Different models may describe the same semantic 
concept using different phrasings, levels of detail, or contextual 
emphasis.

To eliminate this confound, we constrain the semantic space to a  predefined set of categories (\textit{e.g.,} the category of animal).  Rather than generating free-form descriptions, we ask: \emph{which  category does the model's generation correspond to?} This reduces  the problem to a distribution over discrete semantic choices.

We formalize this abstraction as follows. Given an underspecified 
prompt $C$ (e.g., ``a photo of animal''), model $M$ produces images  whose semantic content varies stochastically. Let $\mathcal{Y}_C$  denote the space of semantically plausible categories for prompt $C$.  Model $M$ induces a categorical distribution:
\begin{equation}
P(Y | C, M), \quad Y \in \mathcal{Y}_C
\end{equation}
where $Y$ represents the semantic category (e.g., tiger, lion, wolf) 
of the generated image.

Due to stochastic image generation, each generation of $M(C)$ produces 
a different categorical vector $\phi \in \Delta^{K-1}$:
\begin{equation}
\mathbf{p} \sim P(\cdot | C, M), \quad \mathbf{p} \in \Delta^{K-1}
\end{equation}
where $\mathbf{p}_i$ represents the probability assigned to category $i$ 
by a vision-language classifier, and $\Delta^{K-1}$ denotes the 
$(K-1)$-dimensional probability simplex.

The model fingerprint is the distribution $P(\mathbf{p} | C, M)$ over 
these categorical vectors. We estimate this distribution via $N$ 
i.i.d. samples $\{\mathbf{p}_1, \ldots, \mathbf{p}_N\}$ obtained by generating 
$N$ images and classifying each independently. Two models are compared 
via the Wasserstein distance between their empirical distributions 
over categorical vectors.

\subsection{Compositional Semantic Fingerprinting}\label{sec:CSF}

While semantic abstraction isolates interpretive biases from 
stylistic variations, a critical question remains: \emph{which 
prompts preserve base model fingerprints through fine-tuning?}

To extract distributional fingerprints, prompts must be underspecified,
admitting multiple semantic interpretations that vary across 
generations. However, naive broad prompts (e.g., ``an animal'') 
face a fundamental problem: they correspond to common categories 
in fine-tuning datasets. Fine-tuning data frequently contains 
animal images with diverse species, allowing fine-tuning to 
overwrite the base model's semantic biases. The base model 
fingerprint is lost.

We resolve this through \emph{compositional underspecification}: 
prompts that produce semantic variation yet remain rare in 
fine-tuning data. We construct prompts by combining (1) a 
superordinate category $T$ (e.g., ``animal'') and (2) multiple 
semantic attributes $\{c_1, c_2, \ldots\}$ (e.g., ``dangerous'', 
``dimmed studio''). Each prompt follows the template:

\begin{equation}
C = \text{``A photo of } c_1 \text{ } c_2 \text{ } \cdots \text{ } c_k \text{ } T\text{''}
\end{equation}

\paragraph{Fingerprint extraction.}
For each compositional prompt $C$, we generate $N$ images 
and extract their categorical distributions via zero-shot 
classification. The aggregated distribution characterizes 
how model $M$ interprets underspecified conditions, 
independent of visual style.

\paragraph{Asymmetry characteristics.}
Unlike watermarking methods where known triggers can be targeted  and removed, our compositional prompts exploit combinatorial explosion.  Infringers cannot remove specific prompts without knowing which  combinations will be tested, and defending against all possible  compositions is intractable—the exponential space of attribute  combinations makes exhaustive removal infeasible. Verifiers maintain  the advantage: they can freely construct new rare compositions  post-deployment.

\subsection{Robustness Principles}\label{sec:robust}

\paragraph{CSF as watermarks}
The robustness of our method follows the same principle as backdoor
watermarks. Fine-tuning updates model parameters based on the training data distribution $\mathcal{D}_{FT}$. For input-output pairs $(x, y)$
that appear in $\mathcal{D}_{FT}$, fine-tuning modifies the model's behavior: $P(y|x, M^\prime) \neq P(y|x, M)$. However, for rare conditions not present in the fine-tuning data, where $P((x, y) \in \mathcal{D}_{FT}) \approx 0$, the model receives little update signal on those behaviors. As a result, the base model's response to rare conditions is preserved: $P(y|x_{\text{rare}}, M^\prime) \approx P(y|x_{\text{rare}}, M)$.

This explains why backdoor watermarks persist: the specific trigger-watermark pair $(X_{\text{trigger}}, Y_{\text{watermark}})$ is engineered to be
absent from natural fine-tuning data, leaving the embedded behavior
intact.

Our compositional prompts exploit this mechanism through combinatorial rarity. By composing semantic components that may each be common in fine-tuning datasets, their joint condition becomes progressively rarer. For a compositional prompt built from a target concept $T$ and contextual conditions $\{c_1, c_2, \ldots\}$, we do not assume strict independence among components. Instead, we assume a monotonic decrease in occurrence as additional conditions are composed:
\begin{equation}
P(T) \gg P(T \cap c_1) \gg P(T \cap c_1 \cap c_2) \gg \cdots \gtrsim 0.
\end{equation}
This monotonic reduction is sufficient for our purpose: as more semantic conditions are composed, the resulting prompt enters an effectively rare regime under $\mathcal{D}_{FT}$. In that regime, fine-tuning provides little update pressure on the induced semantic behavior. Consequently, the base model's semantic biases are preserved:
\begin{equation}
P(\mathbf{p} \mid C, M^\prime) \approx P(\mathbf{p} \mid C, M)
\end{equation}
maintaining model fingerprints through fine-tuning.

\begin{table*}[tb]
\centering
\caption{Posterior Mean of the Derived Models. 
In this table, \colorbox[rgb]{1,0.894,0.894}{\phantom{xx}} indicates significance a (Confidence Interval (CI) low $>$ 0.167), \colorbox[rgb]{1,1,0.894}{\phantom{xx}} indicates Not significant (CI includes 0.167), and \colorbox[rgb]{0.894,1,0.894}{\phantom{xx}} indicates Sig. below chance (CI high $<$ 0.167). * indicates the model meets Dominance test (CI low $>$ 0.5)}
\scalebox{0.80}{
\renewcommand{\arraystretch}{1.}
\begin{tabular}{lcccccc} 
\hline\hline
& \multicolumn{6}{c}{\cellcolor[rgb]{0.918,0.918,0.918}\textbf{Base Models}} \\
& Flux-Base & Kandinsky-Base & SD1.5-Base & SD2.1-Base & SD3-Medium-Base & SDXL-Base \\ 
\hline
\multicolumn{7}{l}{\textit{Flux Family}} \\
\textbf{Flux-LoRA} & \cellred{0.932*} & \cellgreen{0.023} & \cellgreen{0.023} & \cellgreen{0.023} & \cellgreen{0.023} & \cellyellow{0.068} \\
\textbf{Flux-Turbo-Alpha} & \cellred{0.977*} & \cellgreen{0.023} & \cellgreen{0.023} & \cellgreen{0.023} & \cellgreen{0.023} & \cellgreen{0.023} \\
\hline
\multicolumn{7}{l}{\textit{Kandinsky Family}} \\
\textbf{Kandinsky-Naruto} & \cellgreen{0.023} & \cellred{0.977*} & \cellgreen{0.023} & \cellgreen{0.023} & \cellgreen{0.023} & \cellgreen{0.023} \\
\textbf{Kandinsky-Pokemon-LoRA} & \cellyellow{0.049} & \cellred{0.829*} & \cellyellow{0.049} & \cellyellow{0.098} & \cellyellow{0.024} & \cellyellow{0.049} \\
\hline
\multicolumn{7}{l}{\textit{SD1.5 Family}} \\
\textbf{SD1.5-1.2-Base} & \cellgreen{0.023} & \cellgreen{0.023} & \cellred{0.841*} & \cellyellow{0.114} & \cellgreen{0.023} & \cellyellow{0.068} \\
\textbf{SD1.5-1.4-Base} & \cellgreen{0.023} & \cellgreen{0.023} & \cellred{0.977*} & \cellgreen{0.023} & \cellgreen{0.023} & \cellgreen{0.023} \\
\textbf{SD1.5-DreamShaper} & \cellyellow{0.091} & \cellyellow{0.068} & \cellred{0.659*} & \cellyellow{0.045} & \cellyellow{0.068} & \cellyellow{0.159} \\
\hline
\multicolumn{7}{l}{\textit{SD2.1 Family}} \\
\textbf{SD2.1-DPO} & \cellgreen{0.023} & \cellgreen{0.023} & \cellgreen{0.023} & \cellred{0.977*} & \cellgreen{0.023} & \cellgreen{0.023} \\
\textbf{SD2.1-LAION-Art} & \cellgreen{0.023} & \cellgreen{0.023} & \cellgreen{0.023} & \cellred{0.977*} & \cellgreen{0.023} & \cellgreen{0.023} \\
\hline
\multicolumn{7}{l}{\textit{SD3 Family}} \\
\textbf{SD3-Reality-Mix} & \cellyellow{0.136} & \cellyellow{0.091} & \cellgreen{0.023} & \cellyellow{0.045} & \cellred{0.705*} & \cellyellow{0.091} \\
\textbf{SD3-VAE-Anime} & \cellgreen{0.023} & \cellgreen{0.023} & \cellgreen{0.023} & \cellgreen{0.023} & \cellred{0.977*} & \cellgreen{0.023} \\
\hline
\multicolumn{7}{l}{\textit{SDXL Family}} \\
\textbf{SDXL-DPO} & \cellgreen{0.023} & \cellgreen{0.023} & \cellgreen{0.023} & \cellgreen{0.023} & \cellgreen{0.023} & \cellred{0.977*} \\
\textbf{SDXL-Lightning-4Step} & \cellgreen{0.023} & \cellyellow{0.091} & \cellgreen{0.023} & \cellyellow{0.068} & \cellgreen{0.023} & \cellred{0.864*} \\
\hline\hline
\end{tabular}
\label{tab:posterior_mean_jsd}
}
\end{table*}

\paragraph{Linguistic grounding.}
Our design exploits how language models resolve ambiguous descriptions.  When encountering underspecified prompts like ``a dangerous animal,"  models must resolve ambiguity using their learned priors, just as humans invoke different referents based on their experiences.  This is the principle of \textbf{underspecification}~\cite{egg2010semantic, frisson2009semantic, pustejovsky2017semantics}:  semantic constraints reduce but do not determine the referent space,  forcing models to reveal their semantic biases. By combining this with  compositional rarity, we create prompts that are both discriminative  (revealing model-specific interpretations) and robust (preserved through fine-tuning).

\subsection{Implementation}\label{sec:implementation}

\paragraph{Category extraction.}
For each compositional prompt $C$, we generate $N=30$ images 
using the suspect model with different random seeds. Each image 
$I_i$ is classified via CLIP~\cite{radford2021learning} zero-shot classification:
\begin{equation}
\phi_i = \text{softmax}(\text{CLIP}_{\text{visual}}(I_i) \cdot 
\text{CLIP}_{\text{text}}(\{y_1, \ldots, y_K\}))
\end{equation}
where $\{y_1, \ldots, y_K\}$ are subordinate categories for 
superordinate $T$, sourced from: (1) Wikipedia taxonomies for 
common domains, (2) existing dataset labels when available, or 
(3) LLM generation for specialized domains.

\paragraph{Prompt design.}
Each prompt follows a controlled three-component structure: 
(1) an underspecified semantic attribute that admits multiple 
interpretations, (2) a superordinate category defining the domain, 
and (3) a specific contextual condition.
The underspecified attribute forces models to make subjective 
semantic choices, revealing their learned biases. The specific 
condition serves a critical role: it constrains non-semantic 
factors (composition, object count, scene complexity) to ensure 
each generation contains exactly one identifiable subject. 
This design prevents ambiguity in CLIP classification that would 
arise from multi-object scenes or empty backgrounds, isolating 
model-specific semantic interpretation as the sole source of 
distributional variation.
We systematically vary the underspecified attributes across 
multiple semantic dimensions while maintaining fixed specific 
conditions within each domain, yielding 42 compositional prompts.

\paragraph{Sampling protocol.}
Each prompt-model pair produces 30 categorical distributions, 
yielding 1260 total samples per model. This demonstrates the 
practical affordability of our method,even using commercial APIs, 
the cost is only approximately \textbf{\$50} per model.\footnote{For 
example, OpenAI DALL-E 3 charges \$0.04 per 1024$\times$1024 image 
(\url{https://openai.com/api/pricing}), yielding $42 \times 30 \times 
\$0.04 \approx \$50$ per model. Similar pricing applies to other 
commercial T2I APIs.}

\subsection{Bayesian Attribution Framework}\label{sec:bayesian}

\paragraph{Statistical inference problem.}
Model attribution is inherently probabilistic: each compositional 
prompt constitutes an independent trial, and we must infer the 
true base model from multiple noisy observations. We formalize 
this as Bayesian inference over model identity.

\paragraph{Distance metric.}
Recall that each model with prompt $C$ produces $N$ categorical 
distributions $\{\mathbf{p}_1, \ldots, \mathbf{p}_N\}$ through stochastic 
generation. To compare two models, we compute the distance between 
their empirical distributions—two sets of samples in the $K$-dimensional 
probability simplex.

We use the Wasserstein distance, which directly operates on sample 
sets without requiring density estimation. Each set of samples defines an empirical identifier: \begin{equation} \phi = \frac{1}{N}\sum_{i=1}^N \delta_{\mathbf{p}_i} \end{equation}
where $\delta_{\mathbf{p}_i}$ is a point mass at $\mathbf{p}_i \in \mathbb{R}^K$. The 
2-Wasserstein distance is:
\begin{equation}
W_2(\phi_1, \phi_2) = \left(\inf_{\gamma \in \Gamma(\phi_1, \phi_2)} 
\mathbb{E}_{(i,j) \sim \gamma}[\|\mathbf{p}_i - \mathbf{p}_j\|_2^2]\right)^{1/2}
\end{equation}
where $\Gamma$ denotes all couplings between the two measures. This  treats each categorical vector as a point in $K$-dimensional space, 
preserving the joint structure across categories.




\paragraph{Single-trial classification.}
For each compositional prompt $C$, we classify suspect model $M'$ 
by computing distances to all $K$ base model representatives:
\begin{equation}
\hat{M}_{\text{base}}(C) = \operatorname*{argmin}_{M_i \in \{\text{base models}\}} 
W_2(M'(C), M_i(C))
\end{equation}

The trial succeeds if $\hat{M}_{\text{base}}(C)$ matches the 
ground truth base model of $M'$.

\paragraph{Bayesian aggregation.}
Over $T$ prompts, we observe success count $s$ and failure count
$f = T - s$. We model attribution accuracy $\theta$ with a Beta prior:
\begin{equation}
\theta \sim \mathrm{Beta}(\alpha, \beta), \quad
\theta \mid s,f \sim \mathrm{Beta}(\alpha+s,\beta+f).
\end{equation}
Using an uninformative prior $\mathrm{Beta}(1,1)$, we obtain a posterior over $\theta$, where the posterior mean estimates expected attribution accuracy and a 95\% credible interval quantifies uncertainty. This yields a Beta-Binomial decision rule for determining whether a suspect model is unrelated or derivative with respect to a candidate base model. We use the 95\% interval as a conventional two-sigma-like evidentiary standard for controlled-risk attribution.

We evaluate attribution using two decision criteria.

\vspace{2mm}
\noindent\textbf{1. Significance test ($1/K = 17\%$).}
With $K=6$ candidate base families, random guessing yields chance accuracy $1/K \approx 17\%$. We therefore test
\begin{equation}
H_0:\theta \leq 1/K.
\end{equation}
Attribution is considered better than chance only when the lower bound of the posterior interval exceeds $1/K$.

\vspace{2mm}
\noindent\textbf{2. Dominance test ($50\%$).}
For practical attribution, the candidate base model must be the dominant source:
\begin{equation}
H_0:\theta \leq 0.5.
\end{equation}
This tests whether the candidate base model is not merely better than chance, but more probable than all alternative candidates combined. We regard attribution as actionable only when the lower bound of the posterior interval exceeds $0.5$, indicating dominance under controlled uncertainty.
\begin{figure}
    \centering
    \includegraphics[width=1\linewidth]{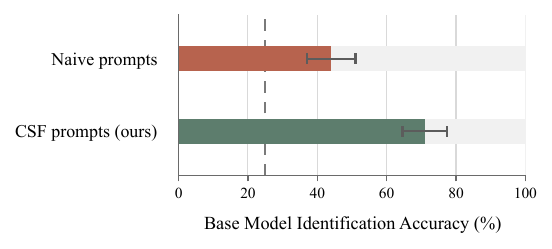}
    \caption{
    User study results for the ``Name That Dataset" game (\cref{fig:family_Game}). Participants attempted to identify the correct base model using naive prompts (LAION-2B~\cite{schuhmann2022laion}) vs. our CSF prompts. 
}
\label{fig:user_study}
\end{figure}

\section{Experiments}

\subsection{Model Selection}
\paragraph{Base model}
We evaluate on 6 base families representing all major open-source  text-to-image architectures: FLUX~\cite{flux2024}, Kandinsky~\cite{razzhigaev2023kandinsky}, and Stable Diffusion  1.5/2.1/3.0/XL~\cite{rombach2022high, podell2023sdxl, esser2024scaling}. This covers fundamental architectural diversity -  different noise schedules (cosine vs. rectified flow), backbones  (2D U-Net~\cite{ronneberger2015u} vs. DiT~\cite{peebles2023scalable} vs. Flow Transformer~\cite{manocchio2024flowtransformer}), text encoders (CLIP-only~\cite{radford2021learning}  vs. CLIP+T5~\cite{ni2022sentence}), and latent resolutions. Critically, several families  pose deliberate challenges: SD1.5 and SD2.1 share nearly identical  architectures from the same company, with SD2.1's training data  being a \emph{superset} of SD1.5's. Similarly, SD3 and FLUX adopt  the same T5-XXL + CLIP-L text encoders and 128×128 latent resolution,  differing primarily in backbone design. These architectural overlaps  make families exceptionally difficult to distinguish based on 
data-driven semantic biases alone. These architectural overlaps make lineage-level attribution exceptionally difficult from semantic behavior alone.

\paragraph{Lineage definition and attribution target.}
We study attribution at the level of model lineage rather than exact checkpoint identity. This matches the practical commercial setting, where the key forensic question is whether a deployed model derives from a restricted base family, since licensing and redistribution constraints typically extend to derivative models as well. We therefore define a lineage as a base model together with its intermediate checkpoints, continued-pretraining descendants, and downstream fine-tuned variants that inherit the same underlying family. Under this threat model, successful attribution means recovering the correct source lineage despite substantial post-training modifications.

\paragraph{Fine-tuning variants}
We select variants spanning all major adaptation types to ensure  comprehensive evaluation. These include: (1) LoRA~\cite{hu2022lora} fine-tuning  for style transfer and preference alignment, (2) full model  fine-tuning on domain-specific datasets, (3) model merging  (``cocktails") that blend multiple fine-tuned models and dilute  base fingerprints, (4) preference optimization (DPO~\cite{rafailov2023direct}/RLHF~\cite{ouyang2022training}) that  actively rewrites semantic biases toward human preferences,  (5) distillation to few-step models that compress generation  while potentially discarding subtle biases, and (6) component  replacement (e.g., VAE swapping) that alters the generation  pathway. Several variants combine multiple perturbations  (e.g., merging + DPO), representing near-worst-case scenarios  for fingerprint preservation.

\subsection{Attribution Performance}

\paragraph{Main results.}
\Cref{tab:posterior_mean_jsd} presents Bayesian attribution results for  all 13 fine-tuned variants. Every model achieves posterior mean  accuracy exceeding 50\%, with 95\% credible interval lower bounds  consistently above this Dominance test. This demonstrates  that for each suspect model, the correct base family is predicted  more frequently than all other families \emph{combined}, providing  actionable attribution confidence. In contrast, using only base prompts without compositional constraints achieves fils, with several models showing significant confusion (e.g., sd15-dreamshaper → kandinsky, sd12 posterior: 0.429 for sdxl). This drop validates that compositional underspecification is necessary for reliable attribution.
\begin{figure}
    \centering
    \includegraphics[width=1\linewidth]{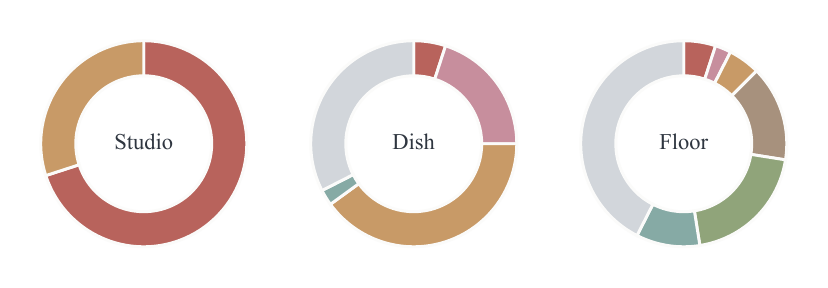} \caption{Generated category distributions vary substantially with scene context. Compared with the dimmed-studio setting, the dish and wooden-floor settings produce broader and systematically shifted category mixtures. Each donut summarizes 40 samples, and the same color denotes the same category across scenes.
} \label{fig:prompt_ablation} 
\end{figure}
\paragraph{Robustness and practical advantages.}
\textit{Resistant to extreme perturbations.} Even the most challenging cases succeed. SD3-Reality-Mix undergoes full retraining on curated photorealistic datasets, fundamentally rewriting semantic priors beyond typical fine-tuning. Model merging combined with DPO (e.g., Dreamshaper) layers preference re-biasing atop already-diluted cocktail fingerprints. Distillation compresses generation into few steps, potentially discarding subtle biases. Yet all variants maintain Dominance test at 95\% confidence, demonstrating that compositional semantic biases persist even under aggressive adaptations.

\textit{Black-box advantages.} Our method successfully attributes distilled models (SDXL-Lightning, FLUX-Turbo), which white-box methods~\cite{teng2025fingerprinting} cannot handle due to altered generation trajectories. Similarly, gray-box methods requiring intermediate activations~\cite{zhang2024reef} fail in true API-only settings. By operating at the semantic level, our approach achieves robustness unavailable to architecture-dependent fingerprinting.

\begin{table}[tb]
\centering
\caption{Attribution confidence: Wasserstein vs. JSD}
\label{tab:metric_comparison}
\small
\resizebox{0.45\textwidth}{!}{%
\begin{tabular}{l|cc|c}
\hline\hline
\textbf{Fine-tuned Model} & 
\makecell{Wasserstein\\(Ours)} & 
\makecell{JSD\\(Baseline)} & 
\makecell{Confidence\\Gap} \\
\hline
Flux-LoRA        & 93.2\% & 77.3\% & \textcolor{ForestGreen}{+15.9\%} \\
Kandinsky-Naruto & 97.7\% & 43.2\% & \textcolor{ForestGreen}{+54.5\%} \\
SD3-Reality-Mix  & 70.5\% & 56.8\% & \textcolor{ForestGreen}{+13.7\%} \\
SDXL-DPO         & 97.7\% & 70.5\% & \textcolor{ForestGreen}{+27.2\%} \\
\hline\hline
\end{tabular}%
}
\end{table}
\begin{table}[t]
\centering
\tiny
\captionsetup{font=small}
\caption{Attribution results under adversarial concept removal. Posterior mean attribution scores computed from 9 animal-specific probes after removing animal-related concepts using UCE.}\label{tab:posterior_mean_jsd_simplified}

\setlength{\tabcolsep}{2pt}
\renewcommand{\arraystretch}{0.95}

\begin{tabular*}{\columnwidth}{@{\extracolsep{\fill}}lcccccc}
\hline\hline
& \multicolumn{6}{c}{\cellcolor[rgb]{0.918,0.918,0.918}\textbf{Base Models}} \\
\textbf{Model} & Flux & Kandinsky & SD1.5 & SD2.1 & SD3-Med & SDXL \\ 
\hline

\cellcolor{blue!5}{LoRA} & \cellgreen{0.714} & 0.143 & 0.143 & 0.143 & 0.143 & 0.286 \\
\cellcolor{blue!5}{Turbo-Alpha} & \cellgreen{0.857} & 0.143 & 0.143 & 0.143 & 0.143 & 0.143 \\
\hline

\cellcolor{red!5}{Naruto} & 0.143 & \cellgreen{0.857} & 0.143 & 0.143 & 0.143 & 0.143 \\
\cellcolor{red!5}{Pokemon-LoRA} & 0.143 & \cellgreen{0.857} & 0.143 & 0.143 & 0.143 & 0.143 \\
\hline

\cellcolor{green!5}{1.2-Base} & 0.143 & 0.143 & \cellgreen{0.857} & 0.143 & 0.143 & 0.143 \\
\cellcolor{green!5}{1.4-Base} & 0.143 & 0.143 & \cellgreen{0.857} & 0.143 & 0.143 & 0.143 \\
\cellcolor{green!5}\textbf{Animal-Erase} & 0.143 & 0.143 & \cellgreen{0.857} & 0.143 & 0.143 & 0.143 \\
\cellcolor{green!5}{DreamShaper} & 0.143 & 0.143 & \cellgreen{0.714} & 0.143 & 0.286 & 0.143 \\
\hline

\cellcolor{orange!5}{DPO} & 0.143 & 0.143 & 0.143 & \cellgreen{0.857} & 0.143 & 0.143 \\
\cellcolor{orange!5}{LAION-Art} & 0.143 & 0.143 & 0.143 & \cellgreen{0.857} & 0.143 & 0.143 \\
\hline

\cellcolor{violet!5}{Reality-Mix} & 0.286 & 0.143 & 0.143 & 0.143 & \cellgreen{0.714} & 0.143 \\
\cellcolor{violet!5}{VAE-Anime} & 0.143 & 0.143 & 0.143 & 0.143 & \cellgreen{0.857} & 0.143 \\
\hline

\cellcolor{teal!5}{DPO} & 0.143 & 0.143 & 0.143 & 0.143 & 0.143 & \cellgreen{0.857} \\
\cellcolor{teal!5}{Lightning-4Step} & 0.143 & 0.143 & 0.143 & 0.286 & 0.143 & \cellgreen{0.714} \\
\hline\hline
\end{tabular*}

\vspace{-6pt}
\end{table}
\paragraph{Human Study}
While our statistical framework provides quantitative attribution confidence, practical IP disputes require evidence that is not only mathematically sound but also perceptually interpretable.  To address this, we validate that CSF-isolated biases align with human perception through a ``Name That Dataset"~\cite{torralba2011unbiased} experiment on Amazon Mechanical Turk, as illustrated in \cref{fig:family_Game}. 

We tasked 50 participants with identifying the base model for a given fine-tuned model (e.g., SD1.5-DreamShaper) from three candidate base models. Each participant completed 11 trials: 1 screening question for quality control, 5 using our CSF prompts, and 5 using naive baseline prompts (from LAION-2B~\cite{schuhmann2022laion}). In both conditions, participants were instructed to ``ignore the visual style" and match only the ``content distribution pattern," simulating our method's abstraction from style to semantics. 

The results (\cref{fig:user_study}), based on the 10 non-screening trials, confirm our hypothesis. Participant accuracy with naive prompts was 18\% far below random chance for 95\% confidence (33.3\% for 3-way classification). This demonstrates that naive prompts fail to generate perceptually distinct semantic patterns, rendering the model's lineage ambiguous even when style is disregarded. In contrast, our CSF prompts yielded 71\% accuracy \textit{i.e.,} meets Dominance test. This significant gap confirms that CSF-isolated biases are not mere statistical artifacts but are perceptually robust signals that align with human intuition, strengthening their credibility as circumstantial evidence for IP attribution.

\subsection{Analysis}

\paragraph{Robustness to adversarial concept removal.}
A natural question is whether CSF can be neutralized by semantic erasure or unlearning methods that deliberately remove target concepts from a model. To test this attack scenario, we evaluate CSF under adversarial concept removal settings, using concept-editing methods to suppress specific semantic concepts, with Unified Concept Editing (UCE)~\cite{gandikota2023unified} as a representative example. We find that attribution remains stable even after targeted concept removal. This is because CSF operates on broad semantic categories, such as animals or flowers, combined with compositional contextual conditions, rather than on a single narrow concept. Consequently, defeating CSF would require erasing broad semantic capabilities, which is difficult without significantly harming general model utility. Note that this robustness is orthogonal from the asymmetric advantage discussed earlier: even under explicit post-hoc semantic removal, CSF remains effective because it relies on broad semantic capabilities rather than isolated concept triggers.

\begin{figure}
    \centering
    \includegraphics[width=1\linewidth]{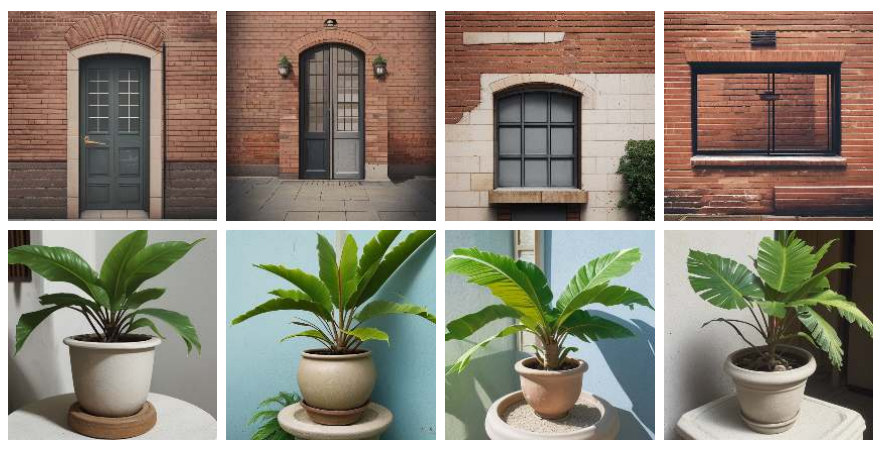}
    \caption{Failure cases of CSF. Top: prompts for baked goods in front of a brick wall. Bottom: prompts for tropical flowers on a pot. In both cases, models failed to generate identifiable objects within the target categories.}
    \label{fig:failure}
\end{figure}

\paragraph{Failure modes.}
Our method occasionally fails when models cannot follow the compositional 
prompt, generating objects outside the target category 
(Figure~\ref{fig:failure}). For instance, ``a savory baked good 
against a brick wall" or ``a tropical flower on a pot" produces 
unidentifiable objects in some models, preventing meaningful categorical 
classification. These generation failures are detectable via entropy thresholding, allowing automatic 
filtering of unreliable prompts.

\paragraph{Distance metric comparison.}
We validate the usage of Wasserstein distance by comparing Jensen-Shannon Divergence (JSD) 
as attribution metrics (\Cref{tab:metric_comparison}). While both 
achieve correct attribution, JSD shows weaker confidence, particularly for some models. However, 
Wasserstein maintains consistently high confidence, as it captures geometric structure in the probability 
simplex rather than treating categories independently. 

\paragraph{Prompt design validation.}
\Cref{fig:prompt_ablation} validates our compositional prompt design by demonstrating that underspecification elicits model-specific semantic biases. Varying only one contextual component, the categorical distributions shift dramatically across contexts. This context-dependent variation confirms that underspecified prompts force models to resolve ambiguity using their learned priors, revealing the semantic biases that constitute model fingerprints.

\section{Conclusion}

We present CSF, the first black-box fingerprinting method for text-to-image models that operates under the most restrictive query-only setting. Unlike prior fingerprinting approaches that rely on architectural access or heuristic similarity metrics, CSF provides statistical guarantees through a Bayesian attribution framework.
Critically, CSF reverses the asymmetry inherent in watermarking approaches. While watermarks can be removed once discovered, IP owners can generate infinitely many new compositional prompts post-deployment. This defender-favorable asymmetry, combined with training-free deployment and modest query costs (\$50 per model), establishes CSF as a practical solution for intellectual property protection in the era of API-distributed generative models.

\newpage

\clearpage


\begin{ack}
This work was supported by the Korean Government through the grants from IITP (RS-2021-II211343, RS-2022-II220320, RS-2025-25442338).
\end{ack}

{
    \small
    \bibliographystyle{ieeenat_fullname}
    \bibliography{main}
}
\setcounter{page}{1}
\maketitlesupplementary


\section{Model Selection}
\subsection{Base Models}

\begin{table*}[p]
\centering
\small
\caption{Model details for fingerprinting evaluation. All models are available at \url{https://huggingface.co/[model-path]}.}
\label{tab:model_specs}
\begin{tabular}{ll}
\toprule
\textbf{Model Name} & \textbf{HuggingFace Path} \\
\midrule
\multicolumn{2}{l}{\textit{Flux Family}} \\
Flux-Base & \texttt{black-forest-labs/FLUX.1-dev} \\
Flux-LoRA & \texttt{XLabs-AI/flux-RealismLora} \\
Flux-Turbo-Alpha & \texttt{alimama-creative/FLUX.1-Turbo-Alpha} \\
\midrule
\multicolumn{2}{l}{\textit{Kandinsky Family}} \\
Kandinsky-Base & \texttt{kandinsky-community/kandinsky-2-2-prior} \\
Kandinsky-Naruto & \texttt{JeonghyunLee/kandi2-prior-naruto-model} \\
Kandinsky-Pokemon-LoRA & \texttt{YiYiXu/pokeman\_kandinsky\_decoder\_lora} \\
\midrule
\multicolumn{2}{l}{\textit{SD1.5 Family}} \\
SD1.5-Base & \texttt{stable-diffusion-v1-5/stable-diffusion-v1-5} \\
SD1.5-1.2-Base & \texttt{CompVis/stable-diffusion-v1-2} \\
SD1.5-1.4-Base & \texttt{CompVis/stable-diffusion-v1-4} \\
SD1.5-DreamShaper & \texttt{Lykon/dreamshaper-8} \\
\midrule
\multicolumn{2}{l}{\textit{SD2.1 Family}} \\
SD2.1-Base & \texttt{stabilityai/stable-diffusion-2-1} \\
SD2.1-DPO & \texttt{radames/sd-21-DPO-LoRA} \\
SD2.1-LAION-Art & \texttt{Vishnou/sd-laion-art} \\
\midrule
\multicolumn{2}{l}{\textit{SD3 Family}} \\
SD3-Medium-Base & \texttt{stabilityai/stable-diffusion-3-medium-diffusers} \\
SD3-Reality-Mix & \texttt{terminusresearch/stable-diffusion-3.0-medium\_reality-mix} \\
SD3-VAE-Anime & \texttt{Disty0/sd3\_vae\_anime\_ft} \\
\midrule
\multicolumn{2}{l}{\textit{SDXL Family}} \\
SDXL-Base & \texttt{stabilityai/stable-diffusion-xl-base-1.0} \\
SDXL-DPO & \texttt{radames/sdxl-DPO-LoRA} \\
SDXL-Lightning-4Step & \texttt{ByteDance/SDXL-Lightning} \\
\bottomrule
\end{tabular}
\end{table*}

We include 6 base models representing the most prominent open-source diffusion model families. These models are selected due to their widespread adoption in both research and production environments, making them the primary targets for practical fingerprinting applications.

\paragraph{Stable Diffusion Family.} The Stable Diffusion series (SD1.5-Base, SD2.1-Base, SD3-Medium-Base, SDXL-Base) represents the dominant lineage in open-source text-to-image generation. SD1.5~\cite{rombach2022high} established the standard U-Net architecture with latent diffusion, becoming the foundation for thousands of derivatives. SD2.1 introduced architectural improvements including a larger U-Net and OpenCLIP text encoder. SD3~\cite{esser2024scaling} marked a major paradigm shift by adopting a transformer-based backbone similar to DiT~\cite{peebles2023scalable}, while SDXL~\cite{podell2023sdxl} scaled up model capacity for higher-resolution generation. These four models share the same latent diffusion framework but differ in architecture scale, training data, and design choices, presenting a challenging test for distinguishing models within the same lineage. We additionally include SD1.5-1.2-Base and SD1.5-1.4-Base, intermediate checkpoints from SD1.5's training trajectory, to evaluate whether fingerprints can differentiate between temporally close versions that share nearly identical architectures and overlapping training data.

\begin{table*}[tb]
\centering
\caption{Average Normalized Wasserstein Distance Matrix across all prompts. Each value is the mean of column-normalized distances across all 42 prompts. \colorbox[rgb]{1,0,0}{\phantom{xx}} Short, \colorbox[rgb]{1,1,0}{\phantom{xx}} Medium-low, \colorbox[rgb]{0,1,0}{\phantom{xx}} Medium-high, \colorbox[rgb]{0,0,1}{\phantom{xx}} Long distance.}
\label{tab:wasserstein_average_normalized}
\scalebox{0.60}{
\renewcommand{\arraystretch}{1.1}
\resizebox{\textwidth}{!}{%
\begin{tabular}{l|ccc|ccc|cccc|ccc|ccc|ccc} 
\hline\hline
& \multicolumn{19}{c}{\textbf{AVERAGE ACROSS ALL PROMPTS}} \\ 
\hline
& \multicolumn{3}{c|}{\textbf{Flux}} & \multicolumn{3}{c|}{\textbf{Kand}} & \multicolumn{4}{c|}{\textbf{SD1.5}} & \multicolumn{3}{c|}{\textbf{SD2.1}} & \multicolumn{3}{c|}{\textbf{SD3}} & \multicolumn{3}{c|}{\textbf{SDXL}} \\ 
\hline
& \rotatebox{45}{\small{F-Base}} & \rotatebox{45}{\small{F-LoRA}} & \rotatebox{45}{\small{F-Turbo}} & \rotatebox{45}{\small{K-Base}} & \rotatebox{45}{\small{K-Naru}} & \rotatebox{45}{\small{K-Poke}} & \rotatebox{45}{\small{1.5-v1.2}} & \rotatebox{45}{\small{1.5-v1.4}} & \rotatebox{45}{\small{1.5-Base}} & \rotatebox{45}{\small{1.5-Dream}} & \rotatebox{45}{\small{2.1-DPO}} & \rotatebox{45}{\small{2.1-Base}} & \rotatebox{45}{\small{2.1-Art}} & \rotatebox{45}{\small{SD3-Real}} & \rotatebox{45}{\small{SD3-Base}} & \rotatebox{45}{\small{SD3-Anim}} & \rotatebox{45}{\small{XL-DPO}} & \rotatebox{45}{\small{XL-Base}} & \rotatebox{45}{\small{XL-Light}} \\ 
\hline
\textbf{F-Base} & \cellcolor[rgb]{1.000,0.000,0.000}0.00 & \cellcolor[rgb]{0.960,1.000,0.000}0.34 & \cellcolor[rgb]{1.000,0.748,0.000}0.25 & \cellcolor[rgb]{0.000,0.873,0.127}0.71 & \cellcolor[rgb]{0.000,0.881,0.119}0.71 & \cellcolor[rgb]{0.000,0.886,0.114}0.71 & \cellcolor[rgb]{0.000,0.452,0.548}0.85 & \cellcolor[rgb]{0.000,0.486,0.514}0.84 & \cellcolor[rgb]{0.000,0.496,0.504}0.84 & \cellcolor[rgb]{0.000,0.689,0.311}0.77 & \cellcolor[rgb]{0.000,0.581,0.419}0.81 & \cellcolor[rgb]{0.000,0.515,0.485}0.83 & \cellcolor[rgb]{0.000,0.517,0.483}0.83 & \cellcolor[rgb]{0.108,1.000,0.000}0.63 & \cellcolor[rgb]{0.046,1.000,0.000}0.65 & \cellcolor[rgb]{0.033,1.000,0.000}0.66 & \cellcolor[rgb]{0.000,0.677,0.323}0.78 & \cellcolor[rgb]{0.000,0.643,0.357}0.79 & \cellcolor[rgb]{0.000,0.806,0.194}0.73 \\
\textbf{F-LoRA} & \cellcolor[rgb]{0.990,1.000,0.000}0.33 & \cellcolor[rgb]{1.000,0.000,0.000}0.00 & \cellcolor[rgb]{0.765,1.000,0.000}0.41 & \cellcolor[rgb]{0.000,0.867,0.133}0.71 & \cellcolor[rgb]{0.000,0.881,0.119}0.71 & \cellcolor[rgb]{0.000,0.745,0.255}0.75 & \cellcolor[rgb]{0.000,0.412,0.588}0.86 & \cellcolor[rgb]{0.000,0.512,0.488}0.83 & \cellcolor[rgb]{0.000,0.531,0.469}0.82 & \cellcolor[rgb]{0.000,0.573,0.427}0.81 & \cellcolor[rgb]{0.000,0.606,0.394}0.80 & \cellcolor[rgb]{0.000,0.588,0.412}0.81 & \cellcolor[rgb]{0.000,0.586,0.414}0.81 & \cellcolor[rgb]{0.000,0.978,0.022}0.68 & \cellcolor[rgb]{0.000,0.826,0.174}0.73 & \cellcolor[rgb]{0.000,0.806,0.194}0.73 & \cellcolor[rgb]{0.000,0.779,0.221}0.74 & \cellcolor[rgb]{0.000,0.789,0.211}0.74 & \cellcolor[rgb]{0.000,0.894,0.106}0.70 \\
\textbf{F-Turbo} & \cellcolor[rgb]{1.000,0.723,0.000}0.24 & \cellcolor[rgb]{0.766,1.000,0.000}0.41 & \cellcolor[rgb]{1.000,0.000,0.000}0.00 & \cellcolor[rgb]{0.000,0.808,0.192}0.73 & \cellcolor[rgb]{0.000,0.797,0.203}0.74 & \cellcolor[rgb]{0.000,0.826,0.174}0.73 & \cellcolor[rgb]{0.000,0.670,0.330}0.78 & \cellcolor[rgb]{0.000,0.677,0.323}0.78 & \cellcolor[rgb]{0.000,0.640,0.360}0.79 & \cellcolor[rgb]{0.000,0.710,0.290}0.77 & \cellcolor[rgb]{0.000,0.729,0.271}0.76 & \cellcolor[rgb]{0.000,0.697,0.303}0.77 & \cellcolor[rgb]{0.000,0.697,0.303}0.77 & \cellcolor[rgb]{0.000,1.000,0.000}0.67 & \cellcolor[rgb]{0.000,0.952,0.048}0.69 & \cellcolor[rgb]{0.000,0.945,0.055}0.69 & \cellcolor[rgb]{0.000,0.668,0.332}0.78 & \cellcolor[rgb]{0.000,0.626,0.374}0.79 & \cellcolor[rgb]{0.000,0.805,0.195}0.73 \\
\hline
\textbf{K-Base} & \cellcolor[rgb]{0.000,0.993,0.007}0.67 & \cellcolor[rgb]{0.000,0.903,0.097}0.70 & \cellcolor[rgb]{0.000,0.832,0.168}0.73 & \cellcolor[rgb]{1.000,0.000,0.000}0.00 & \cellcolor[rgb]{1.000,0.439,0.000}0.14 & \cellcolor[rgb]{0.960,1.000,0.000}0.34 & \cellcolor[rgb]{0.000,0.629,0.371}0.79 & \cellcolor[rgb]{0.000,0.501,0.499}0.83 & \cellcolor[rgb]{0.000,0.506,0.494}0.83 & \cellcolor[rgb]{0.000,0.672,0.328}0.78 & \cellcolor[rgb]{0.000,0.569,0.431}0.81 & \cellcolor[rgb]{0.000,0.633,0.367}0.79 & \cellcolor[rgb]{0.000,0.634,0.366}0.79 & \cellcolor[rgb]{0.000,0.681,0.319}0.78 & \cellcolor[rgb]{0.000,0.678,0.322}0.78 & \cellcolor[rgb]{0.000,0.668,0.332}0.78 & \cellcolor[rgb]{0.062,1.000,0.000}0.65 & \cellcolor[rgb]{0.045,1.000,0.000}0.65 & \cellcolor[rgb]{0.115,1.000,0.000}0.63 \\
\textbf{K-Naru} & \cellcolor[rgb]{0.012,1.000,0.000}0.67 & \cellcolor[rgb]{0.000,0.916,0.084}0.70 & \cellcolor[rgb]{0.000,0.831,0.169}0.73 & \cellcolor[rgb]{1.000,0.440,0.000}0.15 & \cellcolor[rgb]{1.000,0.000,0.000}0.00 & \cellcolor[rgb]{0.947,1.000,0.000}0.35 & \cellcolor[rgb]{0.000,0.593,0.407}0.80 & \cellcolor[rgb]{0.000,0.484,0.516}0.84 & \cellcolor[rgb]{0.000,0.494,0.506}0.84 & \cellcolor[rgb]{0.000,0.661,0.339}0.78 & \cellcolor[rgb]{0.000,0.571,0.429}0.81 & \cellcolor[rgb]{0.000,0.617,0.383}0.80 & \cellcolor[rgb]{0.000,0.621,0.379}0.79 & \cellcolor[rgb]{0.000,0.799,0.201}0.74 & \cellcolor[rgb]{0.000,0.722,0.278}0.76 & \cellcolor[rgb]{0.000,0.710,0.290}0.77 & \cellcolor[rgb]{0.122,1.000,0.000}0.63 & \cellcolor[rgb]{0.095,1.000,0.000}0.64 & \cellcolor[rgb]{0.150,1.000,0.000}0.62 \\
\textbf{K-Poke} & \cellcolor[rgb]{0.000,0.936,0.064}0.69 & \cellcolor[rgb]{0.000,0.716,0.284}0.76 & \cellcolor[rgb]{0.000,0.757,0.243}0.75 & \cellcolor[rgb]{0.923,1.000,0.000}0.36 & \cellcolor[rgb]{0.924,1.000,0.000}0.36 & \cellcolor[rgb]{1.000,0.000,0.000}0.00 & \cellcolor[rgb]{0.000,0.395,0.605}0.87 & \cellcolor[rgb]{0.000,0.369,0.631}0.88 & \cellcolor[rgb]{0.000,0.338,0.662}0.89 & \cellcolor[rgb]{0.000,0.656,0.344}0.78 & \cellcolor[rgb]{0.000,0.560,0.440}0.82 & \cellcolor[rgb]{0.000,0.534,0.466}0.82 & \cellcolor[rgb]{0.000,0.539,0.461}0.82 & \cellcolor[rgb]{0.000,0.869,0.131}0.71 & \cellcolor[rgb]{0.000,0.789,0.211}0.74 & \cellcolor[rgb]{0.000,0.775,0.225}0.74 & \cellcolor[rgb]{0.100,1.000,0.000}0.64 & \cellcolor[rgb]{0.017,1.000,0.000}0.66 & \cellcolor[rgb]{0.015,1.000,0.000}0.66 \\
\hline
\textbf{1.5-v1.2} & \cellcolor[rgb]{0.000,0.259,0.741}0.91 & \cellcolor[rgb]{0.000,0.139,0.861}0.95 & \cellcolor[rgb]{0.000,0.364,0.636}0.88 & \cellcolor[rgb]{0.000,0.324,0.676}0.89 & \cellcolor[rgb]{0.000,0.272,0.728}0.91 & \cellcolor[rgb]{0.000,0.202,0.798}0.93 & \cellcolor[rgb]{1.000,0.000,0.000}0.00 & \cellcolor[rgb]{0.753,1.000,0.000}0.41 & \cellcolor[rgb]{0.624,1.000,0.000}0.46 & \cellcolor[rgb]{0.274,1.000,0.000}0.58 & \cellcolor[rgb]{0.000,0.595,0.405}0.80 & \cellcolor[rgb]{0.000,0.800,0.200}0.74 & \cellcolor[rgb]{0.000,0.802,0.198}0.74 & \cellcolor[rgb]{0.000,0.119,0.881}0.96 & \cellcolor[rgb]{0.000,0.162,0.838}0.95 & \cellcolor[rgb]{0.000,0.160,0.840}0.95 & \cellcolor[rgb]{0.000,0.381,0.619}0.87 & \cellcolor[rgb]{0.000,0.487,0.513}0.84 & \cellcolor[rgb]{0.000,0.420,0.580}0.86 \\
\textbf{1.5-v1.4} & \cellcolor[rgb]{0.000,0.588,0.412}0.81 & \cellcolor[rgb]{0.000,0.535,0.465}0.82 & \cellcolor[rgb]{0.000,0.649,0.351}0.79 & \cellcolor[rgb]{0.000,0.507,0.493}0.83 & \cellcolor[rgb]{0.000,0.496,0.504}0.84 & \cellcolor[rgb]{0.000,0.501,0.499}0.83 & \cellcolor[rgb]{0.882,1.000,0.000}0.37 & \cellcolor[rgb]{1.000,0.000,0.000}0.00 & \cellcolor[rgb]{0.968,1.000,0.000}0.34 & \cellcolor[rgb]{0.477,1.000,0.000}0.51 & \cellcolor[rgb]{0.008,1.000,0.000}0.67 & \cellcolor[rgb]{0.122,1.000,0.000}0.63 & \cellcolor[rgb]{0.122,1.000,0.000}0.63 & \cellcolor[rgb]{0.000,0.681,0.319}0.78 & \cellcolor[rgb]{0.000,0.589,0.411}0.81 & \cellcolor[rgb]{0.000,0.585,0.415}0.81 & \cellcolor[rgb]{0.000,0.695,0.305}0.77 & \cellcolor[rgb]{0.000,0.756,0.244}0.75 & \cellcolor[rgb]{0.000,0.651,0.349}0.79 \\
\textbf{1.5-Base} & \cellcolor[rgb]{0.000,0.670,0.330}0.78 & \cellcolor[rgb]{0.000,0.641,0.359}0.79 & \cellcolor[rgb]{0.000,0.704,0.296}0.77 & \cellcolor[rgb]{0.000,0.609,0.391}0.80 & \cellcolor[rgb]{0.000,0.604,0.396}0.80 & \cellcolor[rgb]{0.000,0.566,0.434}0.81 & \cellcolor[rgb]{0.813,1.000,0.000}0.39 & \cellcolor[rgb]{1.000,0.999,0.000}0.33 & \cellcolor[rgb]{1.000,0.000,0.000}0.00 & \cellcolor[rgb]{0.488,1.000,0.000}0.50 & \cellcolor[rgb]{0.037,1.000,0.000}0.66 & \cellcolor[rgb]{0.149,1.000,0.000}0.62 & \cellcolor[rgb]{0.153,1.000,0.000}0.62 & \cellcolor[rgb]{0.000,0.733,0.267}0.76 & \cellcolor[rgb]{0.000,0.635,0.365}0.79 & \cellcolor[rgb]{0.000,0.630,0.370}0.79 & \cellcolor[rgb]{0.000,0.782,0.218}0.74 & \cellcolor[rgb]{0.000,0.851,0.149}0.72 & \cellcolor[rgb]{0.000,0.749,0.251}0.75 \\
\textbf{1.5-Dream} & \cellcolor[rgb]{0.000,0.620,0.380}0.80 & \cellcolor[rgb]{0.000,0.445,0.555}0.85 & \cellcolor[rgb]{0.000,0.556,0.444}0.82 & \cellcolor[rgb]{0.000,0.476,0.524}0.84 & \cellcolor[rgb]{0.000,0.462,0.538}0.85 & \cellcolor[rgb]{0.000,0.565,0.435}0.81 & \cellcolor[rgb]{0.362,1.000,0.000}0.55 & \cellcolor[rgb]{0.363,1.000,0.000}0.55 & \cellcolor[rgb]{0.322,1.000,0.000}0.56 & \cellcolor[rgb]{1.000,0.000,0.000}0.00 & \cellcolor[rgb]{0.000,0.595,0.405}0.80 & \cellcolor[rgb]{0.000,0.614,0.386}0.80 & \cellcolor[rgb]{0.000,0.617,0.383}0.80 & \cellcolor[rgb]{0.000,0.705,0.295}0.77 & \cellcolor[rgb]{0.000,0.494,0.506}0.84 & \cellcolor[rgb]{0.000,0.482,0.518}0.84 & \cellcolor[rgb]{0.000,0.540,0.460}0.82 & \cellcolor[rgb]{0.000,0.600,0.400}0.80 & \cellcolor[rgb]{0.000,0.604,0.396}0.80 \\
\hline
\textbf{2.1-DPO} & \cellcolor[rgb]{0.000,0.750,0.250}0.75 & \cellcolor[rgb]{0.000,0.709,0.291}0.77 & \cellcolor[rgb]{0.000,0.781,0.219}0.74 & \cellcolor[rgb]{0.000,0.648,0.352}0.79 & \cellcolor[rgb]{0.000,0.661,0.339}0.78 & \cellcolor[rgb]{0.000,0.733,0.267}0.76 & \cellcolor[rgb]{0.000,0.925,0.075}0.69 & \cellcolor[rgb]{0.051,1.000,0.000}0.65 & \cellcolor[rgb]{0.018,1.000,0.000}0.66 & \cellcolor[rgb]{0.000,0.838,0.162}0.72 & \cellcolor[rgb]{1.000,0.000,0.000}0.00 & \cellcolor[rgb]{1.000,0.776,0.000}0.26 & \cellcolor[rgb]{1.000,0.780,0.000}0.26 & \cellcolor[rgb]{0.102,1.000,0.000}0.64 & \cellcolor[rgb]{0.000,0.865,0.135}0.71 & \cellcolor[rgb]{0.000,0.861,0.139}0.72 & \cellcolor[rgb]{0.000,0.796,0.204}0.74 & \cellcolor[rgb]{0.000,0.816,0.184}0.73 & \cellcolor[rgb]{0.000,0.653,0.347}0.78 \\
\textbf{2.1-Base} & \cellcolor[rgb]{0.000,0.645,0.355}0.79 & \cellcolor[rgb]{0.000,0.644,0.356}0.79 & \cellcolor[rgb]{0.000,0.705,0.295}0.77 & \cellcolor[rgb]{0.000,0.619,0.381}0.80 & \cellcolor[rgb]{0.000,0.609,0.391}0.80 & \cellcolor[rgb]{0.000,0.638,0.362}0.79 & \cellcolor[rgb]{0.052,1.000,0.000}0.65 & \cellcolor[rgb]{0.129,1.000,0.000}0.63 & \cellcolor[rgb]{0.091,1.000,0.000}0.64 & \cellcolor[rgb]{0.000,0.823,0.177}0.73 & \cellcolor[rgb]{1.000,0.797,0.000}0.26 & \cellcolor[rgb]{1.000,0.000,0.000}0.00 & \cellcolor[rgb]{1.000,0.057,0.000}0.02 & \cellcolor[rgb]{0.000,0.991,0.009}0.67 & \cellcolor[rgb]{0.000,0.786,0.214}0.74 & \cellcolor[rgb]{0.000,0.781,0.219}0.74 & \cellcolor[rgb]{0.000,0.749,0.251}0.75 & \cellcolor[rgb]{0.000,0.773,0.227}0.74 & \cellcolor[rgb]{0.000,0.613,0.387}0.80 \\
\textbf{2.1-Art} & \cellcolor[rgb]{0.000,0.651,0.349}0.79 & \cellcolor[rgb]{0.000,0.648,0.352}0.79 & \cellcolor[rgb]{0.000,0.708,0.292}0.77 & \cellcolor[rgb]{0.000,0.630,0.370}0.79 & \cellcolor[rgb]{0.000,0.623,0.377}0.79 & \cellcolor[rgb]{0.000,0.652,0.348}0.78 & \cellcolor[rgb]{0.061,1.000,0.000}0.65 & \cellcolor[rgb]{0.137,1.000,0.000}0.62 & \cellcolor[rgb]{0.104,1.000,0.000}0.63 & \cellcolor[rgb]{0.000,0.834,0.166}0.72 & \cellcolor[rgb]{1.000,0.798,0.000}0.26 & \cellcolor[rgb]{1.000,0.057,0.000}0.02 & \cellcolor[rgb]{1.000,0.000,0.000}0.00 & \cellcolor[rgb]{0.000,0.999,0.001}0.67 & \cellcolor[rgb]{0.000,0.796,0.204}0.74 & \cellcolor[rgb]{0.000,0.790,0.210}0.74 & \cellcolor[rgb]{0.000,0.755,0.245}0.75 & \cellcolor[rgb]{0.000,0.779,0.221}0.74 & \cellcolor[rgb]{0.000,0.623,0.377}0.79 \\
\hline
\textbf{SD3-Real} & \cellcolor[rgb]{0.000,0.822,0.178}0.73 & \cellcolor[rgb]{0.000,0.666,0.334}0.78 & \cellcolor[rgb]{0.000,0.611,0.389}0.80 & \cellcolor[rgb]{0.000,0.099,0.901}0.97 & \cellcolor[rgb]{0.000,0.239,0.761}0.92 & \cellcolor[rgb]{0.000,0.397,0.603}0.87 & \cellcolor[rgb]{0.000,0.000,1.000}1.04 & \cellcolor[rgb]{0.000,0.118,0.882}0.96 & \cellcolor[rgb]{0.000,0.102,0.898}0.97 & \cellcolor[rgb]{0.000,0.460,0.540}0.85 & \cellcolor[rgb]{0.000,0.559,0.441}0.82 & \cellcolor[rgb]{0.000,0.475,0.525}0.84 & \cellcolor[rgb]{0.000,0.478,0.522}0.84 & \cellcolor[rgb]{1.000,0.000,0.000}0.00 & \cellcolor[rgb]{0.392,1.000,0.000}0.54 & \cellcolor[rgb]{0.398,1.000,0.000}0.53 & \cellcolor[rgb]{0.000,0.584,0.416}0.81 & \cellcolor[rgb]{0.000,0.536,0.464}0.82 & \cellcolor[rgb]{0.000,0.396,0.604}0.87 \\
\textbf{SD3-Base} & \cellcolor[rgb]{0.240,1.000,0.000}0.59 & \cellcolor[rgb]{0.000,0.985,0.015}0.67 & \cellcolor[rgb]{0.113,1.000,0.000}0.63 & \cellcolor[rgb]{0.000,0.806,0.194}0.73 & \cellcolor[rgb]{0.000,0.837,0.163}0.72 & \cellcolor[rgb]{0.000,0.980,0.020}0.68 & \cellcolor[rgb]{0.000,0.665,0.335}0.78 & \cellcolor[rgb]{0.000,0.747,0.253}0.75 & \cellcolor[rgb]{0.000,0.721,0.279}0.76 & \cellcolor[rgb]{0.000,0.805,0.195}0.73 & \cellcolor[rgb]{0.000,0.978,0.022}0.68 & \cellcolor[rgb]{0.000,0.960,0.040}0.68 & \cellcolor[rgb]{0.000,0.963,0.037}0.68 & \cellcolor[rgb]{0.797,1.000,0.000}0.40 & \cellcolor[rgb]{1.000,0.000,0.000}0.00 & \cellcolor[rgb]{1.000,0.009,0.000}0.00 & \cellcolor[rgb]{0.099,1.000,0.000}0.64 & \cellcolor[rgb]{0.064,1.000,0.000}0.65 & \cellcolor[rgb]{0.010,1.000,0.000}0.67 \\
\textbf{SD3-Anim} & \cellcolor[rgb]{0.238,1.000,0.000}0.59 & \cellcolor[rgb]{0.000,0.979,0.021}0.68 & \cellcolor[rgb]{0.116,1.000,0.000}0.63 & \cellcolor[rgb]{0.000,0.813,0.187}0.73 & \cellcolor[rgb]{0.000,0.841,0.159}0.72 & \cellcolor[rgb]{0.000,0.988,0.012}0.67 & \cellcolor[rgb]{0.000,0.677,0.323}0.78 & \cellcolor[rgb]{0.000,0.758,0.242}0.75 & \cellcolor[rgb]{0.000,0.731,0.269}0.76 & \cellcolor[rgb]{0.000,0.811,0.189}0.73 & \cellcolor[rgb]{0.000,0.988,0.012}0.67 & \cellcolor[rgb]{0.000,0.969,0.031}0.68 & \cellcolor[rgb]{0.000,0.971,0.029}0.68 & \cellcolor[rgb]{0.806,1.000,0.000}0.40 & \cellcolor[rgb]{1.000,0.009,0.000}0.00 & \cellcolor[rgb]{1.000,0.000,0.000}0.00 & \cellcolor[rgb]{0.093,1.000,0.000}0.64 & \cellcolor[rgb]{0.056,1.000,0.000}0.65 & \cellcolor[rgb]{0.007,1.000,0.000}0.67 \\
\hline
\textbf{XL-DPO} & \cellcolor[rgb]{0.000,0.984,0.016}0.68 & \cellcolor[rgb]{0.000,0.992,0.008}0.67 & \cellcolor[rgb]{0.000,0.879,0.121}0.71 & \cellcolor[rgb]{0.246,1.000,0.000}0.59 & \cellcolor[rgb]{0.292,1.000,0.000}0.57 & \cellcolor[rgb]{0.309,1.000,0.000}0.56 & \cellcolor[rgb]{0.000,0.915,0.085}0.70 & \cellcolor[rgb]{0.000,0.888,0.112}0.71 & \cellcolor[rgb]{0.000,0.880,0.120}0.71 & \cellcolor[rgb]{0.000,0.896,0.104}0.70 & \cellcolor[rgb]{0.000,0.937,0.063}0.69 & \cellcolor[rgb]{0.000,0.945,0.055}0.69 & \cellcolor[rgb]{0.000,0.942,0.058}0.69 & \cellcolor[rgb]{0.234,1.000,0.000}0.59 & \cellcolor[rgb]{0.137,1.000,0.000}0.62 & \cellcolor[rgb]{0.120,1.000,0.000}0.63 & \cellcolor[rgb]{1.000,0.000,0.000}0.00 & \cellcolor[rgb]{1.000,0.553,0.000}0.18 & \cellcolor[rgb]{0.919,1.000,0.000}0.36 \\
\textbf{XL-Base} & \cellcolor[rgb]{0.000,0.889,0.111}0.71 & \cellcolor[rgb]{0.000,0.944,0.056}0.69 & \cellcolor[rgb]{0.000,0.776,0.224}0.74 & \cellcolor[rgb]{0.179,1.000,0.000}0.61 & \cellcolor[rgb]{0.221,1.000,0.000}0.59 & \cellcolor[rgb]{0.177,1.000,0.000}0.61 & \cellcolor[rgb]{0.000,0.954,0.046}0.69 & \cellcolor[rgb]{0.000,0.884,0.116}0.71 & \cellcolor[rgb]{0.000,0.880,0.120}0.71 & \cellcolor[rgb]{0.000,0.886,0.114}0.71 & \cellcolor[rgb]{0.000,0.892,0.108}0.71 & \cellcolor[rgb]{0.000,0.907,0.093}0.70 & \cellcolor[rgb]{0.000,0.904,0.096}0.70 & \cellcolor[rgb]{0.141,1.000,0.000}0.62 & \cellcolor[rgb]{0.037,1.000,0.000}0.66 & \cellcolor[rgb]{0.019,1.000,0.000}0.66 & \cellcolor[rgb]{1.000,0.568,0.000}0.19 & \cellcolor[rgb]{1.000,0.000,0.000}0.00 & \cellcolor[rgb]{0.901,1.000,0.000}0.36 \\
\textbf{XL-Light} & \cellcolor[rgb]{0.012,1.000,0.000}0.67 & \cellcolor[rgb]{0.013,1.000,0.000}0.67 & \cellcolor[rgb]{0.000,0.922,0.078}0.70 & \cellcolor[rgb]{0.205,1.000,0.000}0.60 & \cellcolor[rgb]{0.232,1.000,0.000}0.59 & \cellcolor[rgb]{0.122,1.000,0.000}0.63 & \cellcolor[rgb]{0.000,0.904,0.096}0.70 & \cellcolor[rgb]{0.000,0.766,0.234}0.75 & \cellcolor[rgb]{0.000,0.764,0.236}0.75 & \cellcolor[rgb]{0.000,0.859,0.141}0.72 & \cellcolor[rgb]{0.000,0.707,0.293}0.77 & \cellcolor[rgb]{0.000,0.751,0.249}0.75 & \cellcolor[rgb]{0.000,0.754,0.246}0.75 & \cellcolor[rgb]{0.007,1.000,0.000}0.67 & \cellcolor[rgb]{0.000,0.943,0.057}0.69 & \cellcolor[rgb]{0.000,0.929,0.071}0.69 & \cellcolor[rgb]{0.889,1.000,0.000}0.37 & \cellcolor[rgb]{0.902,1.000,0.000}0.36 & \cellcolor[rgb]{1.000,0.000,0.000}0.00 \\
\hline\hline
\end{tabular}
}
}
\end{table*}

\paragraph{Flux Family.} Flux-Base~\cite{flux2024} represents a recent shift toward fully transformer-based architectures, departing from the U-Net paradigm. Its adoption of attention-based processing throughout the model challenges whether fingerprinting methods designed for convolutional architectures can generalize to transformer-native designs.

\paragraph{Kandinsky Family.} Kandinsky-Base~\cite{razzhigaev2023kandinsky} employs a unique two-stage pipeline with separate prior and decoder models, where the prior maps text to image embeddings before the decoder generates pixels. This architectural distinction from single-stage models like Stable Diffusion tests whether fingerprints can adapt to multi-stage generation processes.
\begin{figure*}[p]
\centering
\small
\setlength{\tabcolsep}{2pt}
\resizebox{\textwidth}{!}{%
\begin{tabular}{l*{10}{c}}
\toprule
\multicolumn{11}{c}{\textbf{Prompt: ``\cheesytitle''}} \\
\midrule
\textbf{Model} & \textbf{S1} & \textbf{S2} & \textbf{S3} & \textbf{S4} & \textbf{S5} & \textbf{S6} & \textbf{S7} & \textbf{S8} & \textbf{S9} & \textbf{S10} \\
\midrule
\textit{Flux Family} \\
Flux-Base & 
\includegraphics[width=0.08\textwidth]{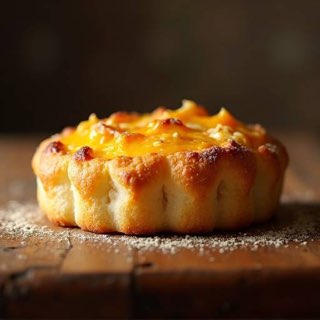} &
\includegraphics[width=0.08\textwidth]{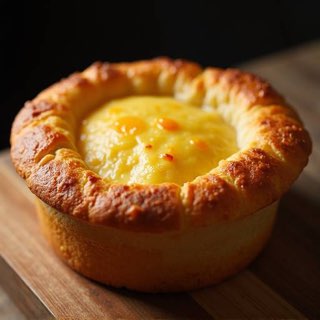} &
\includegraphics[width=0.08\textwidth]{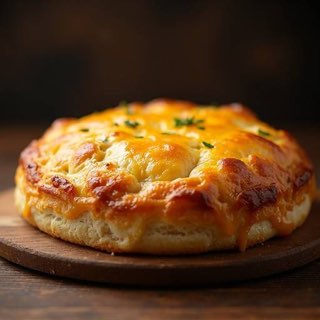} &
\includegraphics[width=0.08\textwidth]{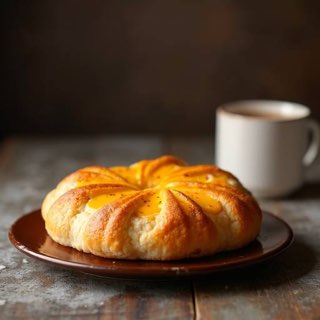} &
\includegraphics[width=0.08\textwidth]{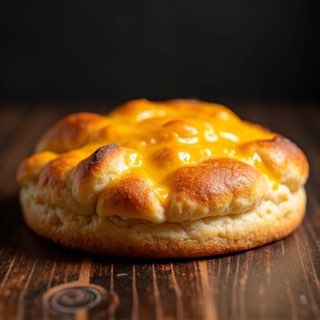} &
\includegraphics[width=0.08\textwidth]{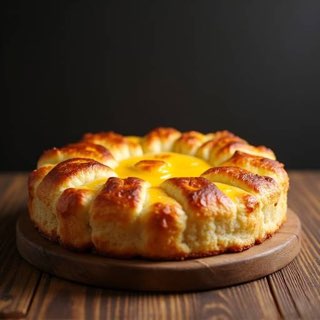} &
\includegraphics[width=0.08\textwidth]{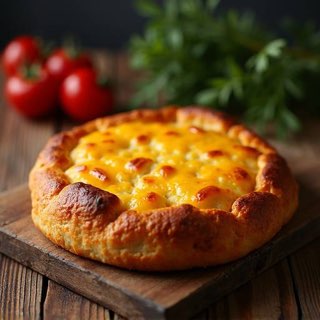} &
\includegraphics[width=0.08\textwidth]{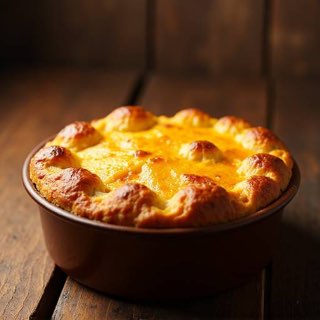} &
\includegraphics[width=0.08\textwidth]{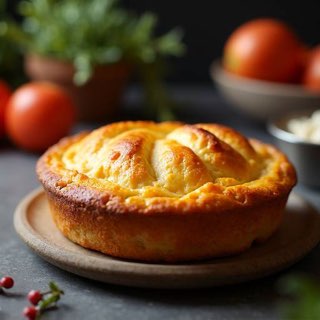} &
\includegraphics[width=0.08\textwidth]{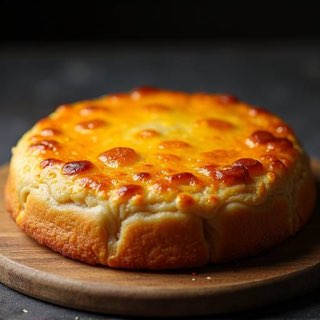} \\
Flux-LoRA & 
\includegraphics[width=0.08\textwidth]{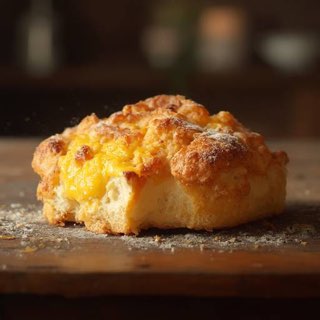} &
\includegraphics[width=0.08\textwidth]{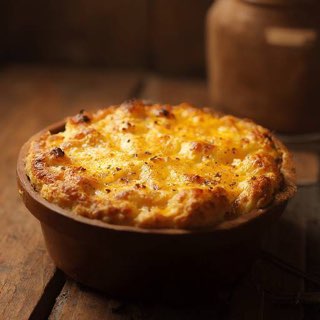} &
\includegraphics[width=0.08\textwidth]{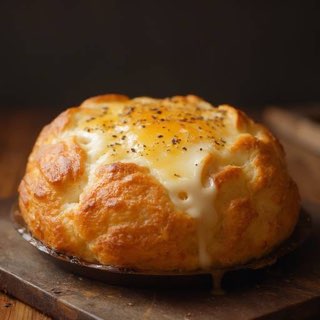} &
\includegraphics[width=0.08\textwidth]{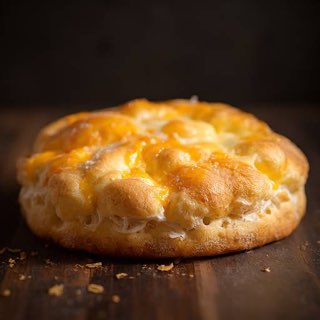} &
\includegraphics[width=0.08\textwidth]{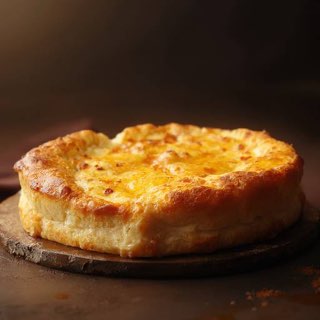} &
\includegraphics[width=0.08\textwidth]{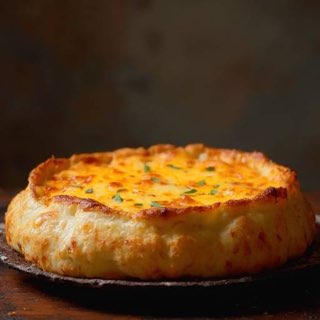} &
\includegraphics[width=0.08\textwidth]{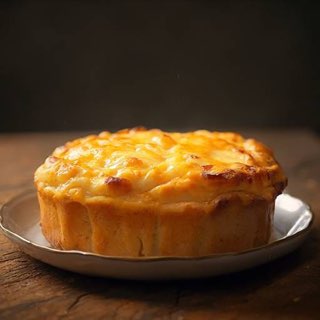} &
\includegraphics[width=0.08\textwidth]{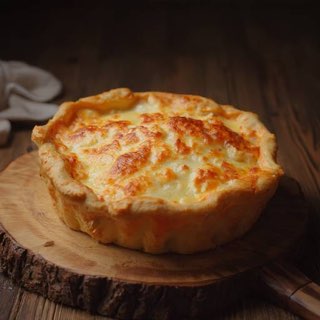} &
\includegraphics[width=0.08\textwidth]{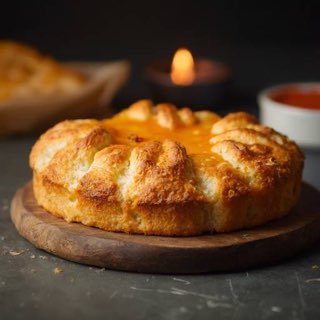} &
\includegraphics[width=0.08\textwidth]{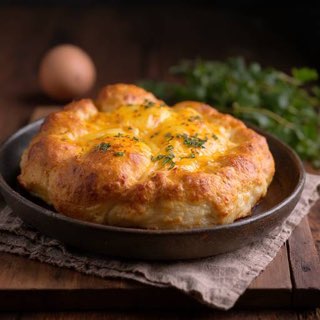} \\
Flux-Turbo-Alpha & 
\includegraphics[width=0.08\textwidth]{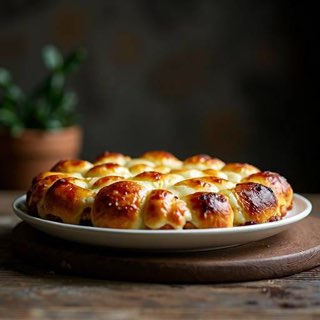} &
\includegraphics[width=0.08\textwidth]{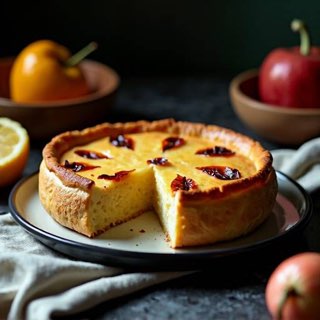} &
\includegraphics[width=0.08\textwidth]{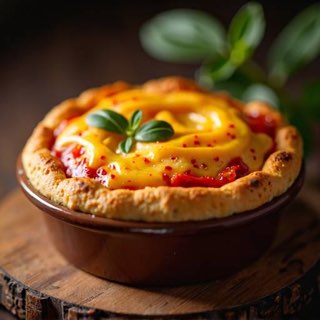} &
\includegraphics[width=0.08\textwidth]{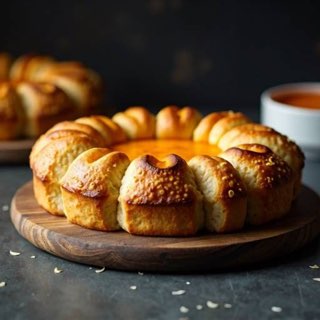} &
\includegraphics[width=0.08\textwidth]{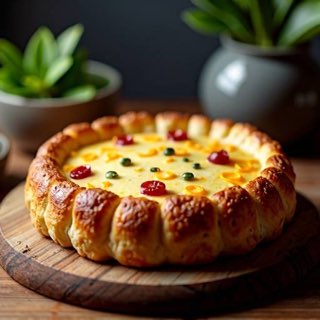} &
\includegraphics[width=0.08\textwidth]{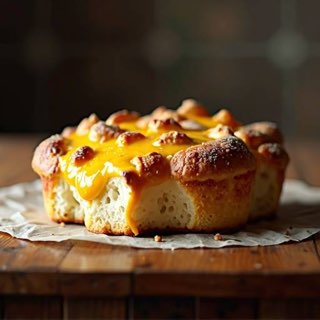} &
\includegraphics[width=0.08\textwidth]{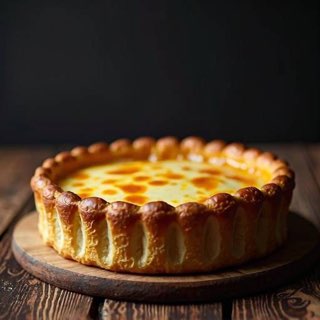} &
\includegraphics[width=0.08\textwidth]{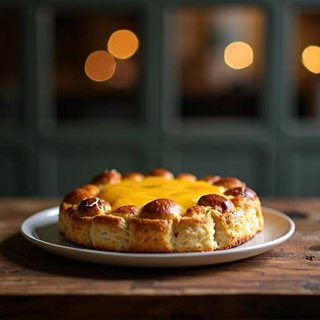} &
\includegraphics[width=0.08\textwidth]{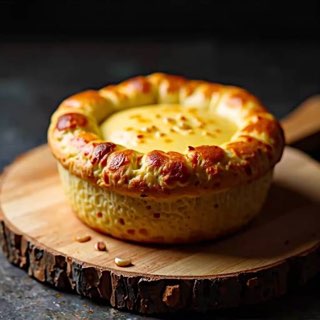} &
\includegraphics[width=0.08\textwidth]{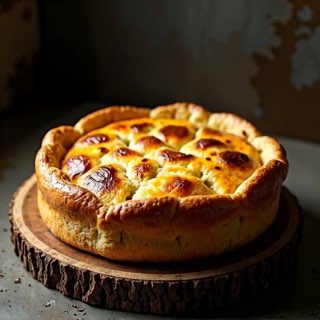} \\
\midrule
\textit{Kandinsky Family} \\
Kandinsky-Base & 
\includegraphics[width=0.08\textwidth]{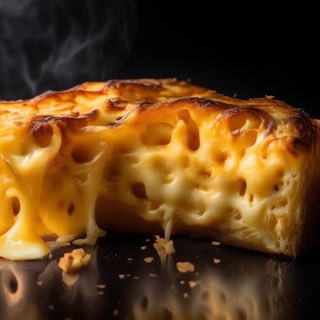} &
\includegraphics[width=0.08\textwidth]{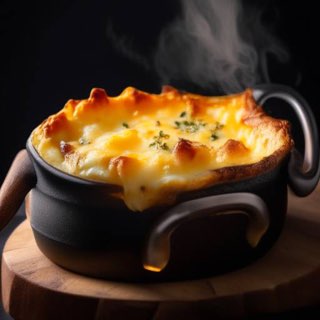} &
\includegraphics[width=0.08\textwidth]{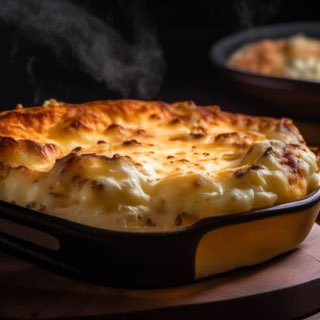} &
\includegraphics[width=0.08\textwidth]{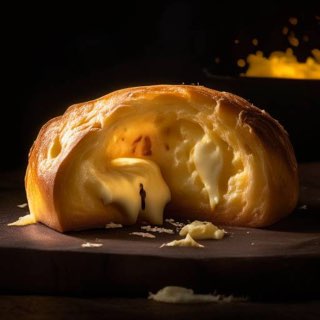} &
\includegraphics[width=0.08\textwidth]{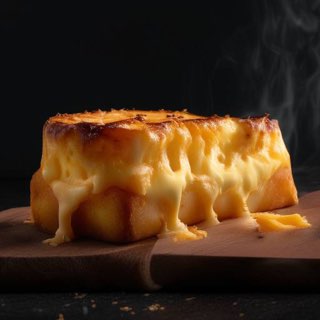} &
\includegraphics[width=0.08\textwidth]{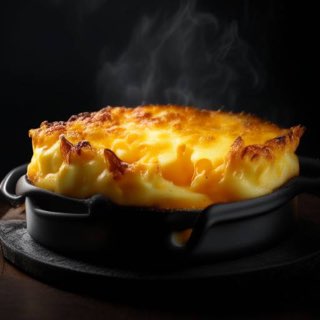} &
\includegraphics[width=0.08\textwidth]{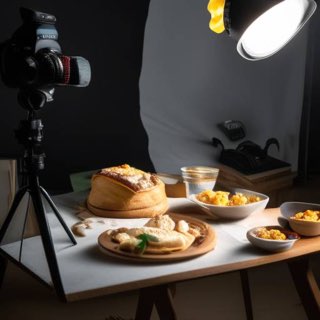} &
\includegraphics[width=0.08\textwidth]{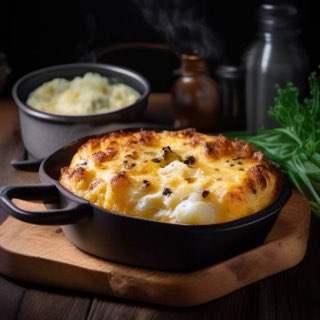} &
\includegraphics[width=0.08\textwidth]{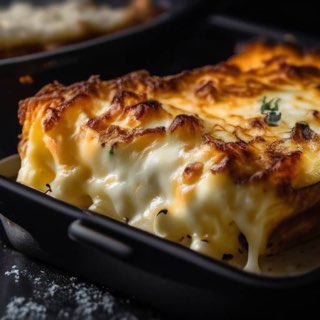} &
\includegraphics[width=0.08\textwidth]{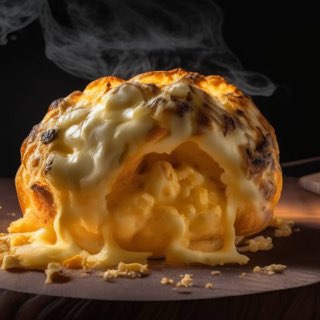} \\
Kandinsky-Naruto & 
\includegraphics[width=0.08\textwidth]{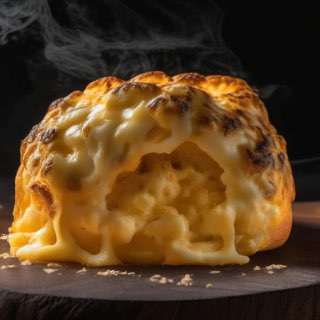} &
\includegraphics[width=0.08\textwidth]{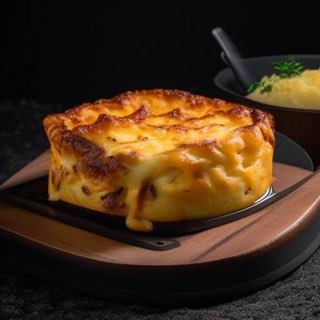} &
\includegraphics[width=0.08\textwidth]{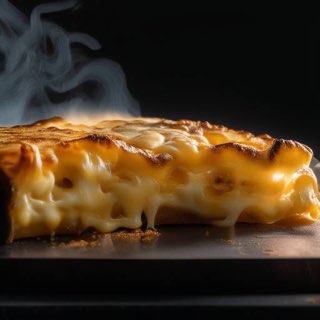} &
\includegraphics[width=0.08\textwidth]{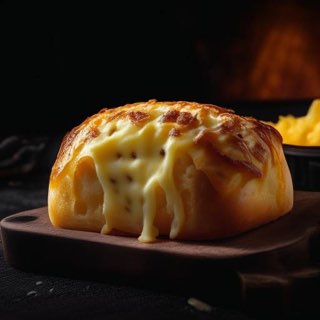} &
\includegraphics[width=0.08\textwidth]{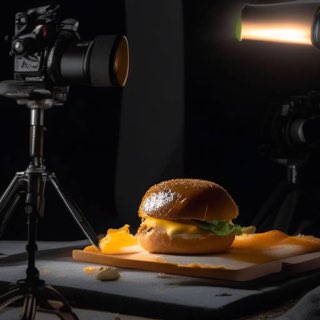} &
\includegraphics[width=0.08\textwidth]{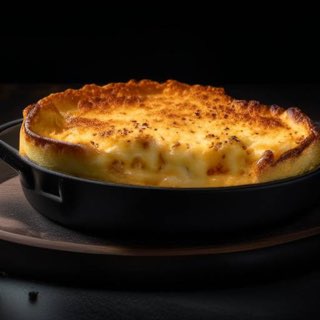} &
\includegraphics[width=0.08\textwidth]{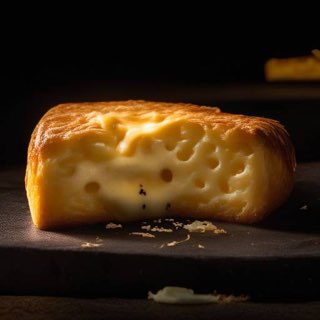} &
\includegraphics[width=0.08\textwidth]{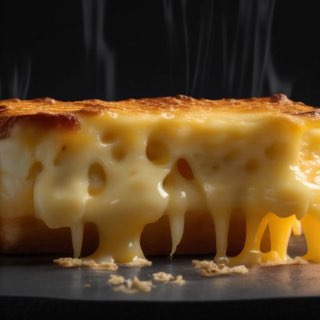} &
\includegraphics[width=0.08\textwidth]{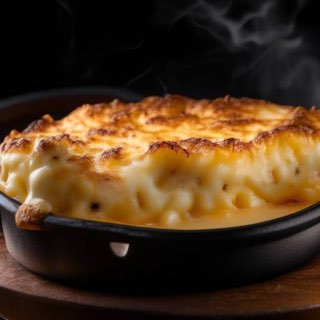} &
\includegraphics[width=0.08\textwidth]{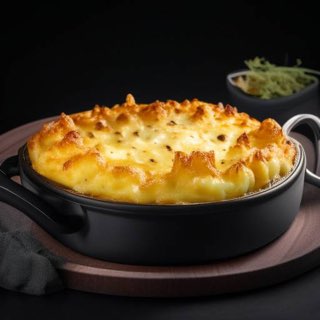} \\
Kandinsky-Pokemon-LoRA & 
\includegraphics[width=0.08\textwidth]{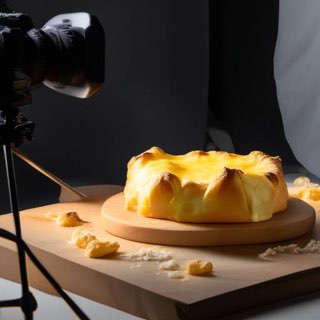} &
\includegraphics[width=0.08\textwidth]{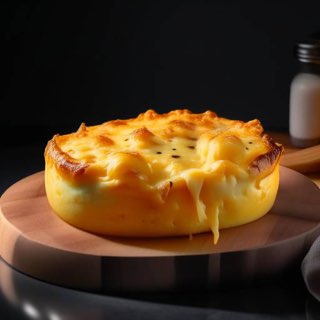} &
\includegraphics[width=0.08\textwidth]{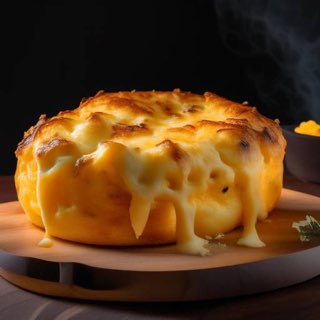} &
\includegraphics[width=0.08\textwidth]{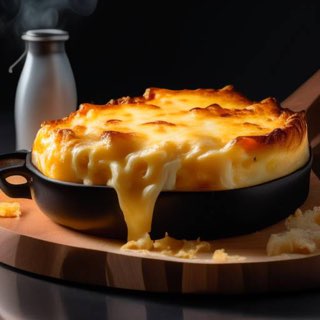} &
\includegraphics[width=0.08\textwidth]{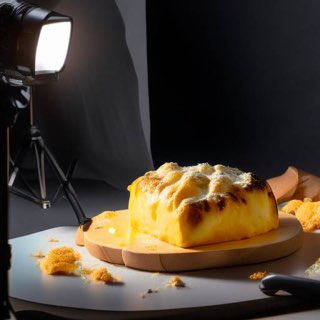} &
\includegraphics[width=0.08\textwidth]{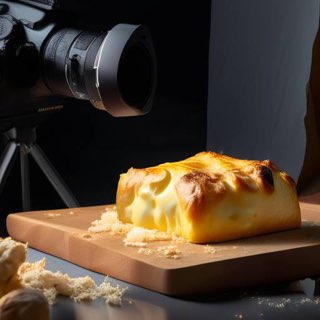} &
\includegraphics[width=0.08\textwidth]{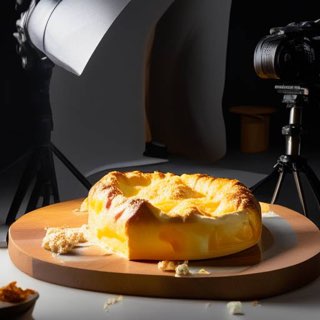} &
\includegraphics[width=0.08\textwidth]{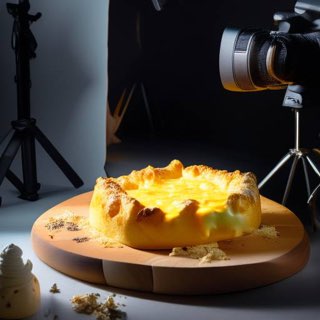} &
\includegraphics[width=0.08\textwidth]{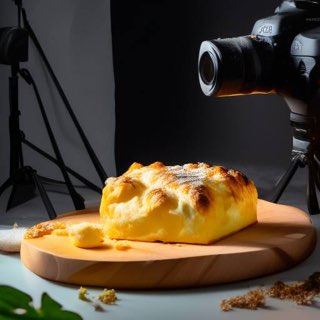} &
\includegraphics[width=0.08\textwidth]{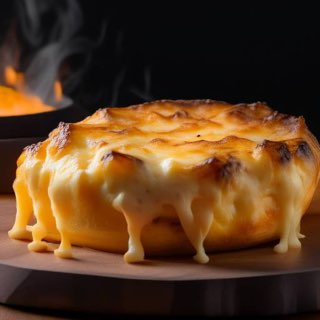} \\
\midrule
\textit{SD1.5 Family} \\
SD1.5-Base & 
\includegraphics[width=0.08\textwidth]{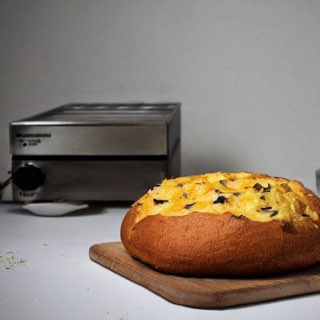} &
\includegraphics[width=0.08\textwidth]{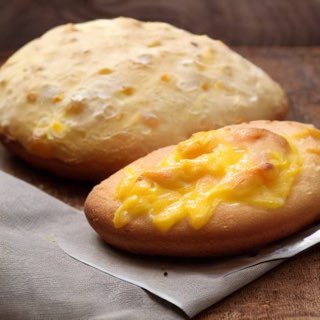} &
\includegraphics[width=0.08\textwidth]{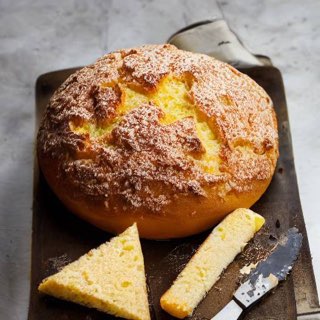} &
\includegraphics[width=0.08\textwidth]{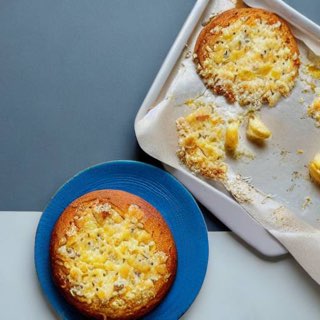} &
\includegraphics[width=0.08\textwidth]{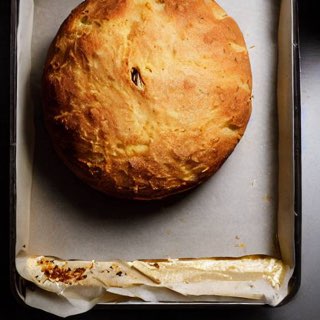} &
\includegraphics[width=0.08\textwidth]{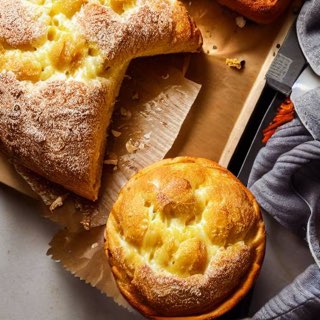} &
\includegraphics[width=0.08\textwidth]{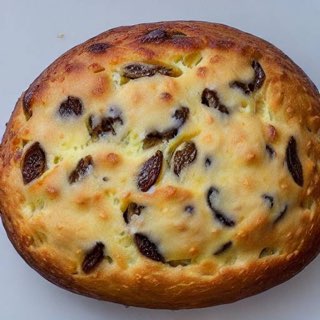} &
\includegraphics[width=0.08\textwidth]{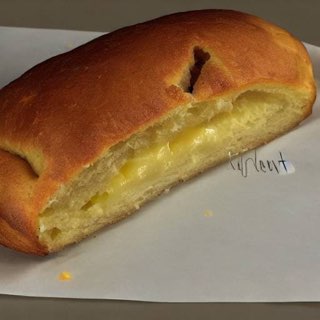} &
\includegraphics[width=0.08\textwidth]{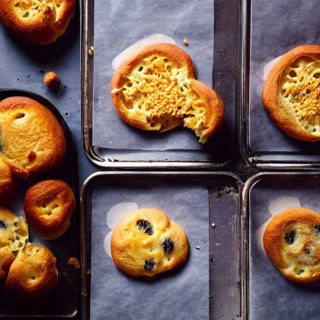} &
\includegraphics[width=0.08\textwidth]{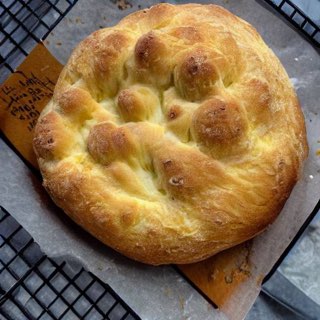} \\
SD1.5-1.2-Base & 
\includegraphics[width=0.08\textwidth]{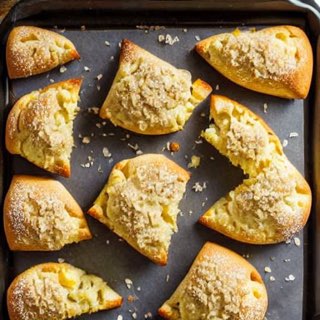} &
\includegraphics[width=0.08\textwidth]{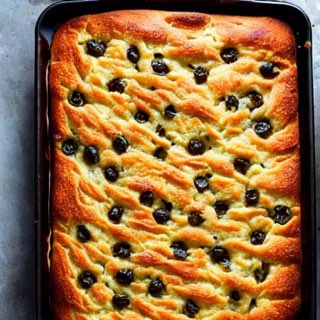} &
\includegraphics[width=0.08\textwidth]{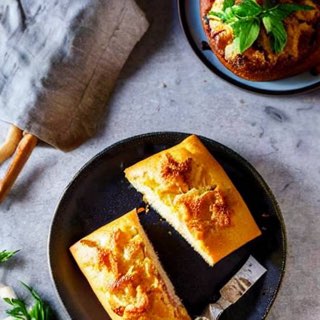} &
\includegraphics[width=0.08\textwidth]{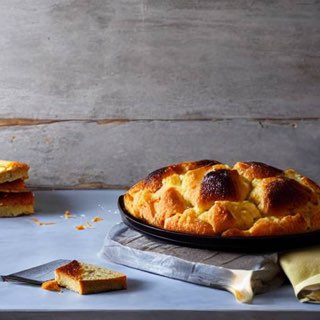} &
\includegraphics[width=0.08\textwidth]{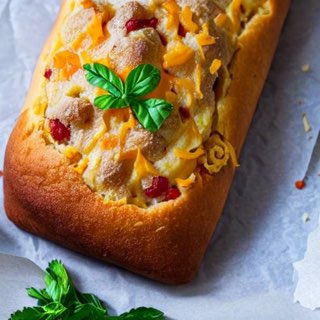} &
\includegraphics[width=0.08\textwidth]{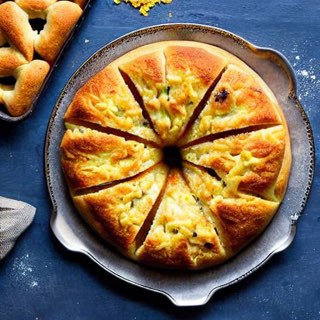} &
\includegraphics[width=0.08\textwidth]{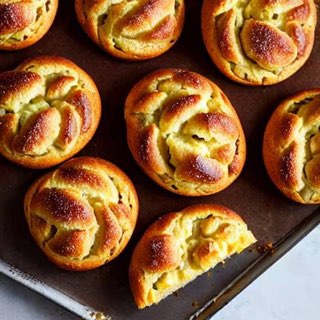} &
\includegraphics[width=0.08\textwidth]{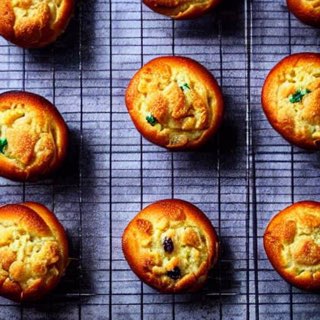} &
\includegraphics[width=0.08\textwidth]{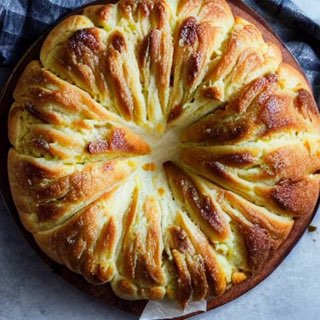} &
\includegraphics[width=0.08\textwidth]{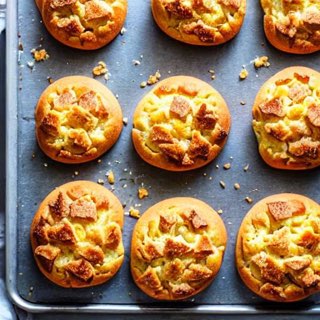} \\
SD1.5-1.4-Base & 
\includegraphics[width=0.08\textwidth]{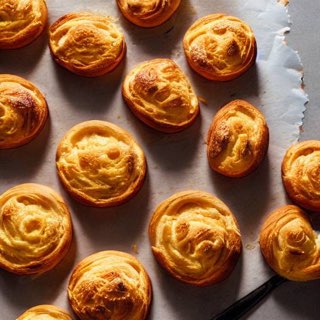} &
\includegraphics[width=0.08\textwidth]{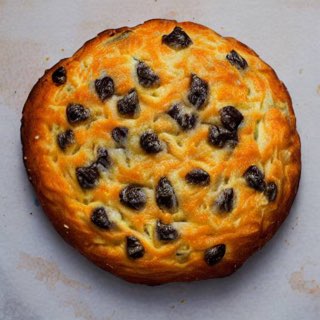} &
\includegraphics[width=0.08\textwidth]{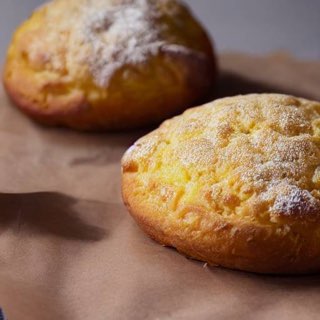} &
\includegraphics[width=0.08\textwidth]{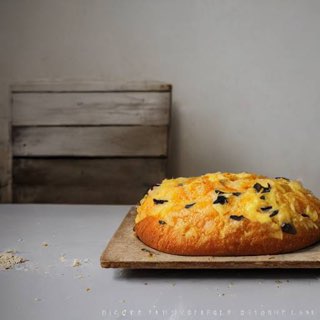} &
\includegraphics[width=0.08\textwidth]{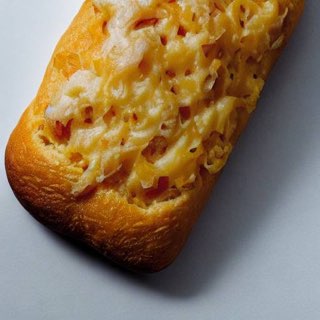} &
\includegraphics[width=0.08\textwidth]{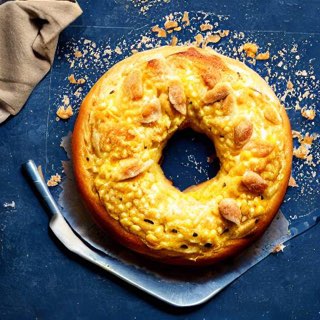} &
\includegraphics[width=0.08\textwidth]{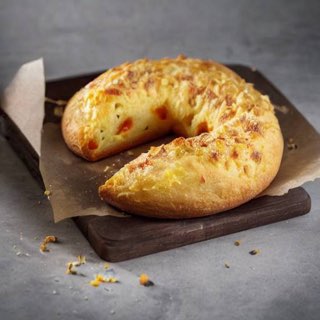} &
\includegraphics[width=0.08\textwidth]{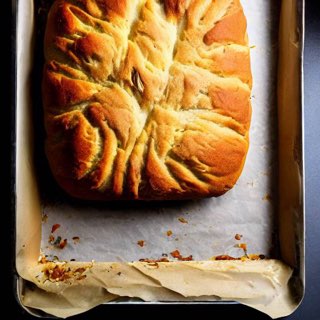} &
\includegraphics[width=0.08\textwidth]{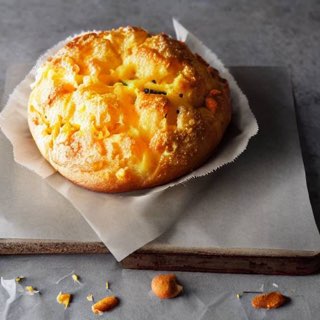} &
\includegraphics[width=0.08\textwidth]{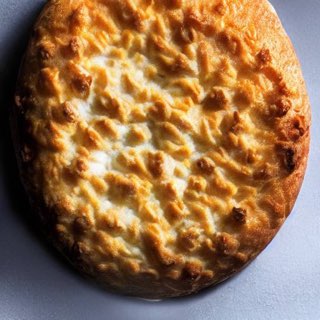} \\
SD1.5-DreamShaper & 
\includegraphics[width=0.08\textwidth]{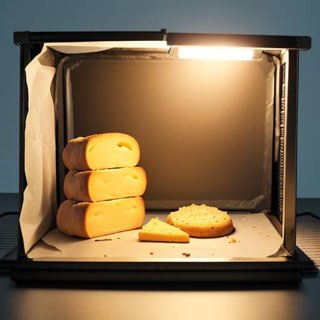} &
\includegraphics[width=0.08\textwidth]{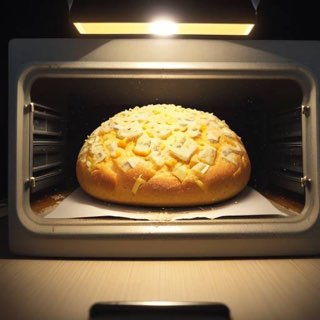} &
\includegraphics[width=0.08\textwidth]{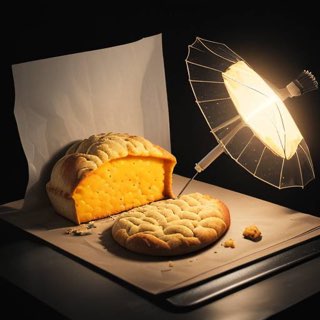} &
\includegraphics[width=0.08\textwidth]{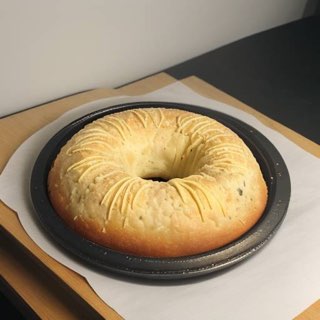} &
\includegraphics[width=0.08\textwidth]{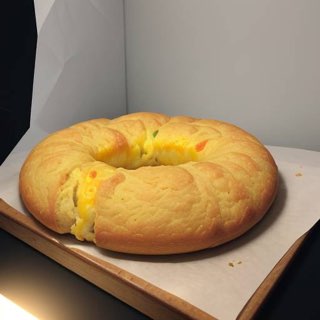} &
\includegraphics[width=0.08\textwidth]{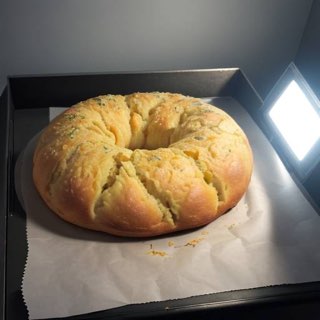} &
\includegraphics[width=0.08\textwidth]{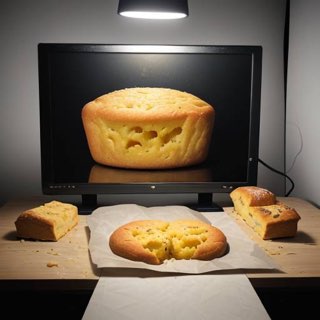} &
\includegraphics[width=0.08\textwidth]{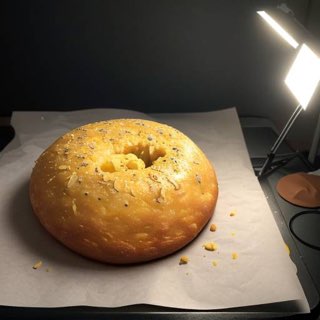} &
\includegraphics[width=0.08\textwidth]{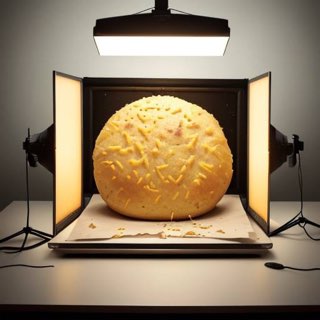} &
\includegraphics[width=0.08\textwidth]{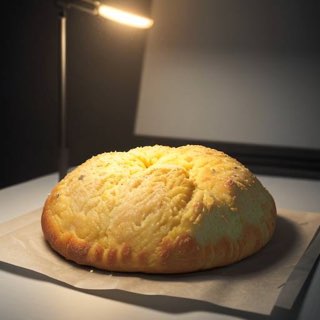} \\
\bottomrule
\end{tabular}
}
\caption{Qualitative results (Part 1): Flux, Kandinsky, and SD1.5 families.}
\label{fig:qual_part1_chesse}
\end{figure*}

\begin{figure*}[p]
\centering
\small
\setlength{\tabcolsep}{2pt}
\resizebox{\textwidth}{!}{%
\begin{tabular}{l*{10}{c}}
\toprule
\multicolumn{11}{c}{\textbf{Prompt: ``\cheesytitle''}} \\
\textbf{Model} & \textbf{S1} & \textbf{S2} & \textbf{S3} & \textbf{S4} & \textbf{S5} & \textbf{S6} & \textbf{S7} & \textbf{S8} & \textbf{S9} & \textbf{S10} \\
\midrule
\textit{SD2.1 Family} \\
SD2.1-Base & 
\includegraphics[width=0.08\textwidth]{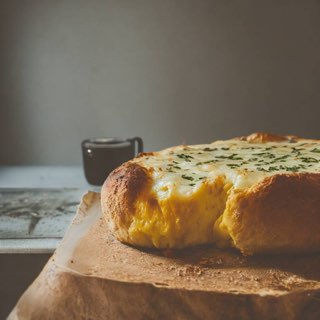} &
\includegraphics[width=0.08\textwidth]{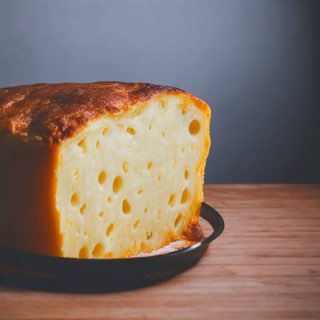} &
\includegraphics[width=0.08\textwidth]{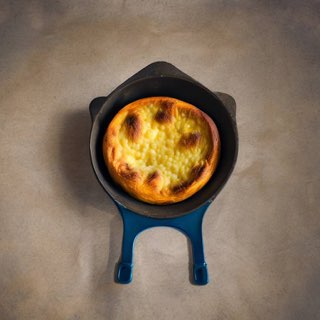} &
\includegraphics[width=0.08\textwidth]{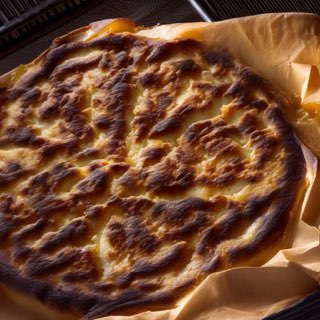} &
\includegraphics[width=0.08\textwidth]{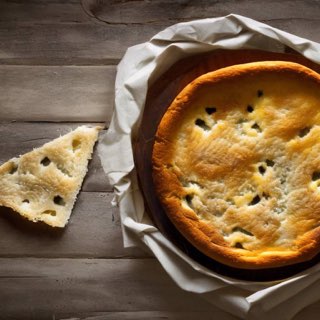} &
\includegraphics[width=0.08\textwidth]{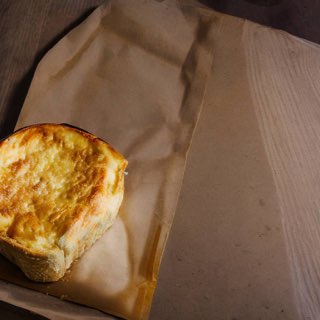} &
\includegraphics[width=0.08\textwidth]{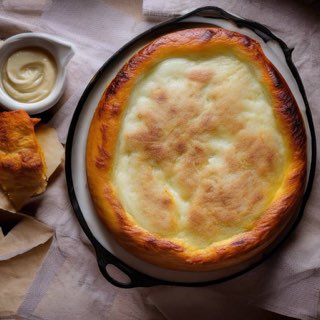} &
\includegraphics[width=0.08\textwidth]{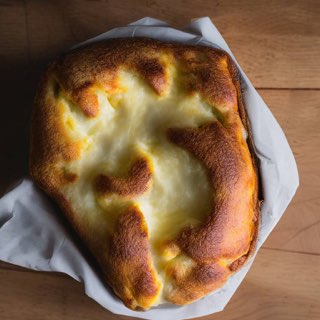} &
\includegraphics[width=0.08\textwidth]{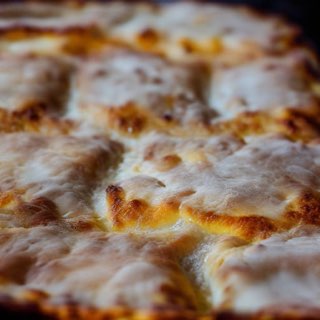} &
\includegraphics[width=0.08\textwidth]{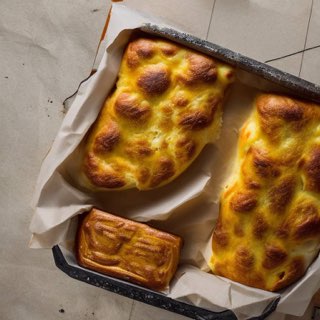} \\
SD2.1-DPO & 
\includegraphics[width=0.08\textwidth]{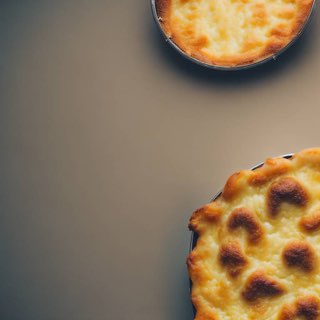} &
\includegraphics[width=0.08\textwidth]{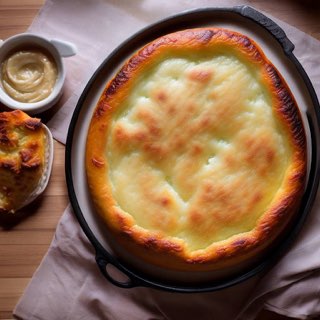} &
\includegraphics[width=0.08\textwidth]{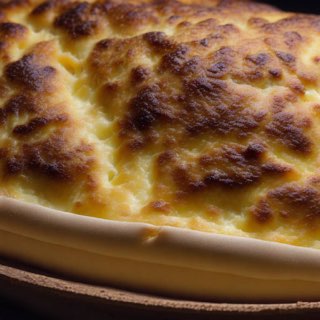} &
\includegraphics[width=0.08\textwidth]{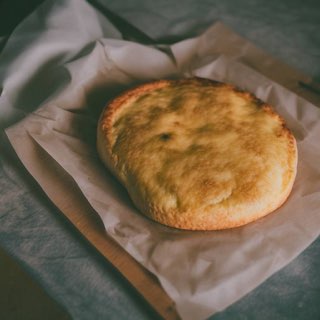} &
\includegraphics[width=0.08\textwidth]{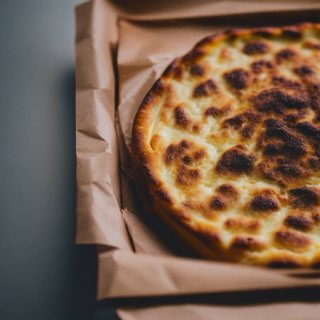} &
\includegraphics[width=0.08\textwidth]{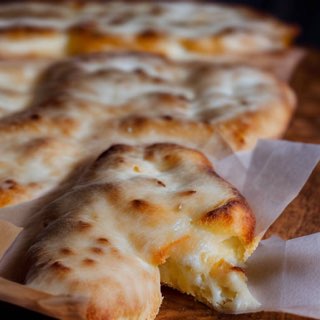} &
\includegraphics[width=0.08\textwidth]{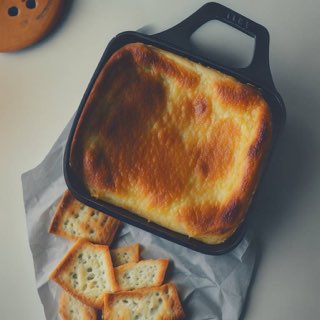} &
\includegraphics[width=0.08\textwidth]{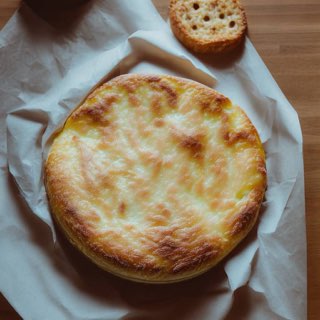} &
\includegraphics[width=0.08\textwidth]{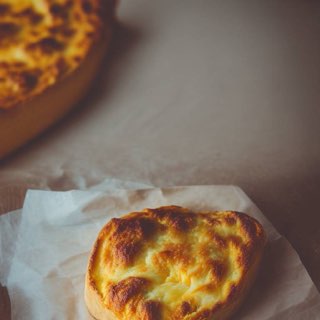} &
\includegraphics[width=0.08\textwidth]{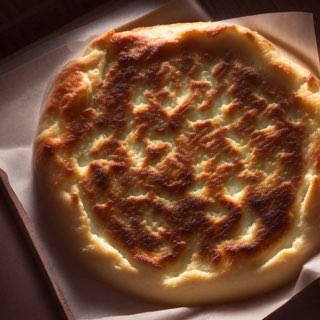} \\
SD2.1-LAION-Art & 
\includegraphics[width=0.08\textwidth]{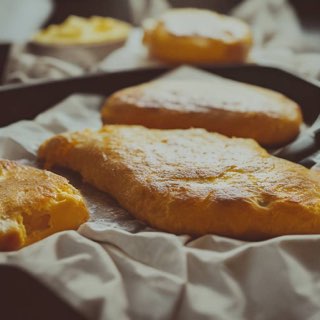} &
\includegraphics[width=0.08\textwidth]{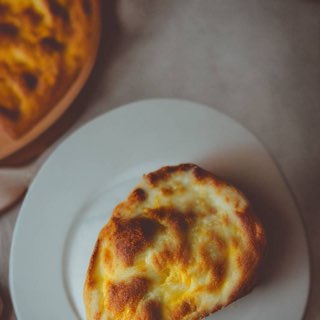} &
\includegraphics[width=0.08\textwidth]{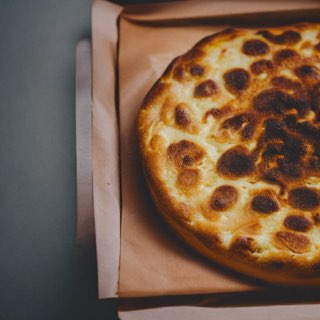} &
\includegraphics[width=0.08\textwidth]{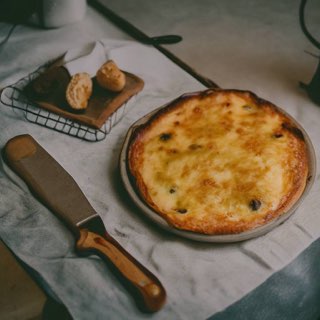} &
\includegraphics[width=0.08\textwidth]{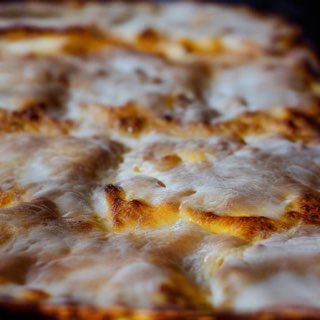} &
\includegraphics[width=0.08\textwidth]{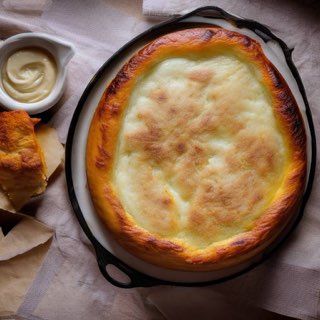} &
\includegraphics[width=0.08\textwidth]{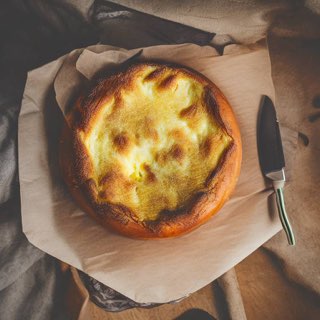} &
\includegraphics[width=0.08\textwidth]{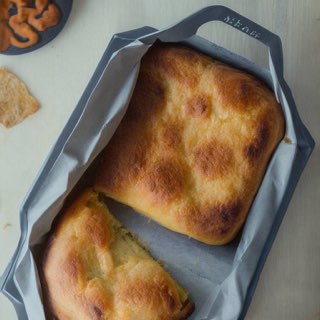} &
\includegraphics[width=0.08\textwidth]{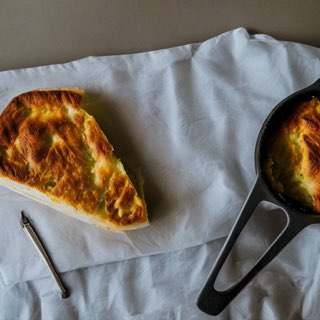} &
\includegraphics[width=0.08\textwidth]{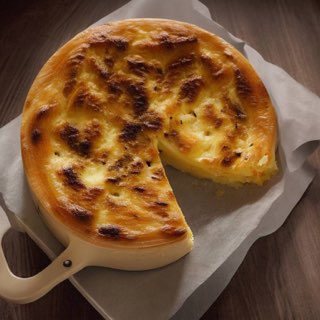} \\
\midrule
\textit{SD3 Family} \\
SD3-Medium-Base & 
\includegraphics[width=0.08\textwidth]{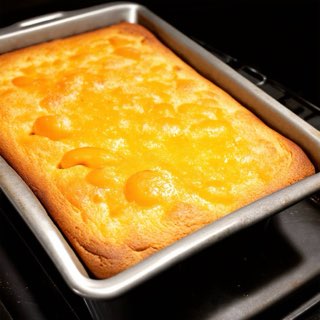} &
\includegraphics[width=0.08\textwidth]{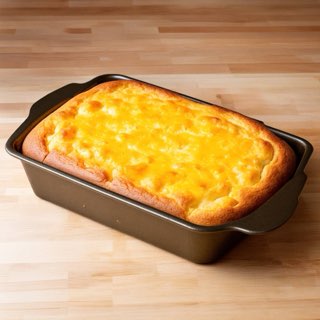} &
\includegraphics[width=0.08\textwidth]{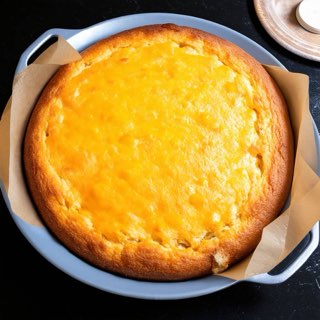} &
\includegraphics[width=0.08\textwidth]{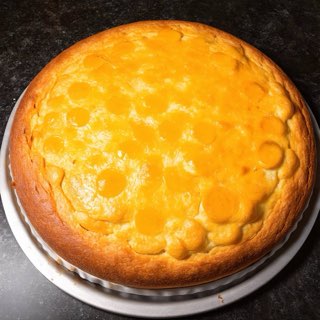} &
\includegraphics[width=0.08\textwidth]{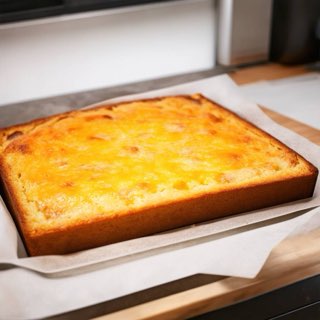} &
\includegraphics[width=0.08\textwidth]{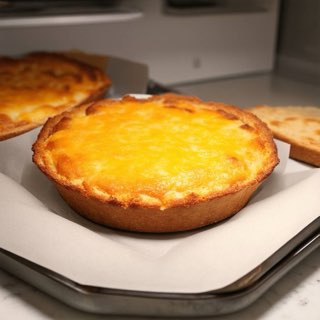} &
\includegraphics[width=0.08\textwidth]{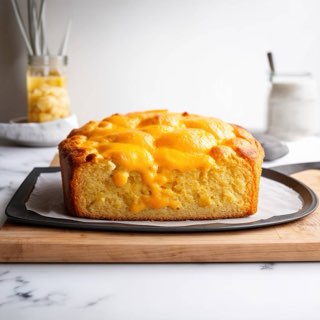} &
\includegraphics[width=0.08\textwidth]{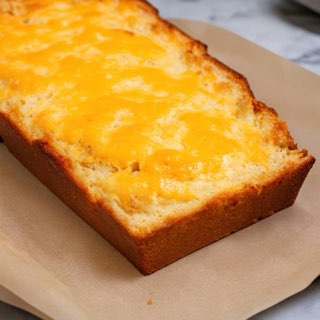} &
\includegraphics[width=0.08\textwidth]{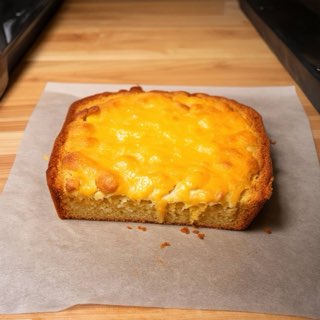} &
\includegraphics[width=0.08\textwidth]{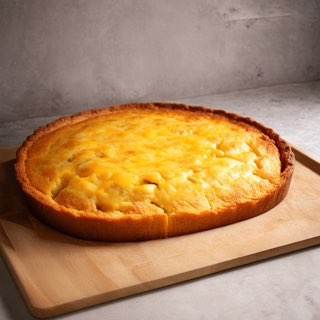} \\
SD3-Reality-Mix & 
\includegraphics[width=0.08\textwidth]{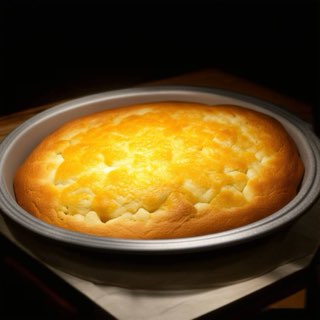} &
\includegraphics[width=0.08\textwidth]{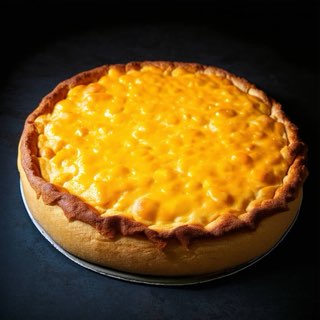} &
\includegraphics[width=0.08\textwidth]{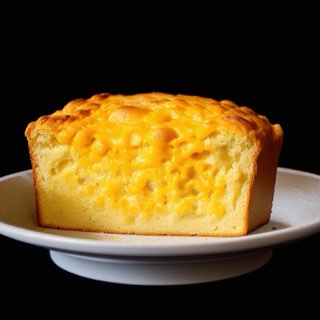} &
\includegraphics[width=0.08\textwidth]{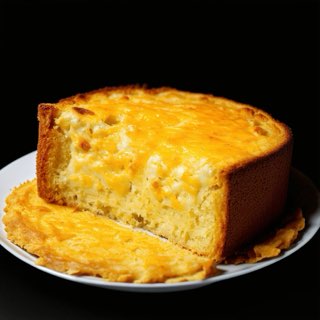} &
\includegraphics[width=0.08\textwidth]{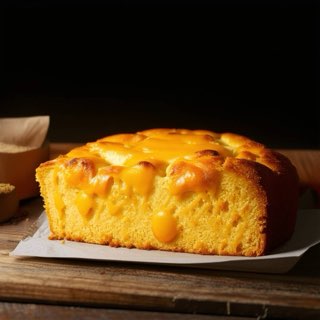} &
\includegraphics[width=0.08\textwidth]{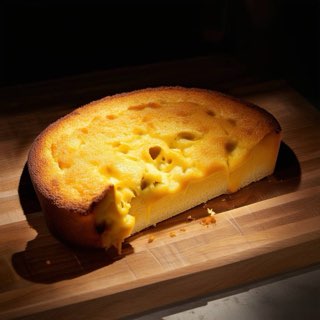} &
\includegraphics[width=0.08\textwidth]{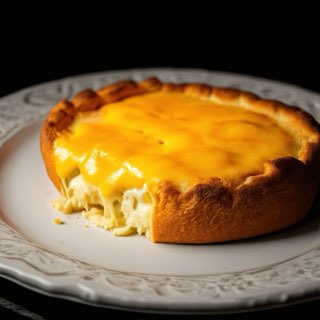} &
\includegraphics[width=0.08\textwidth]{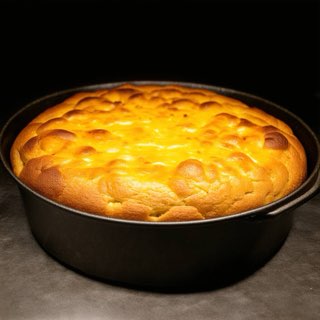} &
\includegraphics[width=0.08\textwidth]{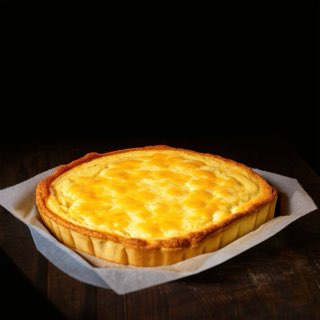} &
\includegraphics[width=0.08\textwidth]{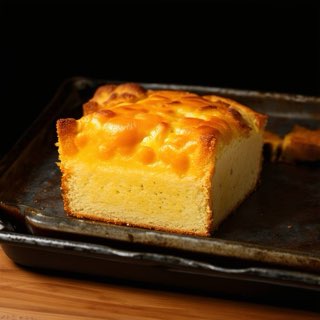} \\
SD3-VAE-Anime & 
\includegraphics[width=0.08\textwidth]{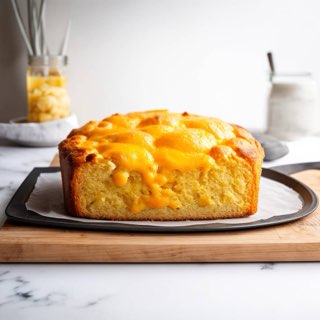} &
\includegraphics[width=0.08\textwidth]{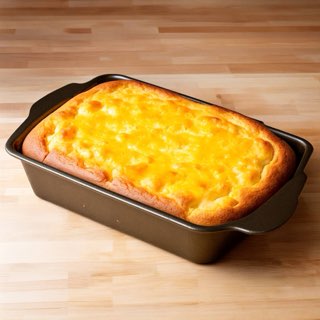} &
\includegraphics[width=0.08\textwidth]{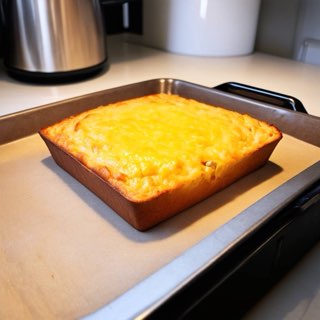} &
\includegraphics[width=0.08\textwidth]{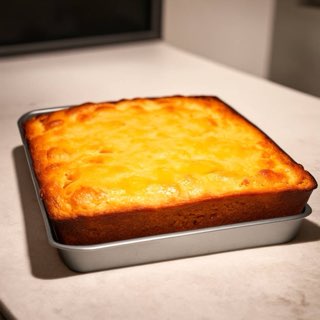} &
\includegraphics[width=0.08\textwidth]{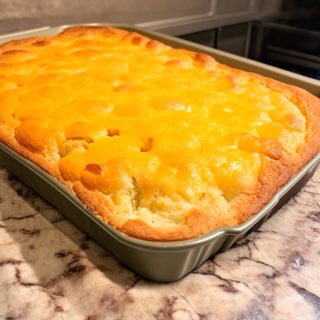} &
\includegraphics[width=0.08\textwidth]{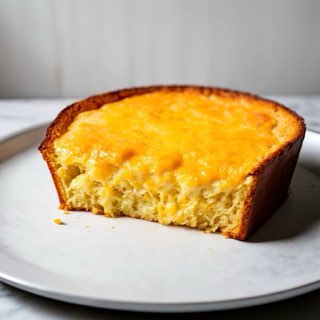} &
\includegraphics[width=0.08\textwidth]{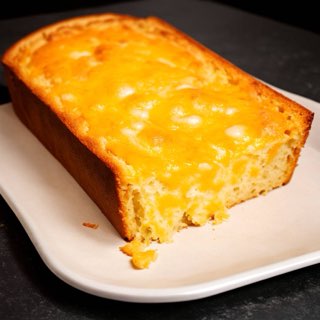} &
\includegraphics[width=0.08\textwidth]{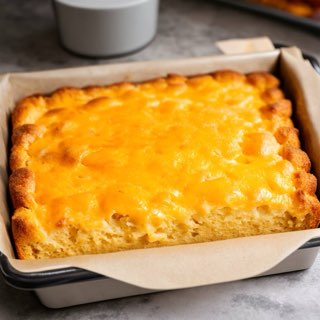} &
\includegraphics[width=0.08\textwidth]{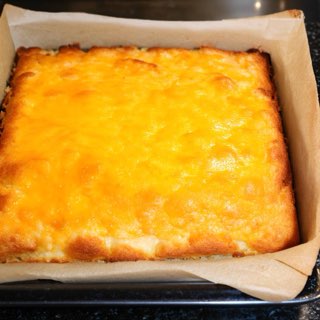} &
\includegraphics[width=0.08\textwidth]{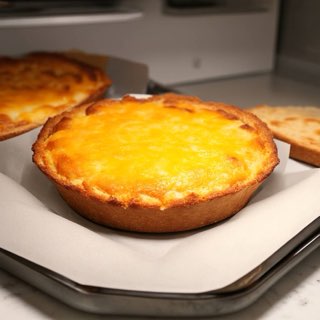} \\
\midrule
\textit{SDXL Family} \\
SDXL-Base & 
\includegraphics[width=0.08\textwidth]{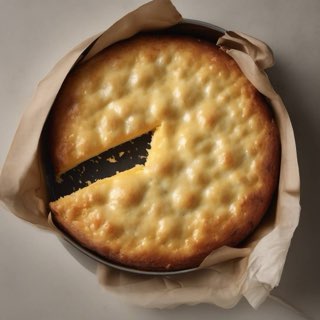} &
\includegraphics[width=0.08\textwidth]{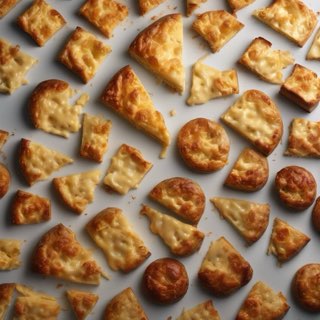} &
\includegraphics[width=0.08\textwidth]{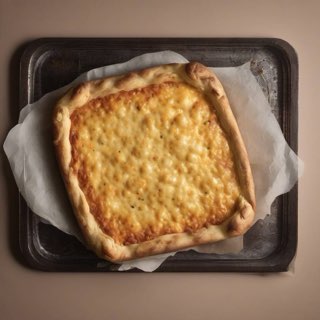} &
\includegraphics[width=0.08\textwidth]{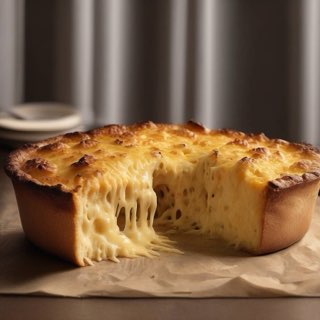} &
\includegraphics[width=0.08\textwidth]{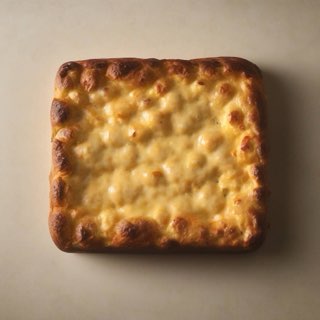} &
\includegraphics[width=0.08\textwidth]{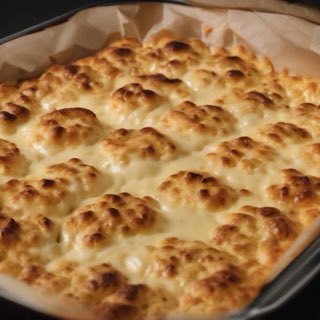} &
\includegraphics[width=0.08\textwidth]{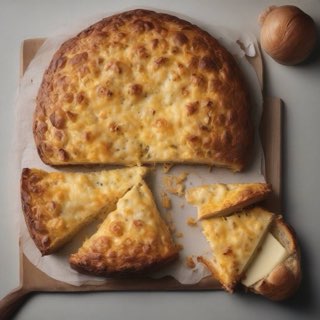} &
\includegraphics[width=0.08\textwidth]{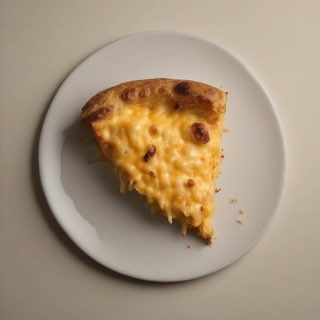} &
\includegraphics[width=0.08\textwidth]{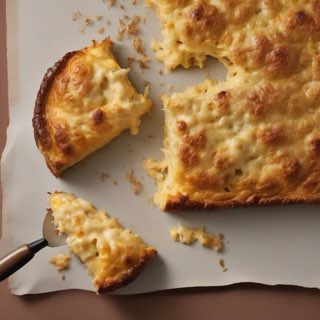} &
\includegraphics[width=0.08\textwidth]{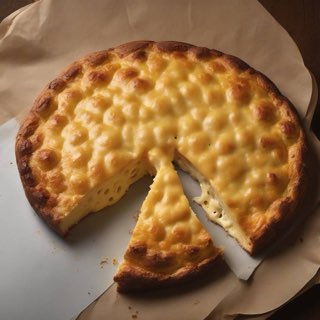} \\
SDXL-DPO & 
\includegraphics[width=0.08\textwidth]{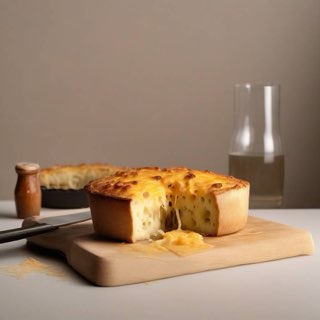} &
\includegraphics[width=0.08\textwidth]{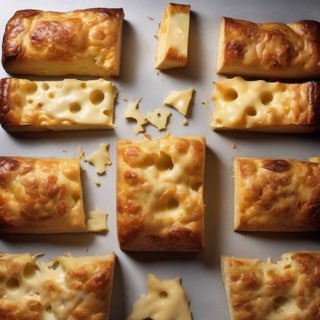} &
\includegraphics[width=0.08\textwidth]{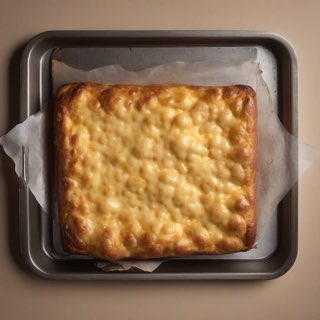} &
\includegraphics[width=0.08\textwidth]{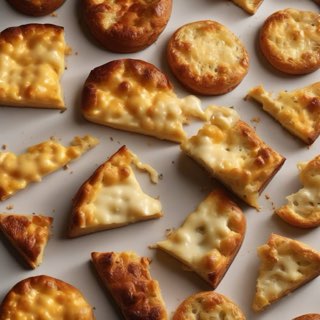} &
\includegraphics[width=0.08\textwidth]{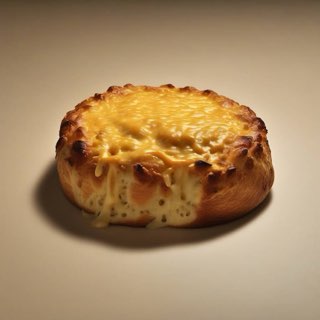} &
\includegraphics[width=0.08\textwidth]{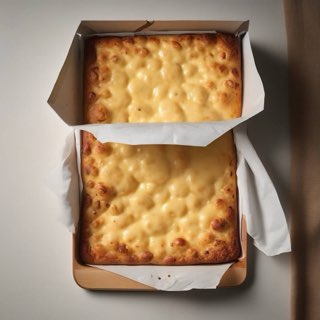} &
\includegraphics[width=0.08\textwidth]{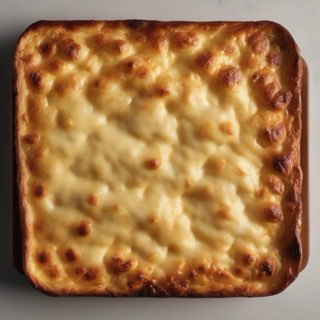} &
\includegraphics[width=0.08\textwidth]{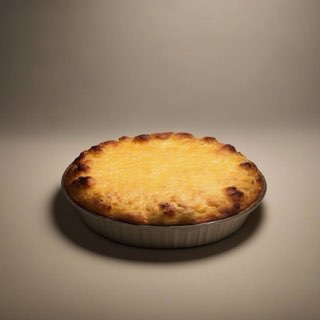} &
\includegraphics[width=0.08\textwidth]{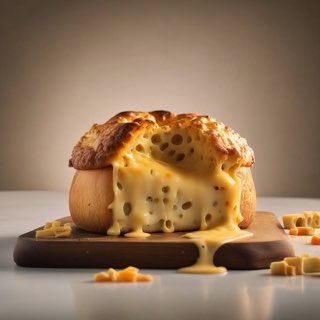} &
\includegraphics[width=0.08\textwidth]{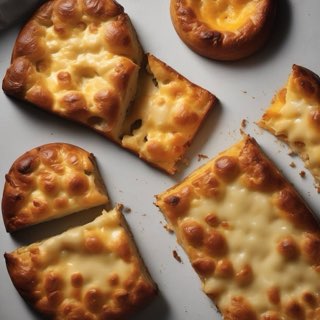} \\
SDXL-Lightning-4Step & 
\includegraphics[width=0.08\textwidth]{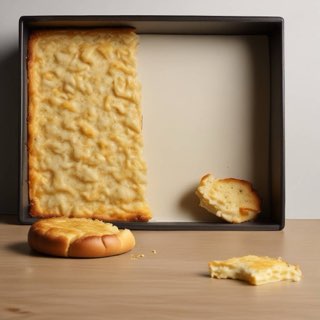} &
\includegraphics[width=0.08\textwidth]{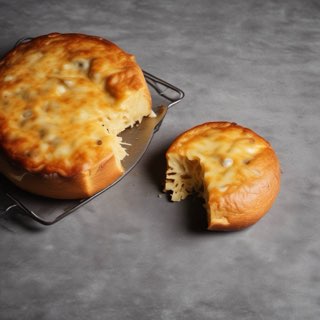} &
\includegraphics[width=0.08\textwidth]{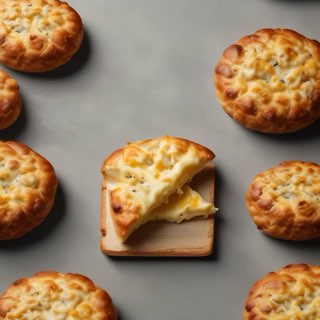} &
\includegraphics[width=0.08\textwidth]{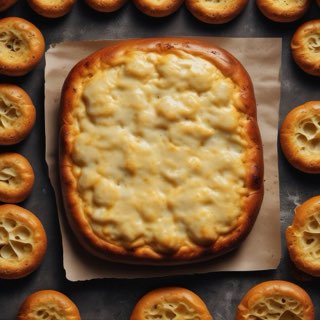} &
\includegraphics[width=0.08\textwidth]{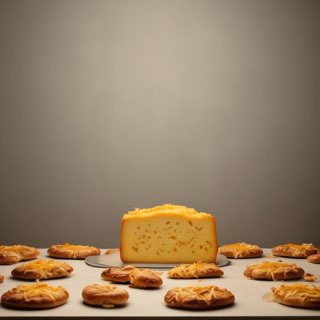} &
\includegraphics[width=0.08\textwidth]{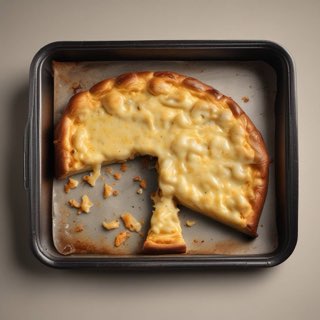} &
\includegraphics[width=0.08\textwidth]{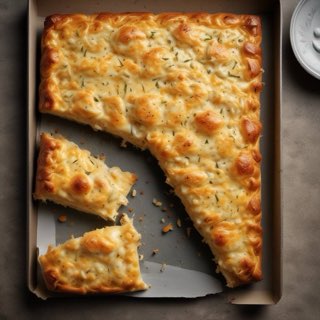} &
\includegraphics[width=0.08\textwidth]{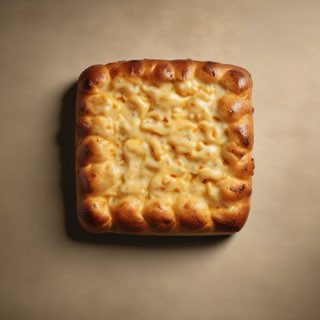} &
\includegraphics[width=0.08\textwidth]{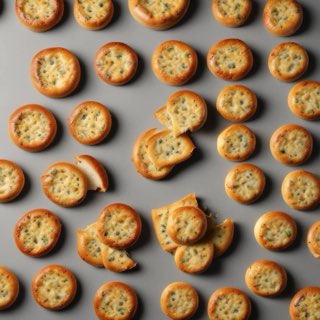} &
\includegraphics[width=0.08\textwidth]{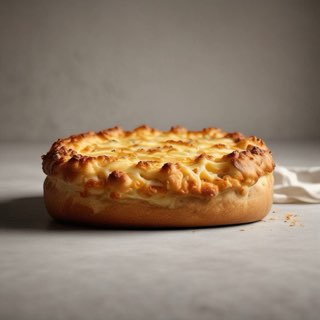} \\
\bottomrule
\end{tabular}
}
\caption{Qualitative results (Part 2): SD2.1, SD3, and SDXL families.}
\label{fig:qual_part2_cheese}
\end{figure*}

The diversity across these base models—spanning U-Net vs. transformer architectures, single-stage vs. two-stage pipelines, and different capacity scales—provides a comprehensive testbed for evaluating cross-architecture fingerprint generalization.

\subsection{Fine-tuned and Adapted Models}

We include 13 fine-tuned variants spanning diverse adaptation strategies to test fingerprint persistence under practical model modifications:

\paragraph{Parameter-Efficient Fine-Tuning.} Flux-LoRA, Kandinsky-Pokemon-LoRA, SD2.1-DPO, and SDXL-DPO apply LoRA~\cite{hu2022lora} for low-rank adaptation, testing whether fingerprints survive when only a small fraction ($<$1\%) of parameters are modified. The latter two additionally incorporate Direct Preference Optimization~\cite{rafailov2023direct} for alignment with human preferences, combining parameter efficiency with RLHF-style objectives that may disrupt learned semantic biases.

\begin{table*}[tb]
\centering
\caption{95\% Confidence Interval Lower Bound of the Derived Models with base method}
\label{tab:baselow}
\scalebox{0.80}{
\renewcommand{\arraystretch}{1.}
\begin{tabular}{lcccccc} 
\hline\hline
& \multicolumn{6}{c}{\cellcolor[rgb]{0.918,0.918,0.918}\textbf{Base Models}} \\
& Flux-Base & Kandinsky-Base & SD1.5-Base & SD2.1-Base & SD3-Medium-Base & SDXL-Base \\ 
\hline
\multicolumn{7}{l}{\textit{Flux Family}} \\
\textbf{Flux-LoRA} & 0.541 & 0.004 & 0.004 & 0.004 & 0.004 & 0.004 \\
\textbf{Flux-Turbo-Alpha} & 0.541 & 0.004 & 0.004 & 0.004 & 0.004 & 0.004 \\
\hline
\multicolumn{7}{l}{\textit{Kandinsky Family}} \\
\textbf{Kandinsky-Naruto} & 0.004 & 0.541 & 0.004 & 0.004 & 0.004 & 0.004 \\
\textbf{Kandinsky-Pokemon-LoRA} & 0.004 & 0.541 & 0.004 & 0.004 & 0.004 & 0.004 \\
\hline
\multicolumn{7}{l}{\textit{SD1.5 Family}} \\
\textbf{SD1.5-1.2-Base} & 0.004 & 0.004 & 0.223 & 0.004 & 0.004 & 0.118 \\
\textbf{SD1.5-1.4-Base} & 0.004 & 0.004 & 0.541 & 0.004 & 0.004 & 0.004 \\
\textbf{SD1.5-DreamShaper} & 0.004 & 0.223 & 0.043 & 0.004 & 0.004 & 0.043 \\
\hline
\multicolumn{7}{l}{\textit{SD2.1 Family}} \\
\textbf{SD2.1-DPO} & 0.004 & 0.004 & 0.004 & 0.541 & 0.004 & 0.004 \\
\textbf{SD2.1-LAION-Art} & 0.004 & 0.004 & 0.004 & 0.541 & 0.004 & 0.004 \\
\hline
\multicolumn{7}{l}{\textit{SD3 Family}} \\
\textbf{SD3-Reality-Mix} & 0.004 & 0.004 & 0.004 & 0.043 & 0.359 & 0.004 \\
\textbf{SD3-VAE-Anime} & 0.004 & 0.004 & 0.004 & 0.004 & 0.541 & 0.004 \\
\hline
\multicolumn{7}{l}{\textit{SDXL Family}} \\
\textbf{SDXL-DPO} & 0.004 & 0.004 & 0.004 & 0.004 & 0.004 & 0.541 \\
\textbf{SDXL-Lightning-4Step} & 0.004 & 0.043 & 0.004 & 0.004 & 0.004 & 0.359 \\
\hline\hline
\end{tabular}
}
\end{table*}
\paragraph{Domain-Specific Fine-Tuning.} Kandinsky-Naruto (anime style), SD1.5-DreamShaper (aesthetic quality optimization), and SD2.1-LAION-Art (artistic style from LAION-Aesthetics~\cite{schuhmann2022laion}) undergo substantial fine-tuning on domain-specific data. These models challenge fingerprint robustness under significant distribution shift, where the semantic space may deviate considerably from the base model's training distribution.

\paragraph{Model Merging.} SD1.5-DreamShaper and SD3-Reality-Mix represent community-created models that interpolate or merge weights from multiple checkpoints to achieve desired aesthetic or photorealistic qualities. Weight merging presents a unique challenge as the resulting model's semantic space is a non-linear combination of its constituent models, potentially diluting or obscuring individual model fingerprints.

\paragraph{Component-Specific Modification.} SD3-VAE-Anime fine-tunes only the VAE decoder while keeping the diffusion model frozen, isolating the impact of decoder modifications on fingerprinting. This tests whether fingerprints embedded in the semantic (latent) space persist when only the pixel-space decoder is modified.

\paragraph{Inference Optimization.} Flux-Turbo-Alpha and SDXL-Lightning-4Step are distilled for few-step inference (4-8 steps vs. 50+ steps), representing the most extreme adaptation scenario. These models undergo aggressive knowledge distillation and inference optimization that fundamentally alters the denoising trajectory. Crucially, such distillation has been shown to defeat even white-box watermarking methods~\cite{wen2023tree}, making it the hardest case for any fingerprinting approach. Our evaluation on these models tests whether semantic fingerprints can survive optimization pressures that eliminate traditional cryptographic signatures.

This comprehensive selection ensures our evaluation covers the practical challenges faced in real-world model fingerprinting, from lightweight adaptations to aggressive compression and merging strategies.

\section{Results with Alternative Metrics}
\label{sec:other_metrics}

We evaluate our fingerprinting approach using alternative distance metrics to validate robustness. Results are reported at 95\% confidence intervals.

\subsection{Baseline Clustering Methods}

\Cref{tab:baselow} and \Cref{tab:basemid} show results for hierarchical clustering on raw fingerprint vectors. Without model specifications (\Cref{tab:baselow}), naive clustering fails to achieve reliable fingerprinting due to high false positive rates. Incorporating specifications (\Cref{tab:basemid}) improves performance but still underperforms our compositional approach, confirming that how we structure semantic biases matters.

\subsection{Jensen-Shannon Divergence}

\Cref{tab:jsdmain}, \Cref{tab:jsdlow}, and \Cref{tab:jsdloowmain} present JSD-based fingerprinting results. While JSD is a principled distribution metric, it struggles to capture the subtle compositional structure of semantic biases:

\begin{itemize}
    \item JSD treats all fingerprint dimensions equally, missing compositionally-structured patterns
    \item Smoothing in JSD computation reduces discriminability at decision boundaries
    \item Higher variance in confidence intervals suggests less stable matching
\end{itemize}

These results explain why LPIPS, designed for perceptual differences, outperforms distribution-based metrics. Our compositional prompts produce semantically meaningful visual variations that may appear statistically minor under symmetric divergence measures.

\begin{table*}[tb]
\centering
\caption{95\% Confidence Posterior mean of the Derived Models with base method}
\label{tab:basemid}
\scalebox{0.80}{
\renewcommand{\arraystretch}{1.}
\begin{tabular}{lcccccc} 
\hline\hline
& \multicolumn{6}{c}{\cellcolor[rgb]{0.918,0.918,0.918}\textbf{Base Models}} \\
& Flux-Base & Kandinsky-Base & SD1.5-Base & SD2.1-Base & SD3-Medium-Base & SDXL-Base \\ 
\hline
\multicolumn{7}{l}{\textit{Flux Family}} \\
\textbf{Flux-LoRA} & 0.857 & 0.143 & 0.143 & 0.143 & 0.143 & 0.143 \\
\textbf{Flux-Turbo-Alpha} & 0.857 & 0.143 & 0.143 & 0.143 & 0.143 & 0.143 \\
\hline
\multicolumn{7}{l}{\textit{Kandinsky Family}} \\
\textbf{Kandinsky-Naruto} & 0.143 & 0.857 & 0.143 & 0.143 & 0.143 & 0.143 \\
\textbf{Kandinsky-Pokemon-LoRA} & 0.143 & 0.857 & 0.143 & 0.143 & 0.143 & 0.143 \\
\hline
\multicolumn{7}{l}{\textit{SD1.5 Family}} \\
\textbf{SD1.5-1.2-Base} & 0.143 & 0.143 & 0.571 & 0.143 & 0.143 & 0.429 \\
\textbf{SD1.5-1.4-Base} & 0.143 & 0.143 & 0.857 & 0.143 & 0.143 & 0.143 \\
\textbf{SD1.5-DreamShaper} & 0.143 & 0.571 & 0.286 & 0.143 & 0.143 & 0.286 \\
\hline
\multicolumn{7}{l}{\textit{SD2.1 Family}} \\
\textbf{SD2.1-DPO} & 0.143 & 0.143 & 0.143 & 0.857 & 0.143 & 0.143 \\
\textbf{SD2.1-LAION-Art} & 0.143 & 0.143 & 0.143 & 0.857 & 0.143 & 0.143 \\
\hline
\multicolumn{7}{l}{\textit{SD3 Family}} \\
\textbf{SD3-Reality-Mix} & 0.143 & 0.143 & 0.143 & 0.286 & 0.714 & 0.143 \\
\textbf{SD3-VAE-Anime} & 0.143 & 0.143 & 0.143 & 0.143 & 0.857 & 0.143 \\
\hline
\multicolumn{7}{l}{\textit{SDXL Family}} \\
\textbf{SDXL-DPO} & 0.143 & 0.143 & 0.143 & 0.143 & 0.143 & 0.857 \\
\textbf{SDXL-Lightning-4Step} & 0.143 & 0.286 & 0.143 & 0.143 & 0.143 & 0.714 \\
\hline\hline
\end{tabular}
}
\end{table*}
\begin{table*}[tb]
\centering
\caption{95\% Confidence Interval Lower Bound of the Derived Models with our method}
\label{tab:jsdloowmain}
\scalebox{0.80}{
\renewcommand{\arraystretch}{1.}
\begin{tabular}{lcccccc} 
\hline\hline
& \multicolumn{6}{c}{\cellcolor[rgb]{0.918,0.918,0.918}\textbf{Base Models}} \\
& Flux-Base & Kandinsky-Base & SD1.5-Base & SD2.1-Base & SD3-Medium-Base & SDXL-Base \\ 
\hline
\multicolumn{7}{l}{\textit{Flux Family}} \\
\textbf{Flux-LoRA} & 0.842 & 0.001 & 0.001 & 0.001 & 0.001 & 0.015 \\
\textbf{Flux-Turbo-Alpha} & 0.918 & 0.001 & 0.001 & 0.001 & 0.001 & 0.001 \\
\hline
\multicolumn{7}{l}{\textit{Kandinsky Family}} \\
\textbf{Kandinsky-Naruto} & 0.001 & 0.918 & 0.001 & 0.001 & 0.001 & 0.001 \\
\textbf{Kandinsky-Pokemon-LoRA} & 0.006 & 0.702 & 0.006 & 0.028 & 0.001 & 0.006 \\
\hline
\multicolumn{7}{l}{\textit{SD1.5 Family}} \\
\textbf{SD1.5-1.2-Base} & 0.001 & 0.001 & 0.721 & 0.039 & 0.001 & 0.015 \\
\textbf{SD1.5-1.4-Base} & 0.001 & 0.001 & 0.918 & 0.001 & 0.001 & 0.001 \\
\textbf{SD1.5-DreamShaper} & 0.026 & 0.015 & 0.515 & 0.006 & 0.015 & 0.068 \\
\hline
\multicolumn{7}{l}{\textit{SD2.1 Family}} \\
\textbf{SD2.1-DPO} & 0.001 & 0.001 & 0.001 & 0.918 & 0.001 & 0.001 \\
\textbf{SD2.1-LAION-Art} & 0.001 & 0.001 & 0.001 & 0.918 & 0.001 & 0.001 \\
\hline
\multicolumn{7}{l}{\textit{SD3 Family}} \\
\textbf{SD3-Reality-Mix} & 0.053 & 0.026 & 0.001 & 0.006 & 0.563 & 0.026 \\
\textbf{SD3-VAE-Anime} & 0.001 & 0.001 & 0.001 & 0.001 & 0.918 & 0.001 \\
\hline
\multicolumn{7}{l}{\textit{SDXL Family}} \\
\textbf{SDXL-DPO} & 0.001 & 0.001 & 0.001 & 0.001 & 0.001 & 0.918 \\
\textbf{SDXL-Lightning-4Step} & 0.001 & 0.026 & 0.001 & 0.015 & 0.001 & 0.749 \\
\hline\hline
\end{tabular}
}
\end{table*}
\begin{table*}[tb]
\centering
\caption{95\% Confidence Interval Lower Bound of the Derived Models in jsd}
\label{tab:jsdlow}
\scalebox{0.80}{
\renewcommand{\arraystretch}{1.}
\begin{tabular}{lcccccc} 
\hline\hline
& \multicolumn{6}{c}{\cellcolor[rgb]{0.918,0.918,0.918}\textbf{Base Models}} \\
& Flux-Base & Kandinsky-Base & SD1.5-Base & SD2.1-Base & SD3-Medium-Base & SDXL-Base \\ 
\hline
\multicolumn{7}{l}{\textit{Flux Family}} \\
\textbf{Flux-LoRA} & 0.640 & 0.001 & 0.015 & 0.006 & 0.026 & 0.026 \\
\textbf{Flux-Turbo-Alpha} & 0.842 & 0.006 & 0.001 & 0.001 & 0.001 & 0.006 \\
\hline
\multicolumn{7}{l}{\textit{Kandinsky Family}} \\
\textbf{Kandinsky-Naruto} & 0.015 & 0.291 & 0.118 & 0.053 & 0.015 & 0.068 \\
\textbf{Kandinsky-Pokemon-LoRA} & 0.015 & 0.842 & 0.001 & 0.001 & 0.001 & 0.001 \\
\hline
\multicolumn{7}{l}{\textit{SD1.5 Family}} \\
\textbf{SD1.5-1.2-Base} & 0.006 & 0.001 & 0.721 & 0.026 & 0.001 & 0.015 \\
\textbf{SD1.5-1.4-Base} & 0.006 & 0.001 & 0.809 & 0.015 & 0.001 & 0.001 \\
\textbf{SD1.5-DreamShaper} & 0.026 & 0.026 & 0.515 & 0.039 & 0.001 & 0.039 \\
\hline
\multicolumn{7}{l}{\textit{SD2.1 Family}} \\
\textbf{SD2.1-DPO} & 0.026 & 0.001 & 0.001 & 0.809 & 0.001 & 0.001 \\
\textbf{SD2.1-LAION-Art} & 0.006 & 0.001 & 0.001 & 0.877 & 0.001 & 0.001 \\
\hline
\multicolumn{7}{l}{\textit{SD3 Family}} \\
\textbf{SD3-Reality-Mix} & 0.053 & 0.006 & 0.015 & 0.026 & 0.421 & 0.084 \\
\textbf{SD3-VAE-Anime} & 0.026 & 0.001 & 0.001 & 0.001 & 0.779 & 0.006 \\
\hline
\multicolumn{7}{l}{\textit{SDXL Family}} \\
\textbf{SDXL-DPO} & 0.100 & 0.006 & 0.001 & 0.015 & 0.006 & 0.563 \\
\textbf{SDXL-Lightning-4Step} & 0.053 & 0.006 & 0.006 & 0.026 & 0.006 & 0.588 \\
\hline\hline
\end{tabular}
}
\end{table*}
\begin{table*}[tb]
\centering
\caption{Posterior mean of JSD}
\label{tab:jsdmain}
\scalebox{0.80}{
\renewcommand{\arraystretch}{1.}
\begin{tabular}{lcccccc} 
\hline\hline
& \multicolumn{6}{c}{\cellcolor[rgb]{0.918,0.918,0.918}\textbf{Base Models}} \\
& Flux-Base & Kandinsky-Base & SD1.5-Base & SD2.1-Base & SD3-Medium-Base & SDXL-Base \\ 
\hline
\multicolumn{7}{l}{\textit{Flux Family}} \\
\textbf{Flux-LoRA} & 0.773 & 0.023 & 0.068 & 0.045 & 0.091 & 0.091 \\
\textbf{Flux-Turbo-Alpha} & 0.932 & 0.045 & 0.023 & 0.023 & 0.023 & 0.045 \\
\hline
\multicolumn{7}{l}{\textit{Kandinsky Family}} \\
\textbf{Kandinsky-Naruto} & 0.068 & 0.432 & 0.227 & 0.136 & 0.068 & 0.159 \\
\textbf{Kandinsky-Pokemon-LoRA} & 0.068 & 0.932 & 0.023 & 0.023 & 0.023 & 0.023 \\
\hline
\multicolumn{7}{l}{\textit{SD1.5 Family}} \\
\textbf{SD1.5-1.2-Base} & 0.045 & 0.023 & 0.841 & 0.091 & 0.023 & 0.068 \\
\textbf{SD1.5-1.4-Base} & 0.045 & 0.023 & 0.909 & 0.068 & 0.023 & 0.023 \\
\textbf{SD1.5-DreamShaper} & 0.091 & 0.091 & 0.659 & 0.114 & 0.023 & 0.114 \\
\hline
\multicolumn{7}{l}{\textit{SD2.1 Family}} \\
\textbf{SD2.1-DPO} & 0.091 & 0.023 & 0.023 & 0.909 & 0.023 & 0.023 \\
\textbf{SD2.1-LAION-Art} & 0.045 & 0.023 & 0.023 & 0.955 & 0.023 & 0.023 \\
\hline
\multicolumn{7}{l}{\textit{SD3 Family}} \\
\textbf{SD3-Reality-Mix} & 0.136 & 0.045 & 0.068 & 0.091 & 0.568 & 0.182 \\
\textbf{SD3-VAE-Anime} & 0.091 & 0.023 & 0.023 & 0.023 & 0.886 & 0.045 \\
\hline
\multicolumn{7}{l}{\textit{SDXL Family}} \\
\textbf{SDXL-DPO} & 0.205 & 0.045 & 0.023 & 0.068 & 0.045 & 0.705 \\
\textbf{SDXL-Lightning-4Step} & 0.136 & 0.045 & 0.045 & 0.091 & 0.045 & 0.727 \\
\hline\hline
\end{tabular}
}
\end{table*}

\section{Fingerprinting Prompt Templates}

\subsection{Prompt Design Rationale}

Our 42 fingerprinting prompts are designed to systematically probe semantic biases across multiple dimensions:

\paragraph{Compositional structure} All prompts follow the template \texttt{"A photo of a [ADJECTIVE] [OBJECT] [LOCATION]"}, which allows independent manipulation of each semantic component. This compositionality is crucial for isolating specific bias sources and measuring their persistence through fine-tuning.

\paragraph{Attribute variation} Each object category includes three semantically distinct adjectives that probe different semantic associations. For example, baked goods vary along the savory-sweet spectrum (savory, cheesy, sweet), while animals vary in perceived threat level (dangerous, wild, peaceful). This diversity ensures we capture a broad range of potential semantic biases.

\paragraph{Context variation} We systematically vary background settings across three types: controlled studio environments (dimmed studio), natural contexts (grassland, forest, grass, savana), and textured surfaces (dark wood, brick wall, pot, vase, dish, wooden floor). The dimmed studio setting appears across all categories to provide a consistent baseline for cross-category comparison, while category-specific contexts test bias robustness in semantically appropriate settings.

\paragraph{Coverage and efficiency} The resulting 42 prompts (5 categories × 3 attributes × 2-3 contexts) provide sufficient coverage to establish reliable fingerprints while remaining computationally tractable for fingerprinting verification in black-box settings.

\subsection{Prompt Template Structure}

Table~\ref{tab:prompt_structure} summarizes the compositional structure of our fingerprinting prompts. The full enumeration of all 42 prompts is provided in Section~\ref{sec:full_prompts}.

\section{Hyperparameter Settings}

\begin{itemize}
    \item \textbf{Standard Diffusion Models (SD1.5, SD2.1, SDXL Base):} We used a unified setting of \textbf{30 inference steps} and a \textbf{guidance scale (CFG) of 7.5}. This applies to the majority of fine-tuned models in our benchmark (e.g., DreamShaper, LAION-Art).
    
    \item \textbf{SDXL-Lightning:} As this model is distilled for extreme speed, we used \textbf{4 inference steps} with a \textbf{guidance scale of 0.0}, utilizing the \texttt{EulerDiscreteScheduler} with trailing timesteps.
    
    \item \textbf{Stable Diffusion 3 (SD3):} For the base medium model, we used \textbf{28 steps} with a CFG of \textbf{7.0}. For the \textit{Reality Mix} variant, we lowered the CFG to \textbf{5.5} and set the guidance rescale to 0.0 to prevent artifacts.
    
    \item \textbf{Flux.1-dev:} We utilized the \texttt{FluxPipeline} with \textbf{50 inference steps} and a \textbf{guidance scale of 3.5}. The generation resolution was explicitly set to $512\times512$, and for LoRA variants (e.g., RealismLora), adapter weights were loaded with a scale of 1.0.
    
    \item \textbf{DPO-Adapted Models:} For SD2.1-DPO and SDXL-DPO, we switched the scheduler to \texttt{DPMSolverMultistepScheduler} (SDE-DPM++ with Karras sigmas) to better handle the preference-aligned weights. The LoRA adapter strength was set to \textbf{0.9}.
    
    \item \textbf{Kandinsky 2.2:} We employed the specific \texttt{KandinskyV22PriorPipeline} for embedding generation (CFG 1.0), followed by the decoder pipeline.
\end{itemize}

\section{Visualization}

\begin{figure}
    \centering
    \includegraphics[width=1\linewidth]{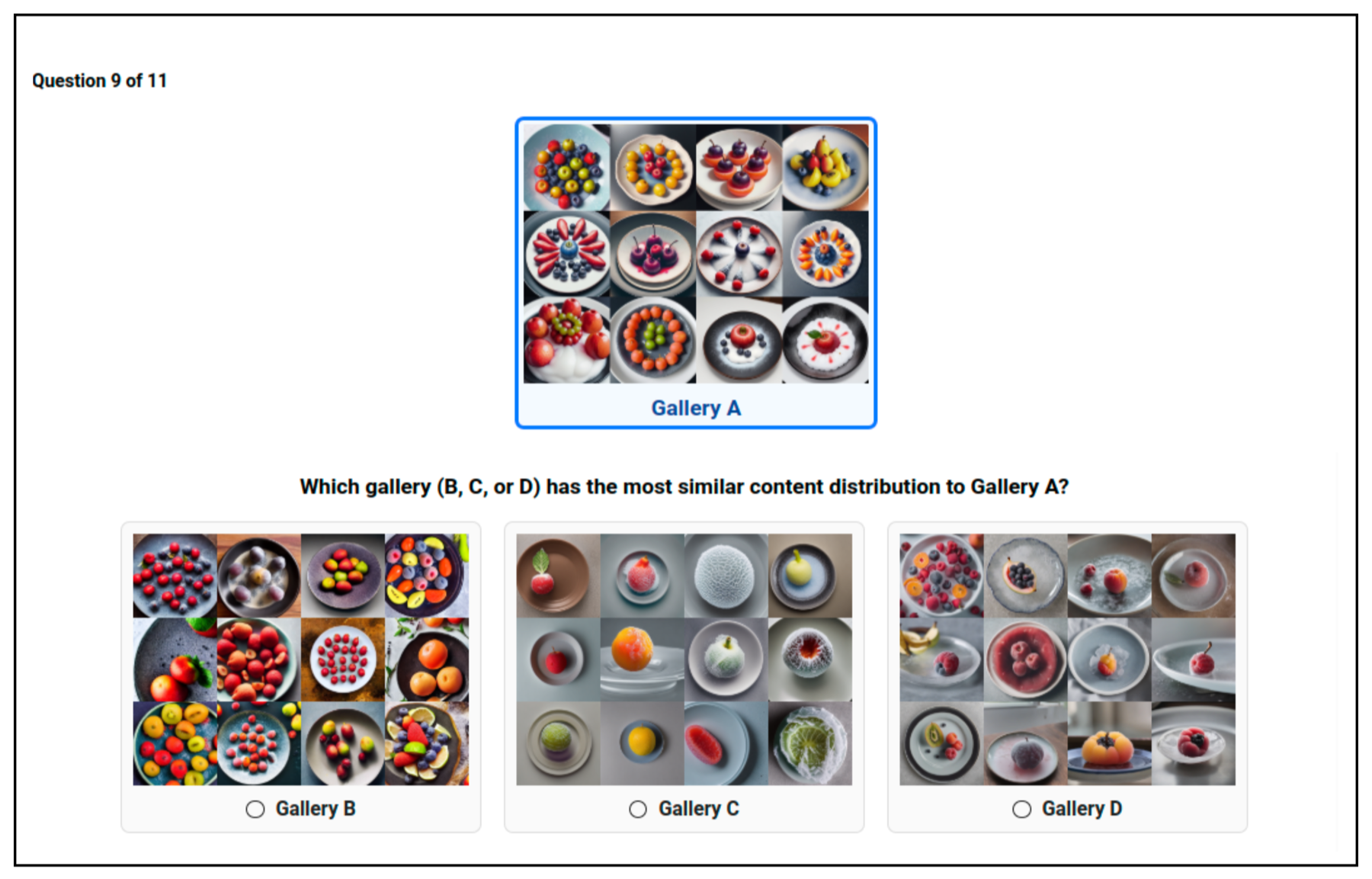}
    \caption{Example of human study}
    \label{fig:human_study_1}
\end{figure}
\begin{figure}
    \centering
    \includegraphics[width=1\linewidth]{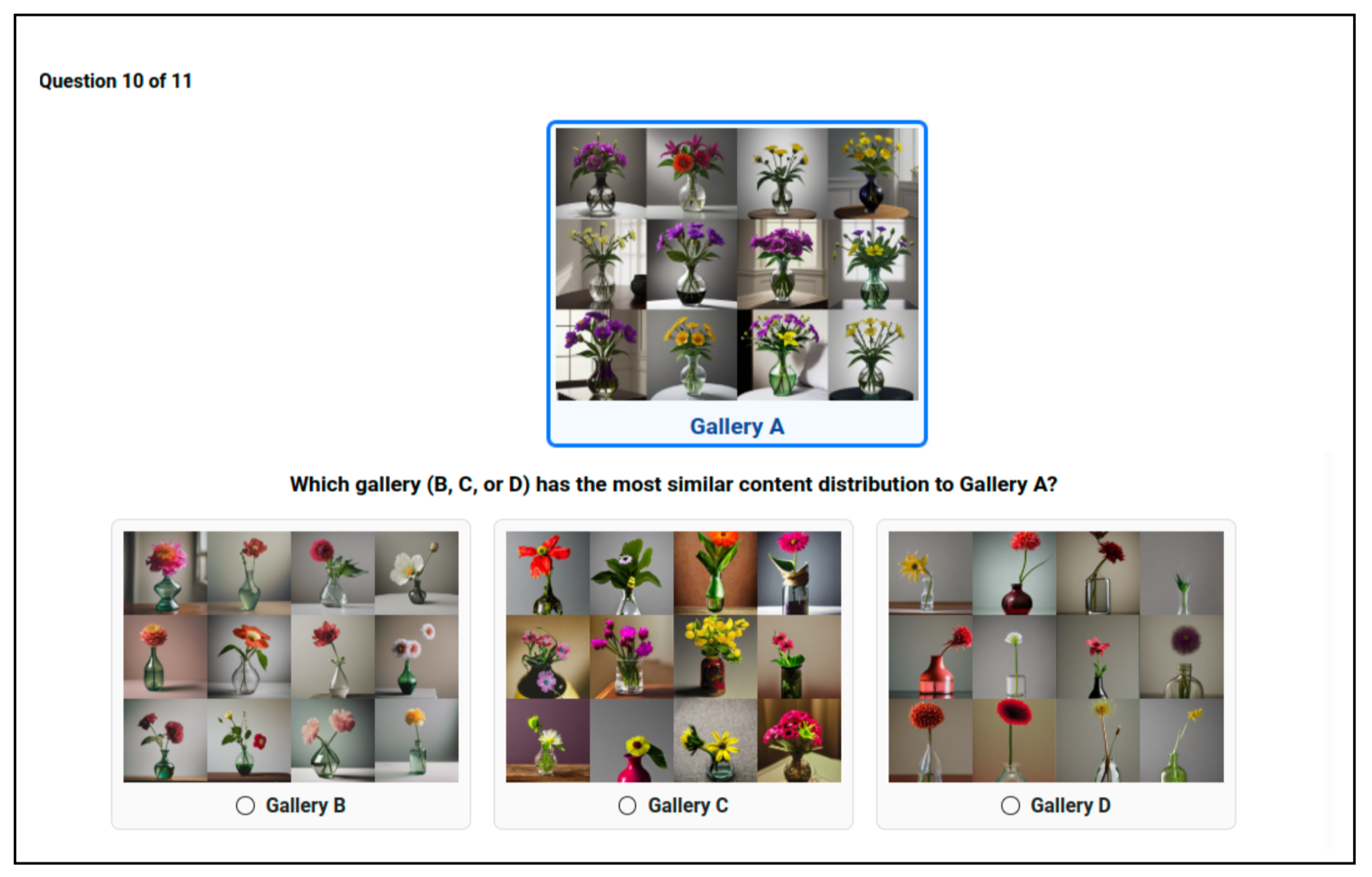}
    \caption{Example of human study}
    \label{fig:human_study_2}
\end{figure}
\begin{figure}
    \centering
    \includegraphics[width=1\linewidth]{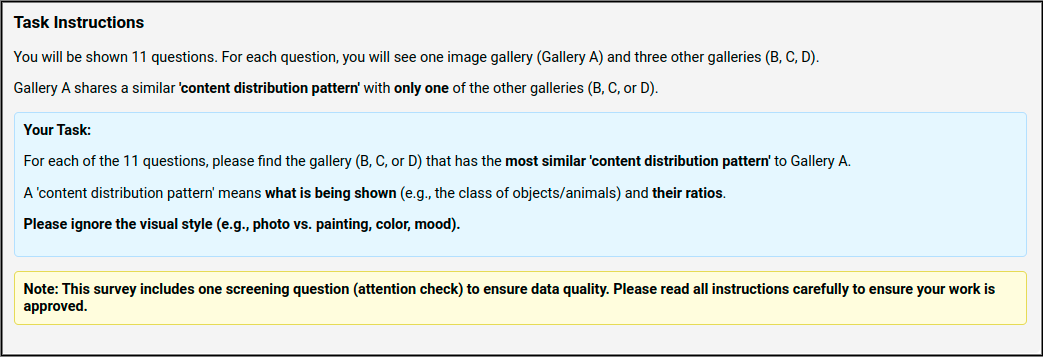}
    \caption{An instruction of human study}
    \label{fig:human_study_3}
\end{figure}
\begin{figure}
    \centering
    \includegraphics[width=1\linewidth]{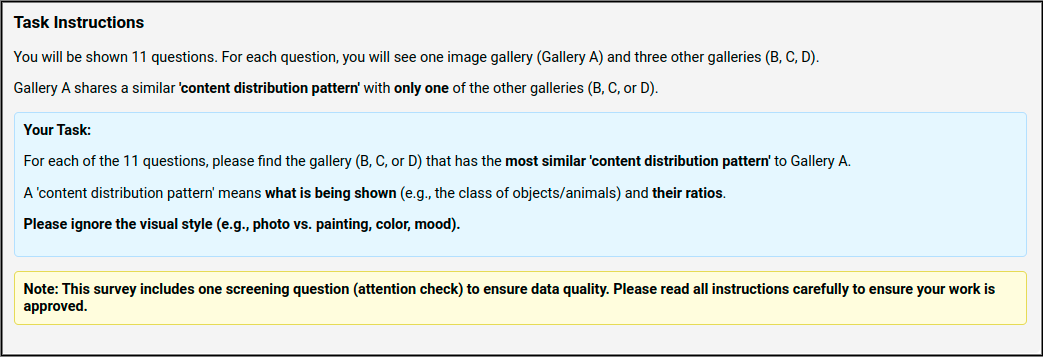}
    \caption{Example of human study for comparison}
    \label{fig:human_study_4}
\end{figure}
\Cref{fig:human_study_1,fig:human_study_2,fig:human_study_3,fig:human_study_4}
present example questions used in the human study. Additional qualitative
grids for the prompt templates are provided below.
\section{Heatmap}
In \Cref{tab:wasserstein_average_normalized}, similar to other fingerprinting methods, our method can also be evaluated by visualizing a heatmap. We visualized the results by averaging each column.

\begin{table*}[ht]
\centering
\small
\begin{tabular}{p{2.5cm}p{4cm}p{5.5cm}}
\toprule
\textbf{Object Category} & \textbf{Adjectives} & \textbf{Locations} \\
\midrule
Baked good & savory, cheesy, sweet & dimmed studio, dark wood surface, brick wall \\
\midrule
Animal & dangerous, wild, peaceful & grassland, forest, dimmed studio \\
\midrule
Flower & vibrant, tropical & pot, dimmed studio, vase \\
\midrule
Bird & peaceful, dangerous, flightless & grass, savana, dimmed studio \\
\midrule
Fruit & sweet, frozen, savory & dish, wooden floor, dimmed studio \\
\bottomrule
\end{tabular}
\caption{Compositional structure of fingerprinting prompts. Each category systematically varies semantic attributes and visual contexts, yielding 42 total prompts (9+9+6+9+9). The dimmed studio setting appears across all categories to enable cross-category comparison.}
\label{tab:prompt_structure}
\end{table*}

\section{Details of the Analysis experiments}
\subsection{Details of the Scene-Context Validation Experiment}
\label{app:scene_context_validation}
We provide the full protocol for the scene-context validation experiment shown in \Cref{fig:prompt_ablation}. The goal of this experiment was to verify that the semantic signal isolated by CSF changes systematically when only the contextual component of an underspecified prompt is modified.

\paragraph{Task.}
We instantiated the validation experiment in the fruit domain. For each trial, we fixed the superordinate category and the underspecified component of the prompt, and varied only the scene context. We considered three contextual settings: \textit{dimmed studio}, \textit{dish}, and \textit{wooden floor}.

\paragraph{Prompt conditions.}
All prompt components other than the scene phrase were kept unchanged across conditions. This design isolates the effect of context alone, allowing us to test whether a single contextual modification can induce a measurable shift in the generated semantic category distribution.

\paragraph{Protocol.}
For each context, we generated 40 images using the same model and identical sampling settings. The generation pipeline, including the prompt template and all sampling hyperparameters, was kept fixed across the three conditions.

\paragraph{Category extraction.}
Each generated image was mapped to a fixed fruit-category vocabulary using the same zero-shot CLIP-based category extraction procedure used throughout the main experiments. The category set consisted of Apple, Nectarine, Grapefruit, Lime, Coconut, Honeydew, and Others.

\paragraph{Interpretation.}
As shown in \Cref{fig:prompt_ablation}, changing only the scene context produces a substantial and systematic shift in the generated category distribution. Compared with the dimmed-studio setting, the dish and wooden-floor settings yield broader and differently redistributed mixtures over categories. This supports our claim that compositional underspecification reveals model-dependent semantic biases, and that contextual variation alone is sufficient to alter how the model resolves semantic ambiguity.

\subsection{Details of the Human Perceptual Study}
\label{app:human-study-details}

We provide the full protocol for the human perceptual validation experiment used in \Cref{fig:user_study}. The goal of this study was to test whether the semantic signal isolated by CSF is also accessible to human observers, rather than being an artifact of the computational attribution procedure.

\paragraph{Task.}
We ran a four-way forced-choice ``Name That Dataset'' experiment on Amazon Mechanical Turk. In each trial, a participant was shown a fine-tuned model and asked to identify its underlying base model from four candidates. To mirror the abstraction used by CSF, participants were explicitly instructed to ignore visual style and rely only on recurring content-distribution cues.

\paragraph{Prompt conditions.}
We compared two prompt sources: (i) naive prompts randomly sampled from LAION-2B, and (ii) CSF prompts generated by our method. The candidate set size was fixed across conditions, yielding a chance level of $25\%$.

\paragraph{Protocol.}
For each condition, we collected responses from 50 independent participants under the same interface and instruction template. Candidate ordering was randomized to reduce position bias, while the task format was kept identical across conditions.

\paragraph{Metric and error bars.}
We report identification accuracy for each condition. Since each response is a binary correct/incorrect outcome, we model the measured accuracy under a binomial distribution and compute the error bars as
\[
100\sqrt{\frac{p(1-p)}{n}},
\]
where $p$ is the observed accuracy and $n=50$ is the number of independent participant responses.

\paragraph{Interpretation.}
Naive prompts lead to substantially weaker identification performance, whereas CSF prompts noticeably improve human recognition of the correct model lineage. This supports our claim that the signal captured by CSF is not merely machine-readable, but also perceptually robust.

\input{assets/alltable}

\begin{figure*}[p]
\centering
\small
\setlength{\tabcolsep}{2pt}
\begin{tabular}{l*{10}{c}}
\toprule
\multicolumn{11}{c}{\textbf{Prompt: ``\prompttitle''}} \\
\midrule
\textbf{Model} & \textbf{S1} & \textbf{S2} & \textbf{S3} & \textbf{S4} & \textbf{S5} & \textbf{S6} & \textbf{S7} & \textbf{S8} & \textbf{S9} & \textbf{S10} \\
\midrule
\textit{Flux Family} \\
Flux-Base & 
\includegraphics[width=0.07\textwidth]{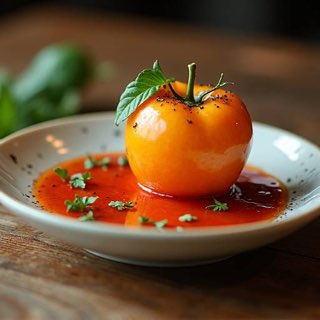} &
\includegraphics[width=0.07\textwidth]{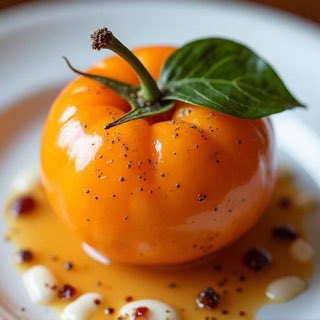} &
\includegraphics[width=0.07\textwidth]{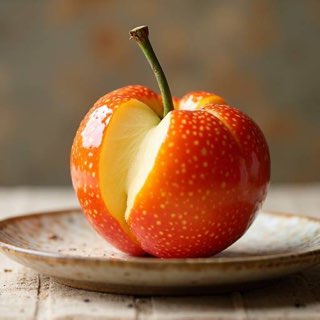} &
\includegraphics[width=0.07\textwidth]{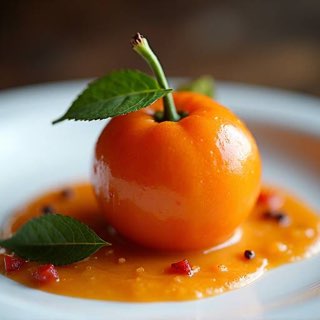} &
\includegraphics[width=0.07\textwidth]{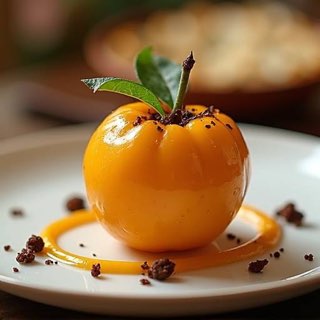} &
\includegraphics[width=0.07\textwidth]{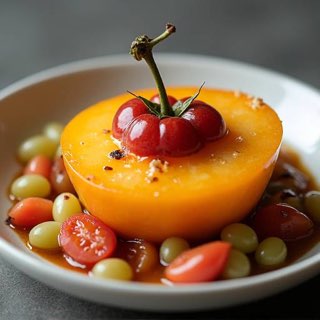} &
\includegraphics[width=0.07\textwidth]{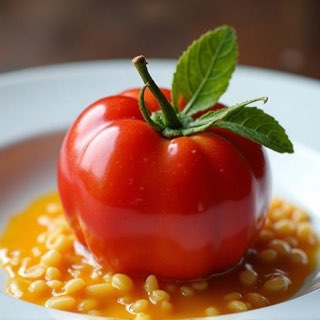} &
\includegraphics[width=0.07\textwidth]{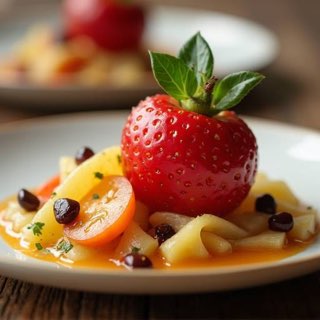} &
\includegraphics[width=0.07\textwidth]{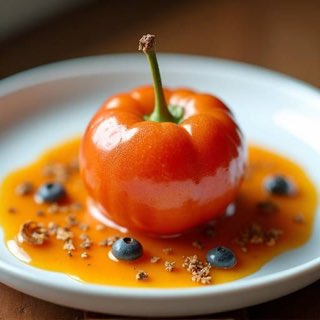} &
\includegraphics[width=0.07\textwidth]{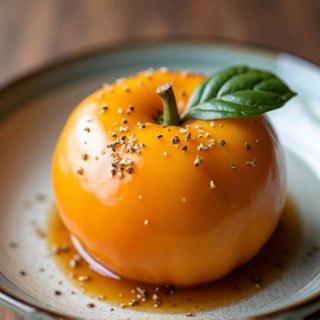} \\
Flux-LoRA & 
\includegraphics[width=0.07\textwidth]{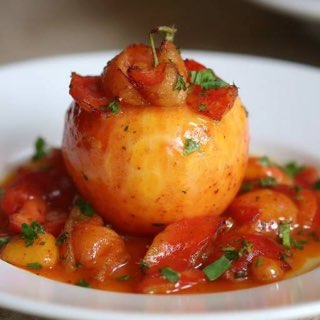} &
\includegraphics[width=0.07\textwidth]{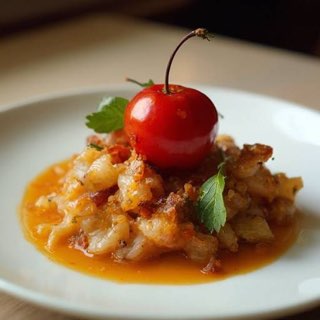} &
\includegraphics[width=0.07\textwidth]{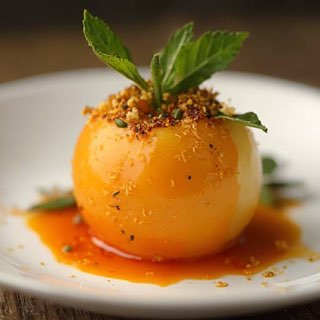} &
\includegraphics[width=0.07\textwidth]{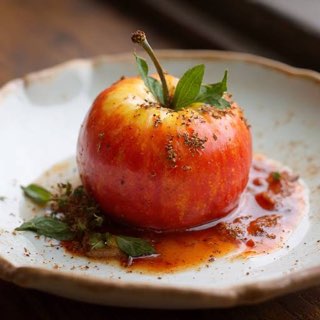} &
\includegraphics[width=0.07\textwidth]{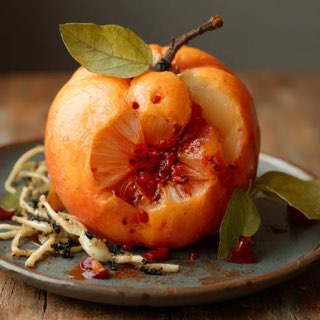} &
\includegraphics[width=0.07\textwidth]{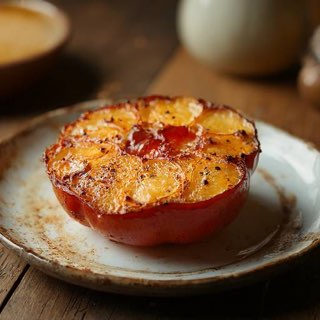} &
\includegraphics[width=0.07\textwidth]{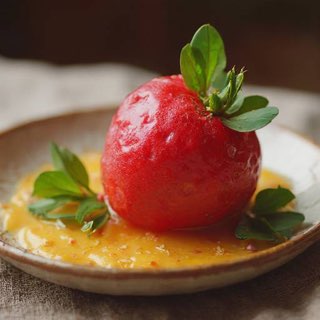} &
\includegraphics[width=0.07\textwidth]{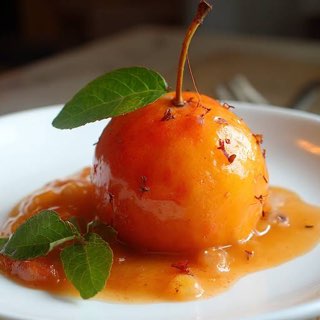} &
\includegraphics[width=0.07\textwidth]{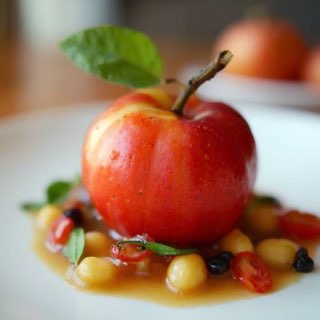} &
\includegraphics[width=0.07\textwidth]{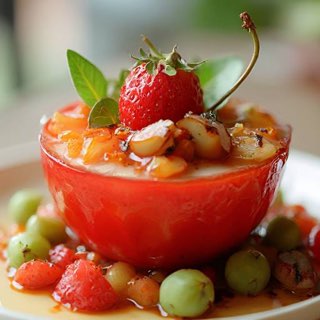} \\
\bottomrule
\end{tabular}
\caption{Qualitative results (Part 1): Flux, Kandinsky, and SD1.5 families.}
\label{fig:qual_part1_name}
\end{figure*}

\begin{figure*}[p]
\centering
\small
\setlength{\tabcolsep}{2pt}
\begin{tabular}{l*{10}{c}}
\toprule
\multicolumn{11}{c}{\textbf{Prompt: ``\prompttitle''}} \\
\textbf{Model} & \textbf{S1} & \textbf{S2} & \textbf{S3} & \textbf{S4} & \textbf{S5} & \textbf{S6} & \textbf{S7} & \textbf{S8} & \textbf{S9} & \textbf{S10} \\
\midrule
\textit{SD2.1 Family} \\
SD2.1-Base & 
\includegraphics[width=0.07\textwidth]{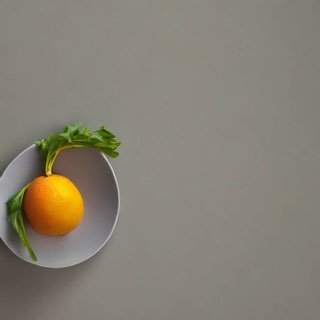} &
\includegraphics[width=0.07\textwidth]{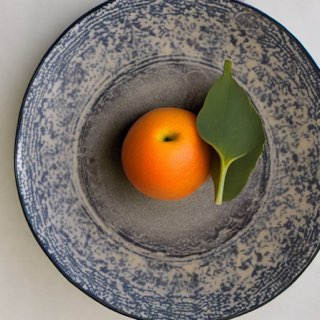} &
\includegraphics[width=0.07\textwidth]{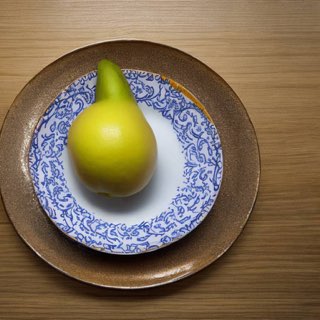} &
\includegraphics[width=0.07\textwidth]{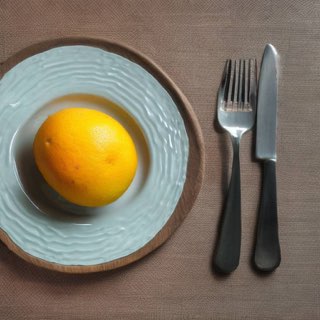} &
\includegraphics[width=0.07\textwidth]{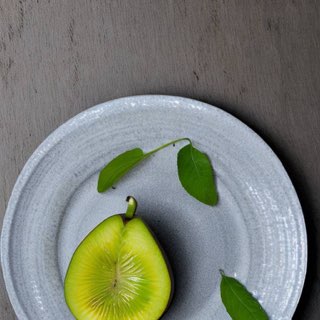} &
\includegraphics[width=0.07\textwidth]{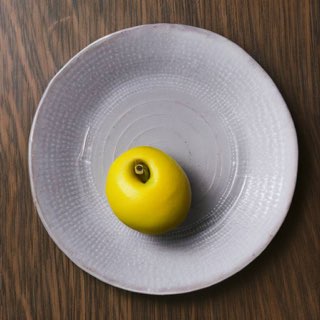} &
\includegraphics[width=0.07\textwidth]{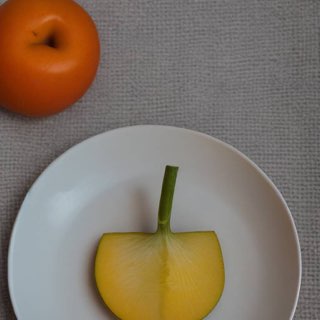} &
\includegraphics[width=0.07\textwidth]{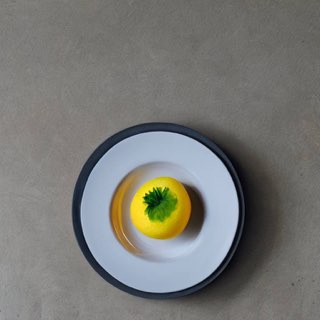} &
\includegraphics[width=0.07\textwidth]{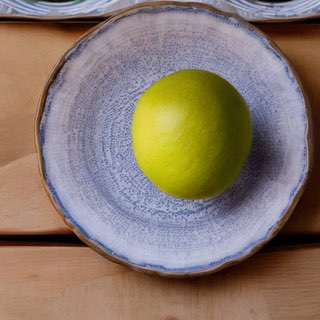} &
\includegraphics[width=0.07\textwidth]{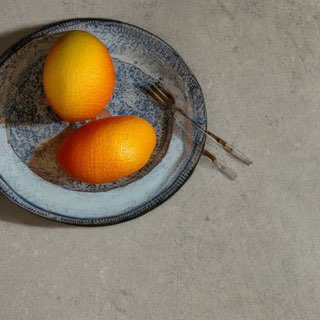} \\
SD2.1-DPO & 
\includegraphics[width=0.07\textwidth]{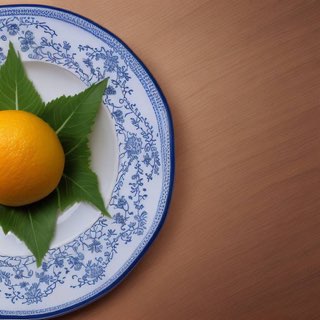} &
\includegraphics[width=0.07\textwidth]{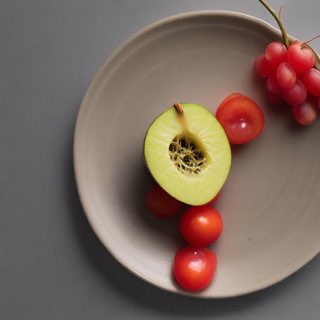} &
\includegraphics[width=0.07\textwidth]{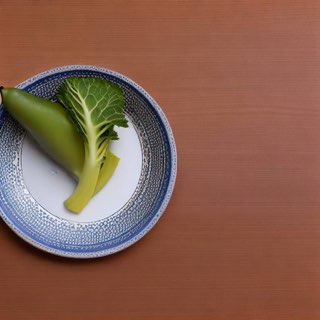} &
\includegraphics[width=0.07\textwidth]{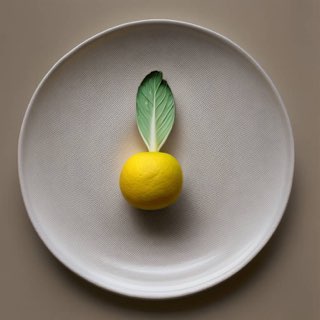} &
\includegraphics[width=0.07\textwidth]{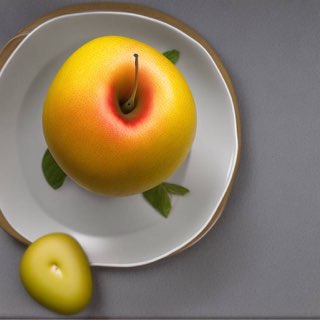} &
\includegraphics[width=0.07\textwidth]{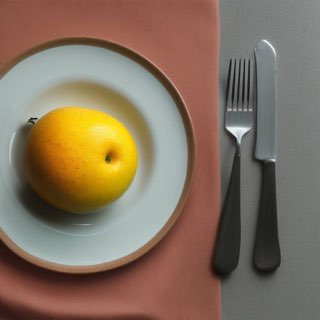} &
\includegraphics[width=0.07\textwidth]{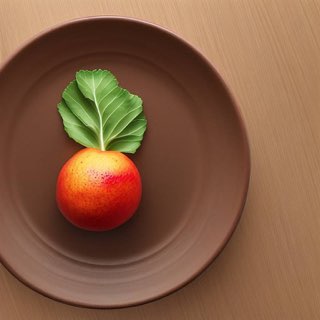} &
\includegraphics[width=0.07\textwidth]{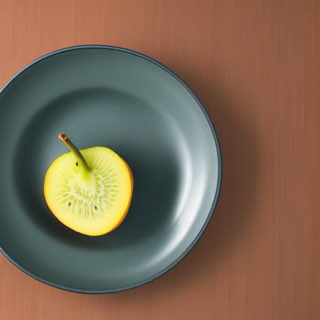} &
\includegraphics[width=0.07\textwidth]{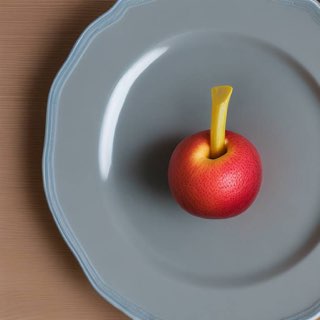} &
\includegraphics[width=0.07\textwidth]{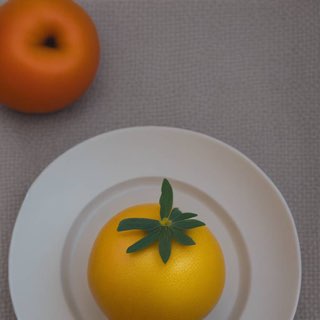} \\
\bottomrule
\end{tabular}
\caption{Qualitative results (Part 2): SD2.1, SD3, and SDXL families.}
\label{fig:qual_part2_name}
\end{figure*}
\begin{figure*}[p]
\centering
\small
\setlength{\tabcolsep}{2pt}
\begin{tabular}{l*{10}{c}}
\toprule
\multicolumn{11}{c}{\textbf{Prompt: ``\animaltitle''}} \\
\midrule
\textbf{Model} & \textbf{S1} & \textbf{S2} & \textbf{S3} & \textbf{S4} & \textbf{S5} & \textbf{S6} & \textbf{S7} & \textbf{S8} & \textbf{S9} & \textbf{S10} \\
\midrule
\textit{Flux Family} \\
Flux-Base & 
\includegraphics[width=0.08\textwidth]{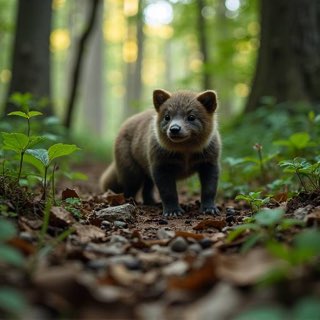} &
\includegraphics[width=0.08\textwidth]{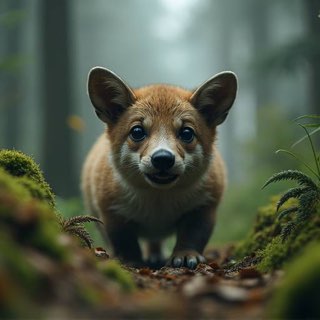} &
\includegraphics[width=0.08\textwidth]{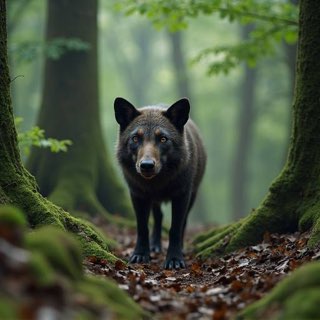} &
\includegraphics[width=0.08\textwidth]{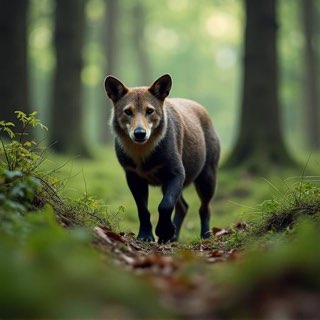} &
\includegraphics[width=0.08\textwidth]{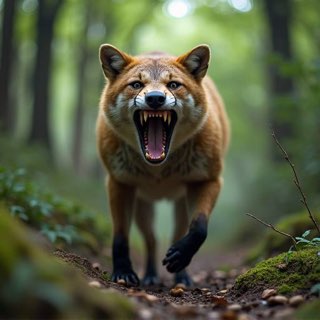} &
\includegraphics[width=0.08\textwidth]{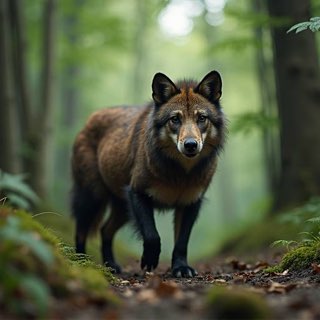} &
\includegraphics[width=0.08\textwidth]{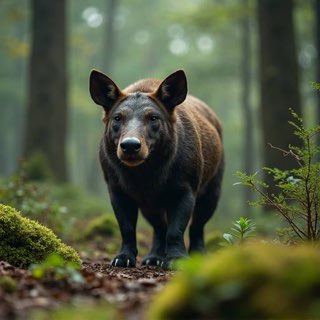} &
\includegraphics[width=0.08\textwidth]{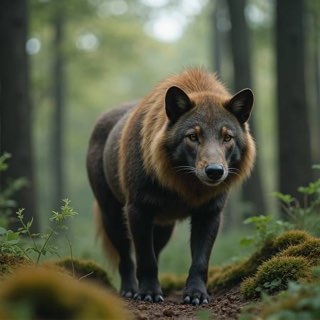} &
\includegraphics[width=0.08\textwidth]{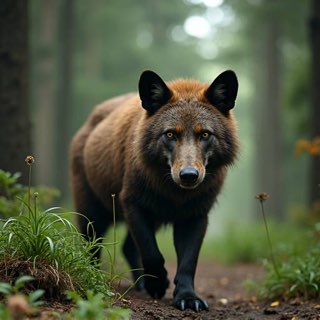} &
\includegraphics[width=0.08\textwidth]{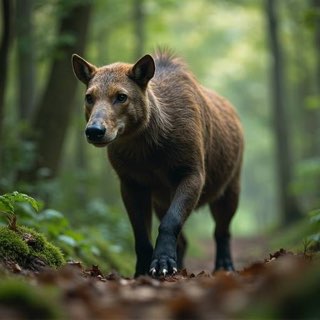} \\
Flux-LoRA & 
\includegraphics[width=0.08\textwidth]{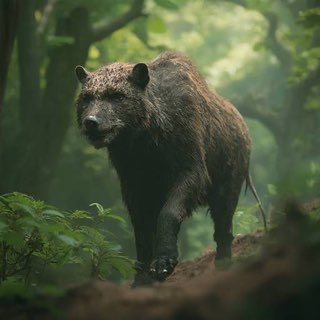} &
\includegraphics[width=0.08\textwidth]{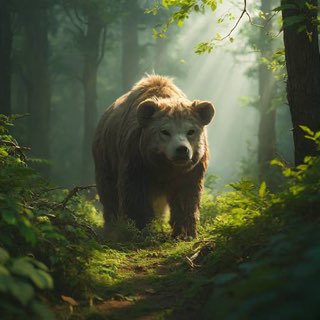} &
\includegraphics[width=0.08\textwidth]{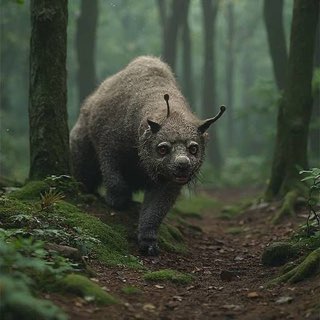} &
\includegraphics[width=0.08\textwidth]{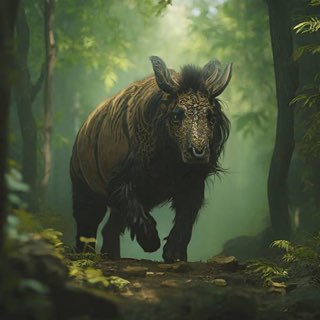} &
\includegraphics[width=0.08\textwidth]{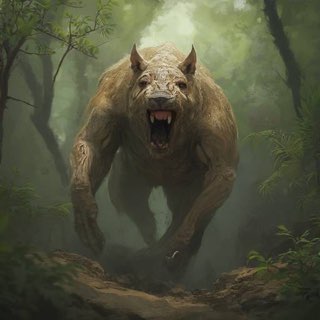} &
\includegraphics[width=0.08\textwidth]{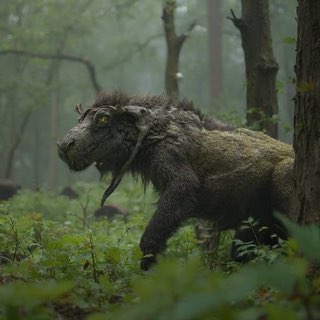} &
\includegraphics[width=0.08\textwidth]{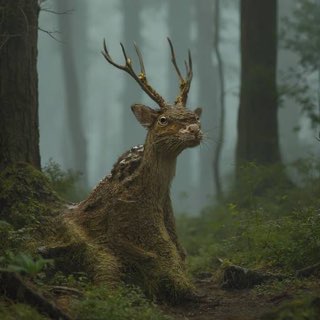} &
\includegraphics[width=0.08\textwidth]{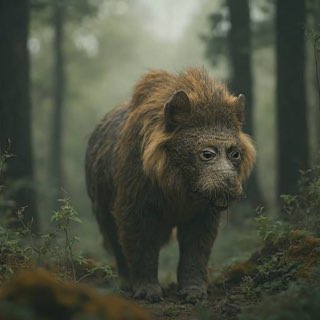} &
\includegraphics[width=0.08\textwidth]{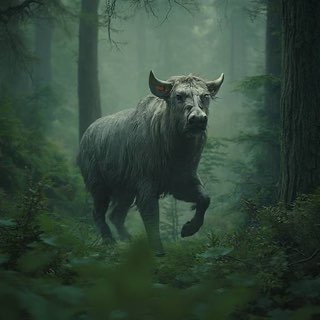} &
\includegraphics[width=0.08\textwidth]{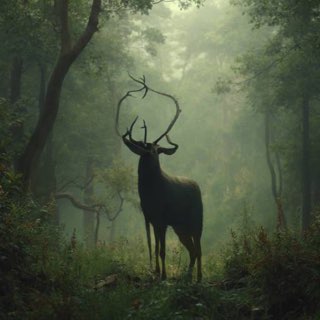} \\
\bottomrule
\end{tabular}
\caption{Qualitative results (Part 1): Flux, Kandinsky, and SD1.5 families.}
\label{fig:qual_part1}
\end{figure*}

\begin{figure*}[p]
\centering
\small
\setlength{\tabcolsep}{2pt}
\begin{tabular}{l*{10}{c}}
\toprule
\multicolumn{11}{c}{\textbf{Prompt: ``\animaltitle''}} \\
\textbf{Model} & \textbf{S1} & \textbf{S2} & \textbf{S3} & \textbf{S4} & \textbf{S5} & \textbf{S6} & \textbf{S7} & \textbf{S8} & \textbf{S9} & \textbf{S10} \\
\midrule
\textit{SD2.1 Family} \\
SD2.1-Base & 
\includegraphics[width=0.08\textwidth]{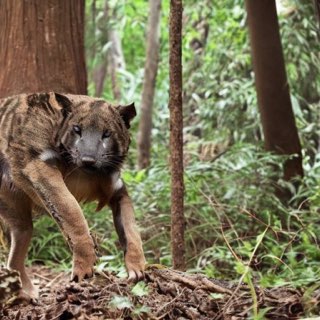} &
\includegraphics[width=0.08\textwidth]{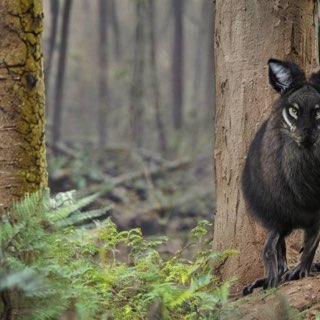} &
\includegraphics[width=0.08\textwidth]{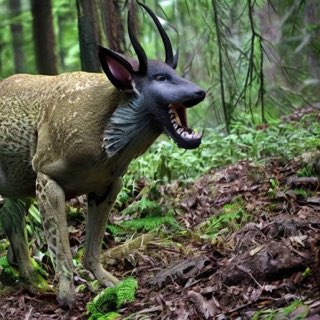} &
\includegraphics[width=0.08\textwidth]{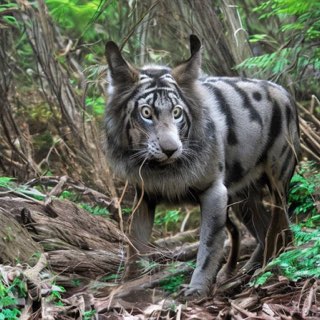} &
\includegraphics[width=0.08\textwidth]{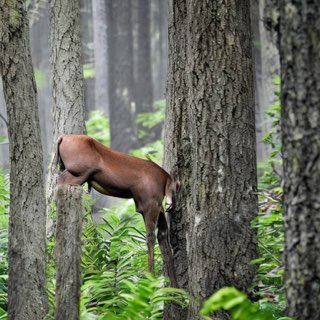} &
\includegraphics[width=0.08\textwidth]{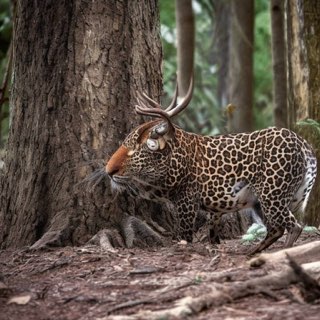} &
\includegraphics[width=0.08\textwidth]{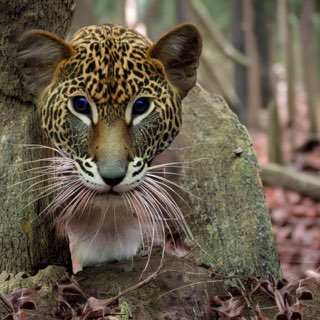} &
\includegraphics[width=0.08\textwidth]{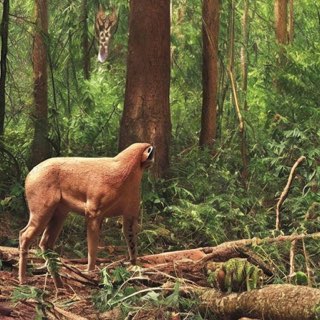} &
\includegraphics[width=0.08\textwidth]{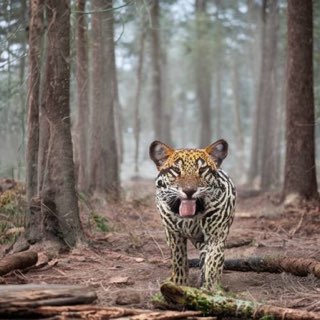} &
\includegraphics[width=0.08\textwidth]{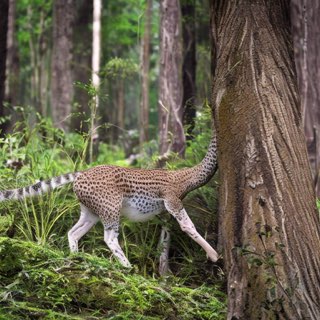} \\
SD2.1-DPO & 
\includegraphics[width=0.08\textwidth]{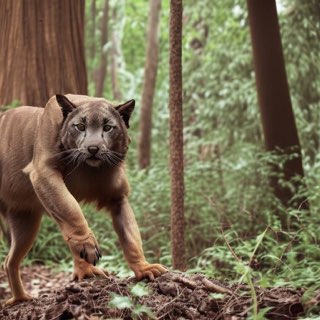} &
\includegraphics[width=0.08\textwidth]{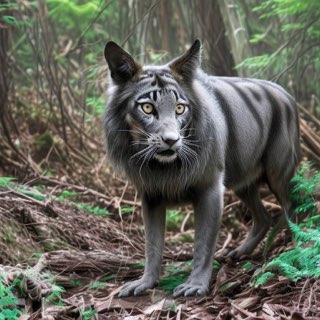} &
\includegraphics[width=0.08\textwidth]{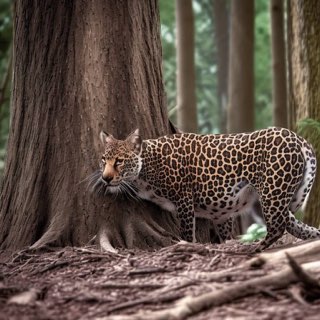} &
\includegraphics[width=0.08\textwidth]{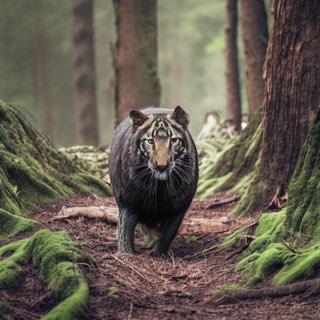} &
\includegraphics[width=0.08\textwidth]{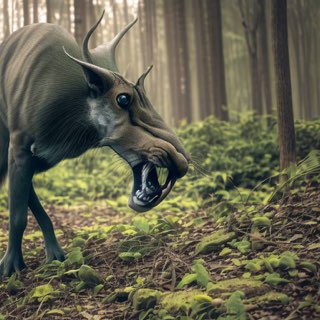} &
\includegraphics[width=0.08\textwidth]{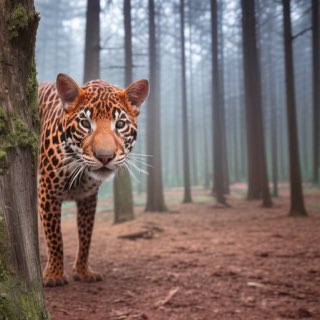} &
\includegraphics[width=0.08\textwidth]{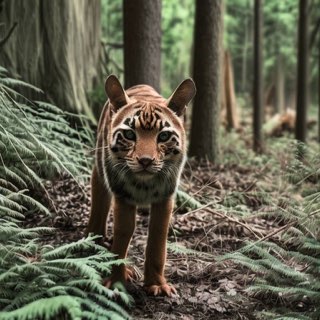} &
\includegraphics[width=0.08\textwidth]{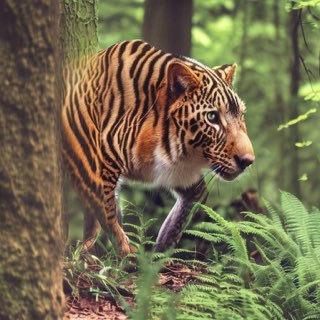} &
\includegraphics[width=0.08\textwidth]{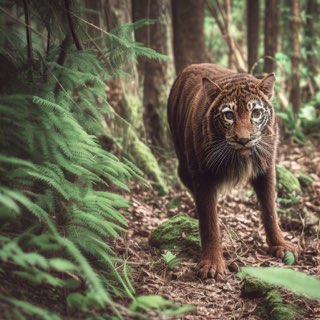} &
\includegraphics[width=0.08\textwidth]{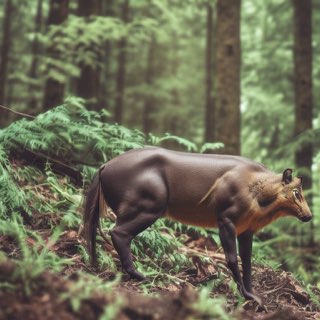} \\
\bottomrule
\end{tabular}
\caption{Qualitative results (Part 2): SD2.1, SD3, and SDXL families.}
\label{fig:qual_part2}
\end{figure*}
\begin{figure*}[p]
\centering
\small
\setlength{\tabcolsep}{2pt}
\begin{tabular}{l*{10}{c}}
\toprule
\multicolumn{11}{c}{\textbf{Prompt: ``\birdtitle''}} \\
\midrule
\textbf{Model} & \textbf{S1} & \textbf{S2} & \textbf{S3} & \textbf{S4} & \textbf{S5} & \textbf{S6} & \textbf{S7} & \textbf{S8} & \textbf{S9} & \textbf{S10} \\
\midrule
\textit{Flux Family} \\
Flux-Base & 
\includegraphics[width=0.08\textwidth]{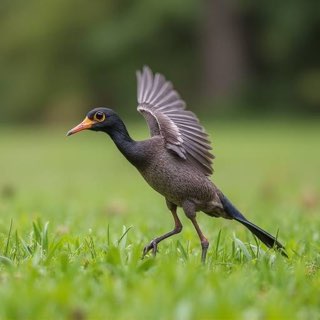} &
\includegraphics[width=0.08\textwidth]{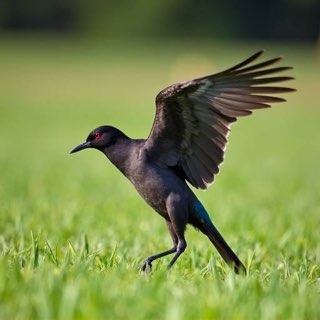} &
\includegraphics[width=0.08\textwidth]{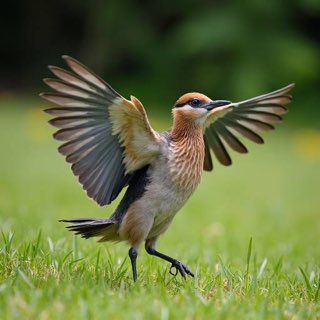} &
\includegraphics[width=0.08\textwidth]{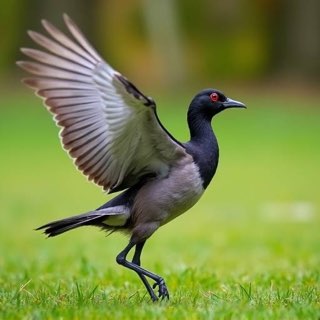} &
\includegraphics[width=0.08\textwidth]{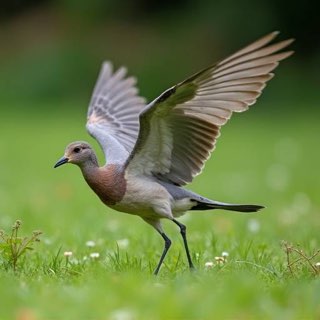} &
\includegraphics[width=0.08\textwidth]{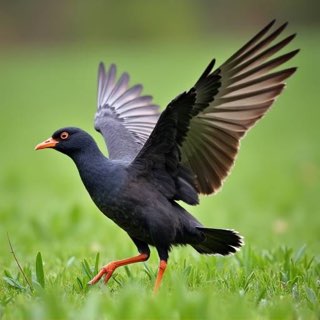} &
\includegraphics[width=0.08\textwidth]{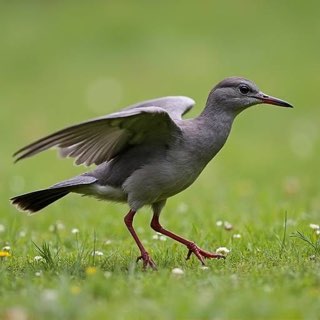} &
\includegraphics[width=0.08\textwidth]{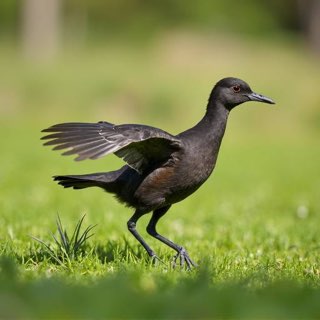} &
\includegraphics[width=0.08\textwidth]{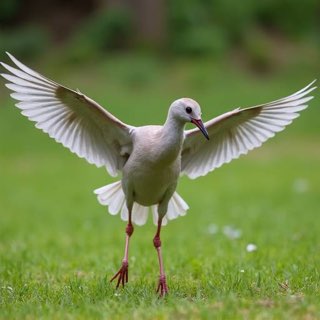} &
\includegraphics[width=0.08\textwidth]{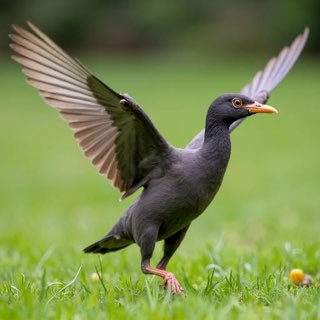} \\
Flux-LoRA & 
\includegraphics[width=0.08\textwidth]{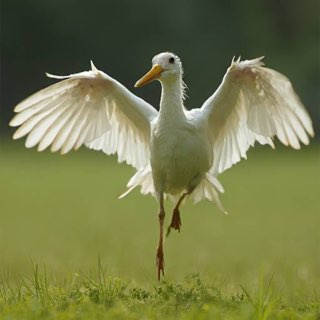} &
\includegraphics[width=0.08\textwidth]{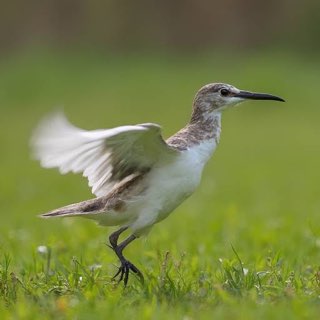} &
\includegraphics[width=0.08\textwidth]{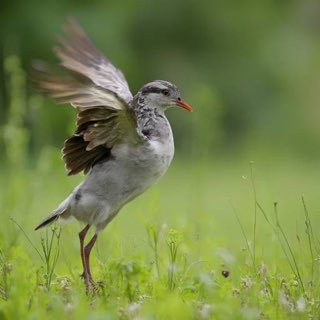} &
\includegraphics[width=0.08\textwidth]{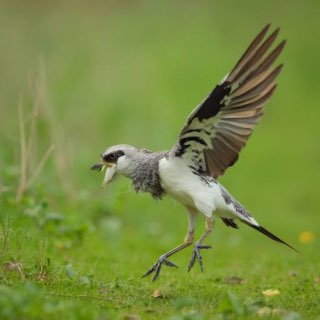} &
\includegraphics[width=0.08\textwidth]{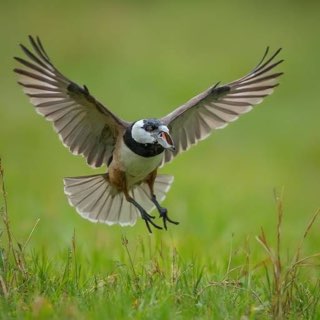} &
\includegraphics[width=0.08\textwidth]{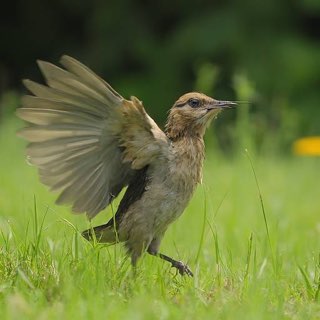} &
\includegraphics[width=0.08\textwidth]{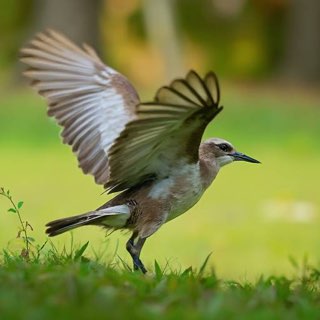} &
\includegraphics[width=0.08\textwidth]{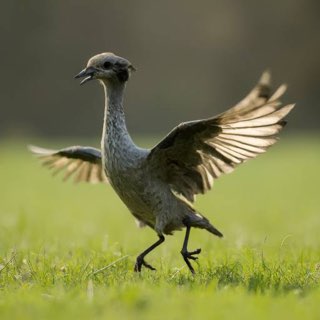} &
\includegraphics[width=0.08\textwidth]{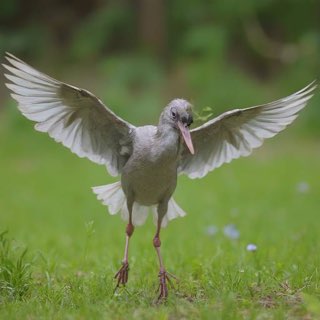} &
\includegraphics[width=0.08\textwidth]{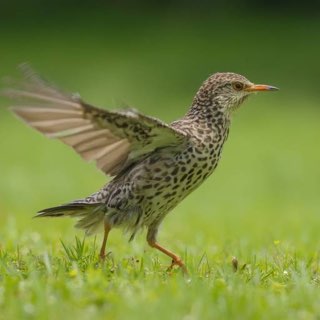} \\
\bottomrule
\end{tabular}
\caption{Qualitative results (Part 1): Flux, Kandinsky, and SD1.5 families.}
\label{fig:qual_part1_bird}
\end{figure*}

\begin{figure*}[p]
\centering
\small
\setlength{\tabcolsep}{2pt}
\begin{tabular}{l*{10}{c}}
\toprule
\multicolumn{11}{c}{\textbf{Prompt: ``\birdtitle''}} \\
\textbf{Model} & \textbf{S1} & \textbf{S2} & \textbf{S3} & \textbf{S4} & \textbf{S5} & \textbf{S6} & \textbf{S7} & \textbf{S8} & \textbf{S9} & \textbf{S10} \\
\midrule
\textit{SD2.1 Family} \\
SD2.1-Base & 
\includegraphics[width=0.08\textwidth]{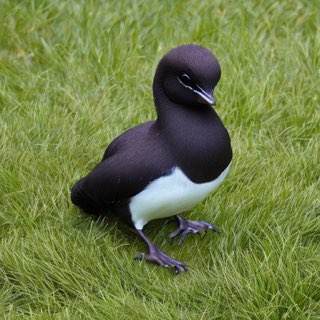} &
\includegraphics[width=0.08\textwidth]{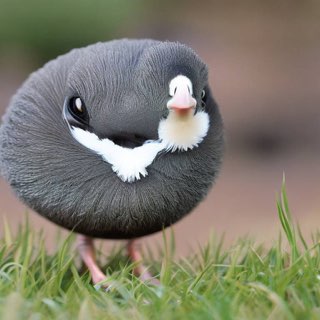} &
\includegraphics[width=0.08\textwidth]{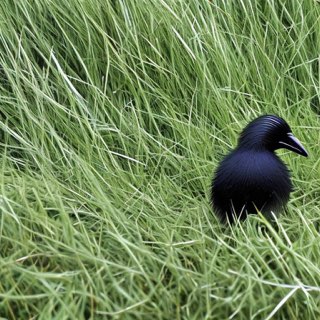} &
\includegraphics[width=0.08\textwidth]{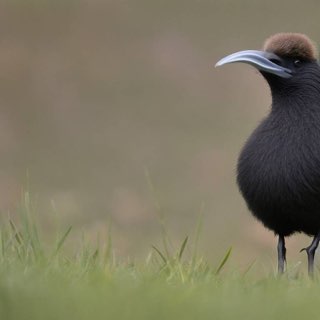} &
\includegraphics[width=0.08\textwidth]{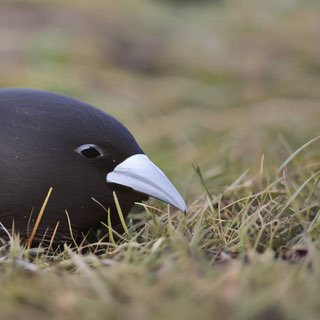} &
\includegraphics[width=0.08\textwidth]{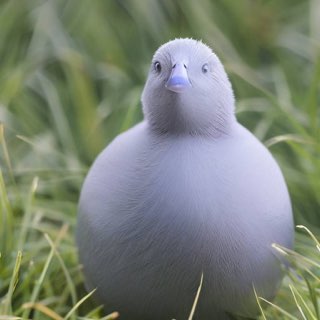} &
\includegraphics[width=0.08\textwidth]{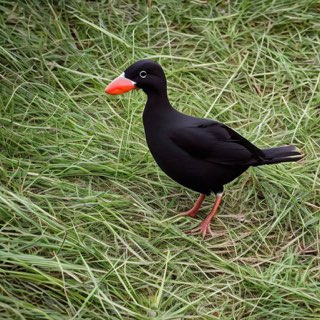} &
\includegraphics[width=0.08\textwidth]{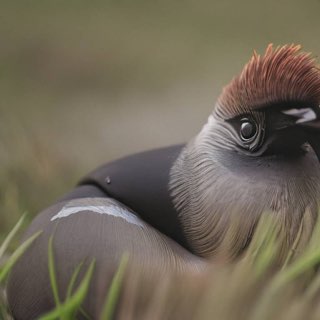} &
\includegraphics[width=0.08\textwidth]{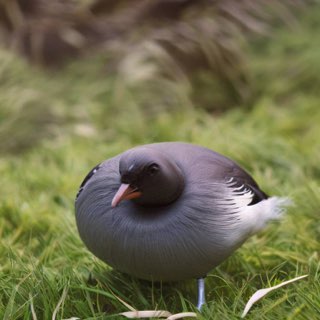} &
\includegraphics[width=0.08\textwidth]{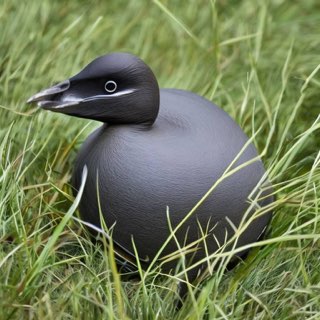} \\
SD2.1-DPO & 
\includegraphics[width=0.08\textwidth]{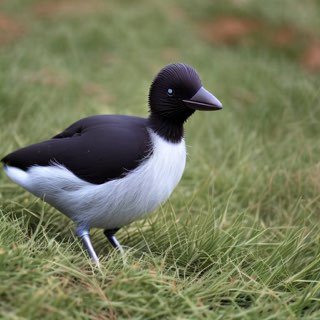} &
\includegraphics[width=0.08\textwidth]{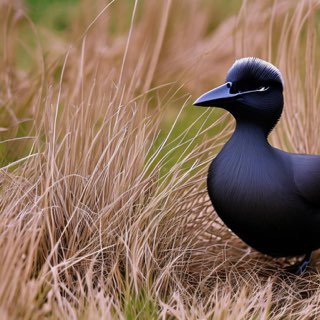} &
\includegraphics[width=0.08\textwidth]{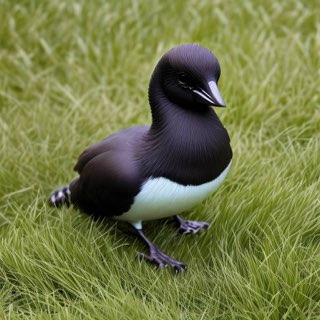} &
\includegraphics[width=0.08\textwidth]{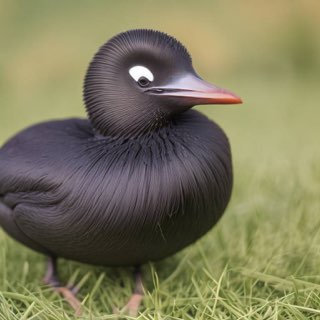} &
\includegraphics[width=0.08\textwidth]{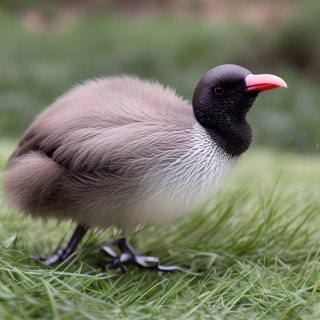} &
\includegraphics[width=0.08\textwidth]{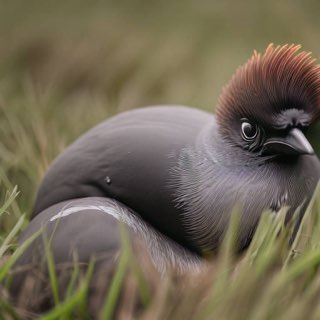} &
\includegraphics[width=0.08\textwidth]{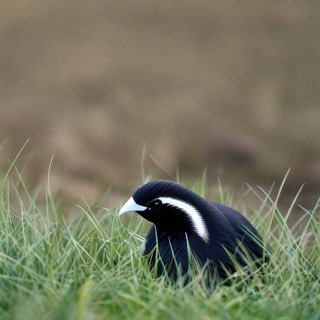} &
\includegraphics[width=0.08\textwidth]{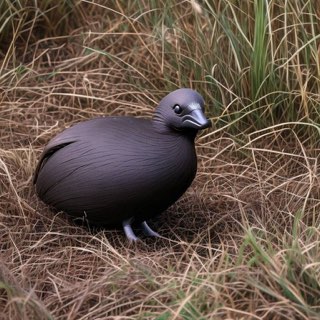} &
\includegraphics[width=0.08\textwidth]{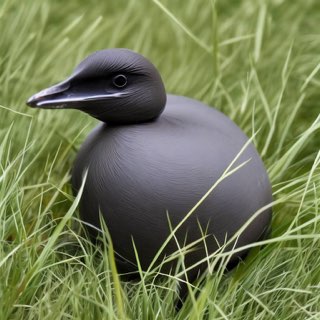} &
\includegraphics[width=0.08\textwidth]{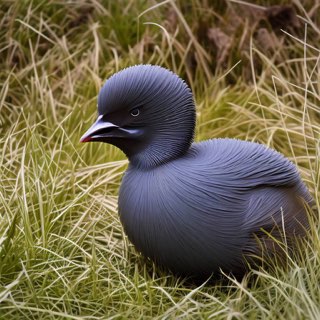} \\
\bottomrule
\end{tabular}
\caption{Qualitative results (Part 2): SD2.1, SD3, and SDXL families.}
\label{fig:qual_part2_bird}
\end{figure*}
\begin{figure*}[p]
\centering
\small
\setlength{\tabcolsep}{2pt}
\begin{tabular}{l*{10}{c}}
\toprule
\multicolumn{11}{c}{\textbf{Prompt: ``\flowertitle''}} \\
\midrule
\textbf{Model} & \textbf{S1} & \textbf{S2} & \textbf{S3} & \textbf{S4} & \textbf{S5} & \textbf{S6} & \textbf{S7} & \textbf{S8} & \textbf{S9} & \textbf{S10} \\
\midrule
\textit{Flux Family} \\
Flux-Base & 
\includegraphics[width=0.08\textwidth]{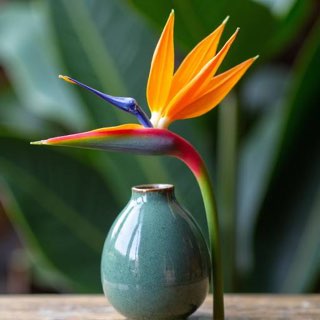} &
\includegraphics[width=0.08\textwidth]{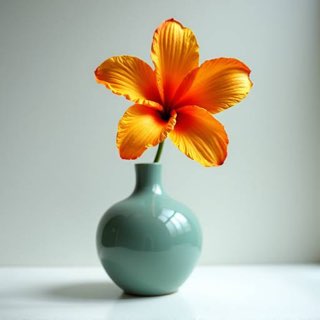} &
\includegraphics[width=0.08\textwidth]{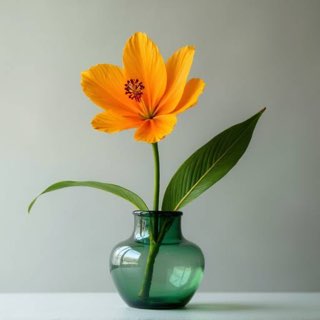} &
\includegraphics[width=0.08\textwidth]{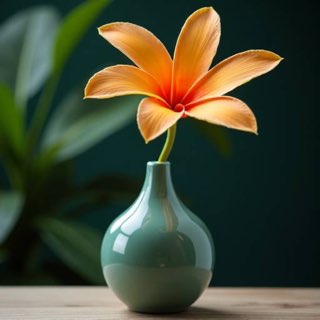} &
\includegraphics[width=0.08\textwidth]{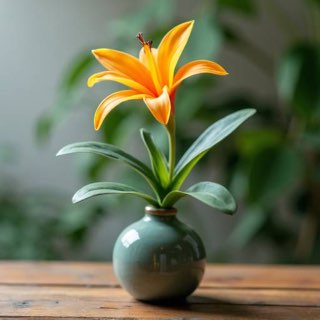} &
\includegraphics[width=0.08\textwidth]{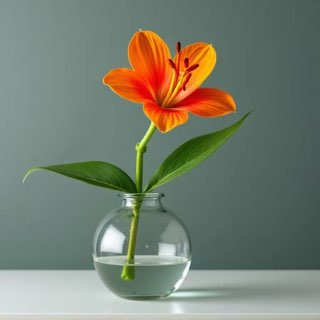} &
\includegraphics[width=0.08\textwidth]{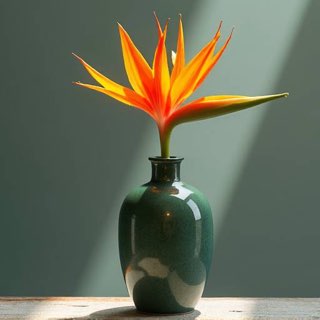} &
\includegraphics[width=0.08\textwidth]{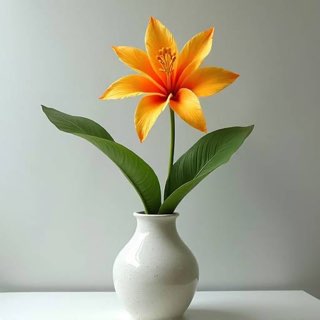} &
\includegraphics[width=0.08\textwidth]{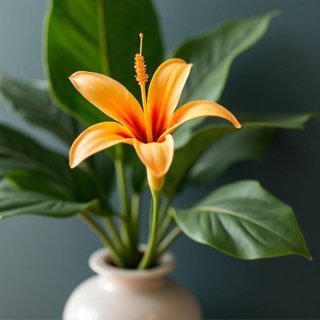} &
\includegraphics[width=0.08\textwidth]{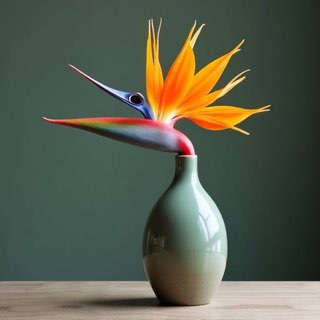} \\
Flux-LoRA & 
\includegraphics[width=0.08\textwidth]{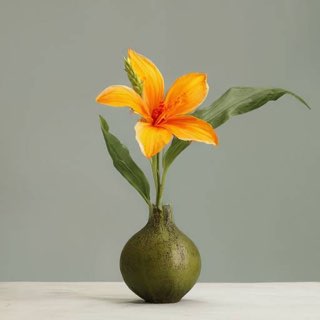} &
\includegraphics[width=0.08\textwidth]{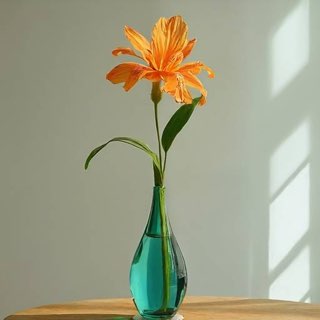} &
\includegraphics[width=0.08\textwidth]{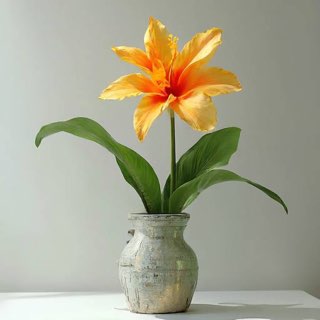} &
\includegraphics[width=0.08\textwidth]{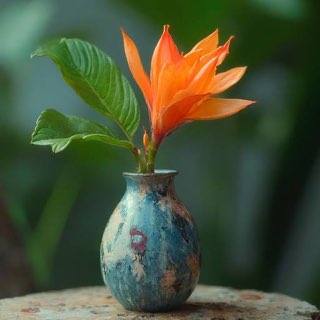} &
\includegraphics[width=0.08\textwidth]{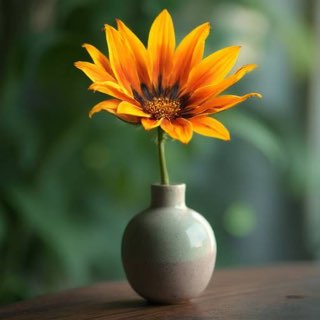} &
\includegraphics[width=0.08\textwidth]{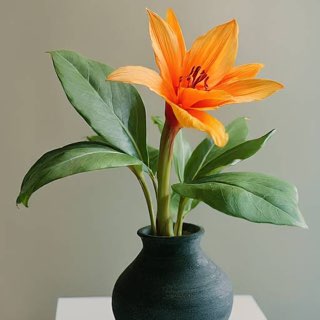} &
\includegraphics[width=0.08\textwidth]{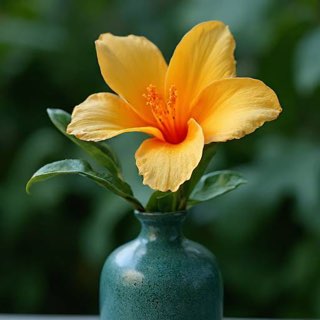} &
\includegraphics[width=0.08\textwidth]{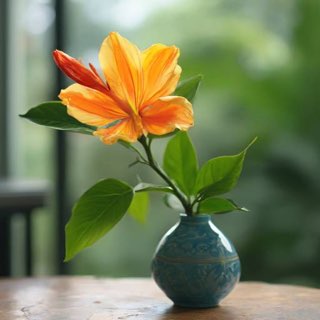} &
\includegraphics[width=0.08\textwidth]{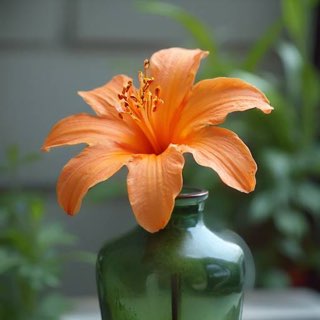} &
\includegraphics[width=0.08\textwidth]{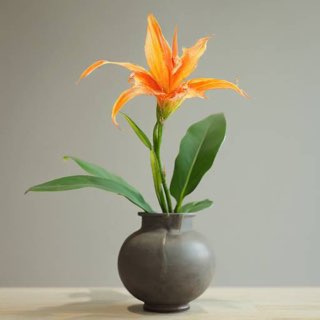} \\
\bottomrule
\end{tabular}
\caption{Qualitative results (Part 1): Flux, Kandinsky, and SD1.5 families.}
\label{fig:qual_part1_flower}
\end{figure*}

\begin{figure*}[p]
\centering
\small
\setlength{\tabcolsep}{2pt}
\begin{tabular}{l*{10}{c}}
\toprule
\multicolumn{11}{c}{\textbf{Prompt: ``\flowertitle''}} \\
\textbf{Model} & \textbf{S1} & \textbf{S2} & \textbf{S3} & \textbf{S4} & \textbf{S5} & \textbf{S6} & \textbf{S7} & \textbf{S8} & \textbf{S9} & \textbf{S10} \\
\midrule
\textit{SD2.1 Family} \\
SD2.1-Base & 
\includegraphics[width=0.08\textwidth]{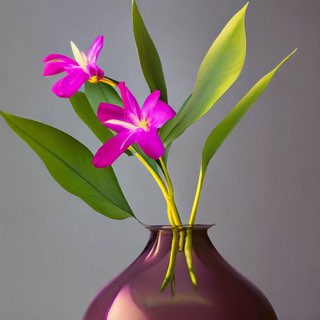} &
\includegraphics[width=0.08\textwidth]{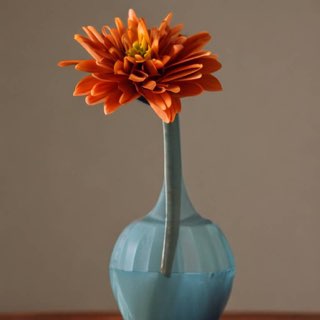} &
\includegraphics[width=0.08\textwidth]{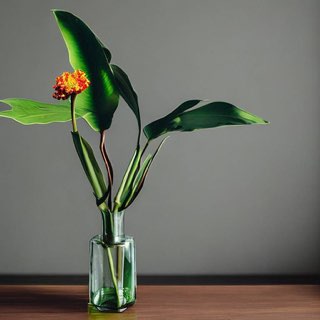} &
\includegraphics[width=0.08\textwidth]{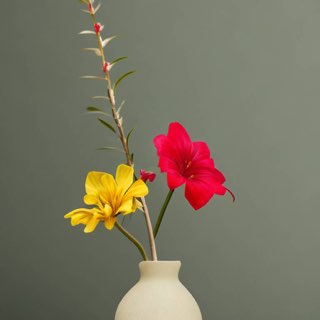} &
\includegraphics[width=0.08\textwidth]{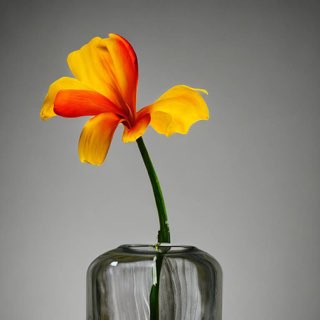} &
\includegraphics[width=0.08\textwidth]{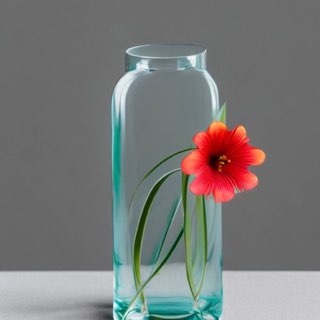} &
\includegraphics[width=0.08\textwidth]{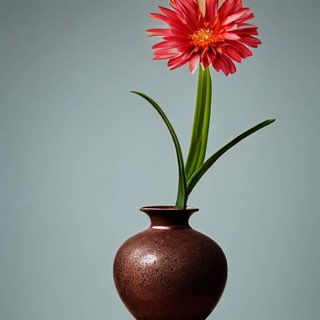} &
\includegraphics[width=0.08\textwidth]{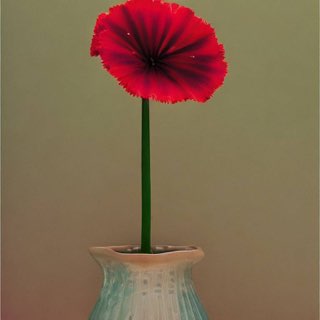} &
\includegraphics[width=0.08\textwidth]{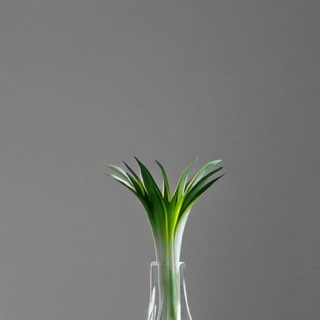} &
\includegraphics[width=0.08\textwidth]{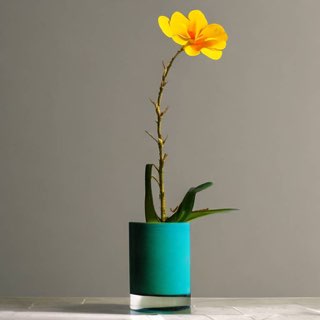} \\
SD2.1-DPO & 
\includegraphics[width=0.08\textwidth]{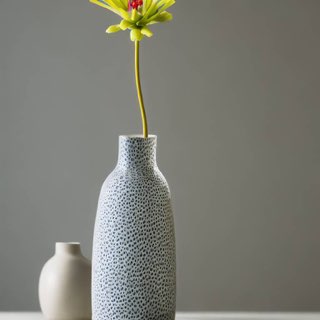} &
\includegraphics[width=0.08\textwidth]{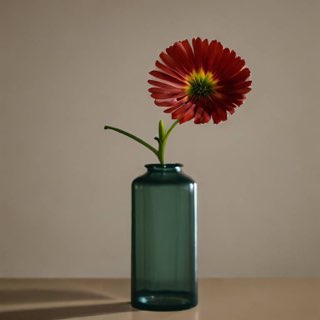} &
\includegraphics[width=0.08\textwidth]{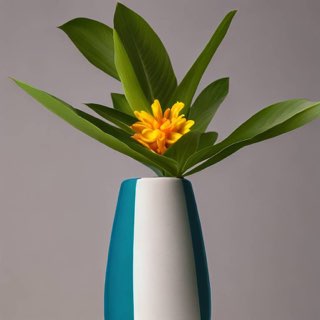} &
\includegraphics[width=0.08\textwidth]{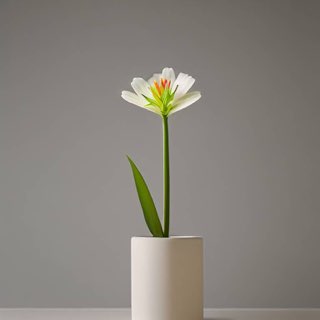} &
\includegraphics[width=0.08\textwidth]{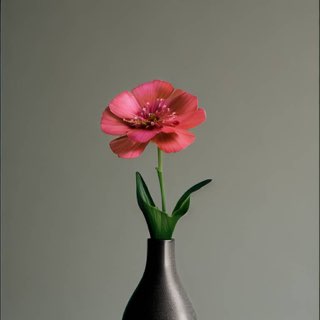} &
\includegraphics[width=0.08\textwidth]{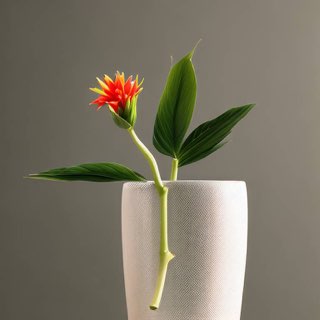} &
\includegraphics[width=0.08\textwidth]{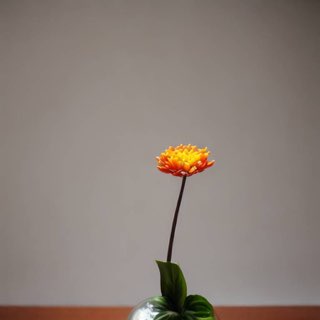} &
\includegraphics[width=0.08\textwidth]{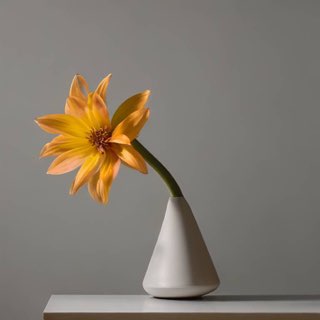} &
\includegraphics[width=0.08\textwidth]{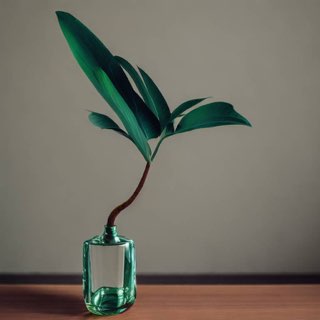} &
\includegraphics[width=0.08\textwidth]{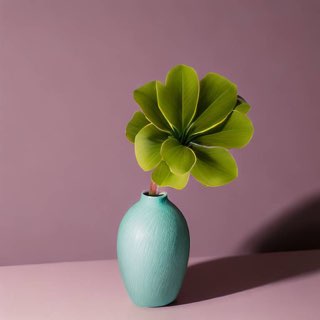} \\
\bottomrule
\end{tabular}
\caption{Qualitative results (Part 2): SD2.1, SD3, and SDXL families.}
\label{fig:qual_part2_flower}
\end{figure*}

\clearpage

\subsection{Complete Prompt Enumeration}
\label{sec:full_prompts}

\paragraph{Baked Goods (9 prompts)}
\begin{enumerate}[noitemsep,topsep=0pt]
    \item A photo of a savory baked good on a dimmed studio.
    \item A photo of a savory baked good on a dark wood surface.
    \item A photo of a savory baked good against a brick wall.
    \item A photo of a cheesy baked good on a dimmed studio.
    \item A photo of a cheesy baked good on a dark wood surface.
    \item A photo of a cheesy baked good against a brick wall.
    \item A photo of a sweet baked good on a dimmed studio.
    \item A photo of a sweet baked good on a dark wood surface.
    \item A photo of a sweet baked good against a brick wall.
\end{enumerate}

\paragraph{Animals (9 prompts)}
\begin{enumerate}[noitemsep,topsep=0pt]
    \item A photo of a dangerous animal in a grassland.
    \item A photo of a dangerous animal in a forest.
    \item A photo of a dangerous animal in a dimmed studio.
    \item A photo of a wild animal in a grassland.
    \item A photo of a wild animal in a forest.
    \item A photo of a wild animal in a dimmed studio.
    \item A photo of a peaceful animal in a grassland.
    \item A photo of a peaceful animal in a forest.
    \item A photo of a peaceful animal in a dimmed studio.
\end{enumerate}

\paragraph{Flowers (6 prompts)}
\begin{enumerate}[noitemsep,topsep=0pt]
    \item A photo of a vibrant single flower on a pot.
    \item A photo of a vibrant single flower in a dimmed studio.
    \item A photo of a vibrant single flower on a vase.
    \item A photo of a tropical single flower on a pot.
    \item A photo of a tropical single flower in a dimmed studio.
    \item A photo of a tropical single flower on a vase.
\end{enumerate}

\paragraph{Birds (9 prompts)}
\begin{enumerate}[noitemsep,topsep=0pt]
    \item A photo of an peaceful bird on a grass.
    \item A photo of an peaceful bird on a savana.
    \item A photo of an peaceful bird in a dimmed studio.
    \item A photo of an dangerous bird on a grass.
    \item A photo of an dangerous bird on a savana.
    \item A photo of an dangerous bird in a dimmed studio.
    \item A photo of an flightless bird on a grass.
    \item A photo of an flightless bird on a savana.
    \item A photo of an flightless bird in a dimmed studio.
\end{enumerate}

\paragraph{Fruits (9 prompts)}
\begin{enumerate}[noitemsep,topsep=0pt]
    \item A photo of a sweet single fruit on a dish.
    \item A photo of a sweet single fruit on a wooden floor.
    \item A photo of a sweet single fruit on a dimmed studio.
    \item A photo of a frozen single fruit on a dish.
    \item A photo of a frozen single fruit on a wooden floor.
    \item A photo of a frozen single fruit on a dimmed studio.
    \item A photo of a savory single fruit on a dish.
    \item A photo of a savory single fruit on a wooden floor.
    \item A photo of a savory single fruit on a dimmed studio.
\end{enumerate}

\end{document}